\documentclass[universe,article,accept,moreauthors,pdftex]{Definitions/mdpi} 

\firstpage{1} 
\pubvolume{8}
\issuenum{1}
\articlenumber{34}
\pubyear{2022}
\copyrightyear{2022}
\externaleditor{Academic Editor: {Veronica Dexheimer and Rodrigo Negreiros}} 
\datereceived{10 November 2021} 
\dateaccepted{31 December 2021} 
\datepublished{5 January 2022} 
\hreflink{https://doi.org/10.3390/\linebreak universe8010034} 

\Title{Radial Oscillations of Quark Stars Admixed with Dark Matter}

\TitleCitation{Radial Oscillations of Quark Stars Admixed with Dark Matter}

\Author{Jos\'{e} C. Jim\'{e}nez $^{1}$\orcidA{} and Eduardo S. Fraga $^{2,}$* \orcidB{}}

\AuthorNames{José C. Jiménez and Eduardo S. Fraga}

\AuthorCitation{Jiménez, J.C.; Fraga, E.S.}

\address{%
$^{1}$ \quad Instituto de F\'isica, Universidade de S\~ao Paulo, S\~ao Paulo 05508-090, Brazil; jimenez@if.usp.br\\

$^{2}$ \quad Instituto de F\'\i sica, Universidade Federal do Rio de Janeiro,  Rio de Janeiro 21941-972, Brazil}

\corres{Correspondence: fraga@if.ufrj.br}

\abstract{We investigated compact stars consisting of cold quark matter and fermionic dark matter treated as two admixed fluids. We computed the stellar structures and fundamental radial oscillation frequencies of different masses of the dark fermion in the cases of weak and strong self-interacting dark matter. We found that the fundamental frequency can be dramatically modified and, in some cases, stable dark strange planets and dark strangelets with very low masses and radii can be formed.}

\keyword{quark stars; dark matter; radial oscillations} 

\begin{document}

\section{Introduction}
	\label{sec:introduction}

Compact stars offer a variety of possibilities for probing the inner structure of matter through astronomical observations. In particular, matter at extremely high densities can only be probed, so far, by investigating unique objects that represent one of  the possible final stages of stellar evolution. The structure of compact stars can be determined by solving the Tolman--Oppenheimer--Volkov (TOV) equations, given the equation of state (EoS) for the matter under consideration \cite{Glendenning:2000,Schaffner-Bielich:2020psc}. 

For high enough central energy densities, one expects to find either hybrid stars, i.e., neutron stars with a quark matter (QM) core, or even more exotic objects, such as quark stars. Quark stars \cite{Ivanenko:1965dg} and their structure \cite{Itoh:1970uw} have been considered for more than half a century, even before the elaboration of quantum chromodynamics (QCD) in the 1970s. Later, after a seminal work by Witten \cite{Witten:1984rs}, a rich phenomenology of self-bound strange stars~\cite{Alcock:1986hz,Haensel:1986qb} and quark (hybrid) stars emerged using the MIT bag model \cite{Farhi:1984qu} as a framework for the EoS at high densities. For a review on quark matter in neutron stars, see Ref. \cite{Buballa:2014jta}.

On the other hand, dark matter (DM) represents about a quarter of the total mass--energy density content of the universe or, equivalently, $\sim$85\% of its matter content. Apart from this, it is needed to explain structure formations without modifying general relativity in the current cosmological standard model \cite{Bertone:2004pz,SDSS:2014iwm,Planck:2013pxb}. Nevertheless, there is still no experimental evidence of DM-constituent particles, and its nature remains one of the greatest mysteries of particle physics. Over the years, many candidates have been proposed as being DM-constituent particles
, with masses ranging from $10^{-33}$ GeV to $10^{15}$ GeV, including weakly interacting massive particles (WIMPs), axions and axion-like particles (ALPs), sterile neutrinos, neutralinos, and so on \cite{Deliyergiyev:2019vti}. In spite of their nature, if DM-constituent particles 
 do not self annihilate and are non-relativistic at freeze-out (cold dark matter (CDM)), the probability of their interaction with ordinary baryonic matter will increase within the extreme densities found in compact stars. In this case, DM can accumulate and thermalize in a small radius. So, if quark stars are to be found in the universe, they have most likely accumulated some amount of dark matter over the course of their lives.

In this paper, we investigate strange quark stars consisting of cold QM and non self-annihilating fermionic cold DM treated as two admixed fluids, attracted only gravitationally. As our main goal, we aim to compute the fundamental radial oscillation frequency of admixed QM and DM {\it two-fluid} stars for a relevant range of masses of the dark fermion. We consider the cases of weak and strong self-interacting DM and also study how the total mass and radius of quark and dark stars are modified by their mutual presence in the admixed star.

We describe the QM component on the framework of the MIT bag model, which represents a choice for simplicity, which was mainly motivated by the possibility of direct comparison to previous work. Moreover, for analyses that depend on a range of values for the dark fermion mass and the intensity of DM self-interaction, it is  convenient to avoid other bands in parameters that would come about naturally in more realistic descriptions of the EoS, such as those relying on perturbative QCD \cite{Fraga:2001id,Fraga:2004gz,Kurkela:2009gj,Fraga:2013qra,Kurkela:2014vha,Fraga:2015xha,Ghisoiu:2016swa,Annala:2017llu,Gorda:2018gpy,Annala:2019puf,Gorda:2021kme}.

Neutron stars and quark stars admixed with DM have  previously been considered. The effects of fermionic and bosonic DM on the equilibrium features and radial oscillations of neutron stars (NS) have been discussed, e.g., in Refs. \cite{Ciarcelluti:2010ji,Leung:2011zz,Leung:2012vea,Li:2012ii,Xiang:2013xwa,Ellis:2018bkr,Karkevandi:2021ygv} (see Ref. \cite{DelPopolo:2020hel} for a more complete list of references). Reference \cite{Tolos:2015qra} considers hybrid NSs with an EoS for neutron star matter that uses perturbative QCD and effective field theory as high and low-density descriptions, respectively, and polytropes as interpolating functions, as discussed in Ref.~\cite{Kurkela:2014vha}\endnote{
In fact, the EoS considered corresponds to the most compact of three cases presented in tabulated format in Ref. \cite{Kurkela:2014vha}.}, in addition to taking into account inner and outer crusts. The authors also consider white dwarfs admixed with asymmetric DM and find dark, compact (Jupiter-like) planets and limits on the DM content of the stars in order to satisfy the two-solar mass observational constraint \cite{Demorest:2010bx,Antoniadis:2013pzd}. A stability analysis is also performed by solving the usual Sturm--Liouville problem for one-fluid stars \citep{Chandrasekhar:1964zza,Shapiro:1983du}. Reference \cite{Deliyergiyev:2019vti} extends these results to a wider range of dark fermion masses, from $1$ GeV to $500$ GeV,  assuming different amounts of DM at the stellar center. From this analysis, the authors inferred that the total mass decreases with $m_D$, putting constraints on $m_D$ and on DM capture. 

Quark stars admixed with dark matter have been discussed in Ref. \cite{Mukhopadhyay:2015xhs}, where the authors use the MIT bag model to describe the EoS for QM admixed with DM made of dark fermions of mass $m_D=100$ GeV (on the typical WIMP mass scale) 
. They considered two cases: free and strongly self-interacting DM. Solving the TOV equations with two fluids that interact only gravitationally, they found minor modifications to the maximum mass and radius, of the order of a few percent, though with higher values of the central energy density due to a greater gravitational pull. 

So far, the stability against radial oscillations of quark stars admixed with dark matter has been studied using one-fluid formalism, usually with simplified interaction terms and unphysical dark-fermion masses  \cite{Panotopoulos:2017eig,Panotopoulos:2018ipq}. A complete treatment of the stability of two-fluid stars requires a non-trivial extension of the Chandrasekhar second-order differential equation, where a coupled system of equations should be solved for the corresponding Lagrangian displacements associated with each fluid, but which also depend on the displacement of the other  fluid 
\cite{Leung:2011zz,Leung:2012vea,Kain:2020zjs,Kain:2021hpk}, as will be discussed in the sequel.

The paper is organized as follows. In Section \ref{sec:framework}, we briefly describe the two-fluid hydrostatic equilibrium equations together with the general-relativistic formalism used to study radial pulsations of the admixed stars. Section \ref{sec:results} contains our main results and discussion. Section \ref{sec:conclusion} presents our summary and perspectives. We adopt natural units, i.e., $\hbar=c=1$.
 
\section{Framework}
	\label{sec:framework}

In this section we summarize the main features of the TOV equations for the admixture of two fluids that interact only gravitationally. The one-fluid radial oscillation equations are conveniently partitioned to analyze either the stability of the quark mater core or dark matter core of the whole compact star. For the bag constant we use $B^{1/4}=145$ MeV, which  implements the Bodmer--Witten--Terazawa hypothesis for strange quark matter \cite{Schaffner-Bielich:2020psc}. This choice yields a maximum mass of $2.01~M_{\odot}$ and a radius of $R=11$ km. As mentioned previously, fermionic dark matter is considered as being either weakly ($y=0.1$) or strongly ($y=10^{3}$) self interacting \cite{Narain:2006kx}. Here, $y\equiv m_D/m_{int}$, where $m_D$ is the dark fermion mass, and $m_{int}$ is the scale of interaction. One can consider that $m_{int} \sim 100$ MeV for strong interactions and $m_{int} \sim 300$ GeV for weak interactions. The values $y=0.1$ and $y=10^{3}$ are commonly adopted as typical illustrations.
  
\subsection{Two-Fluid Hydrostatic Equilibrium Equations}

Since the structure of spherically symmetric, static one-fluid compact stars is determined from the usual TOV equations, they can be separated in order to deal with two fluids that only interact gravitationally. This can be performed as follows. The (perfect) one-fluid energy-momentum tensor is divided into two parts, i.e., $T^{\mu\nu}=T^{\mu\nu}_{1}+T^{\mu\nu}_{2}$. This, in turn, induces a separation of the total pressure and energy density in fluid components, as $p=p_{1}+p_{2}$ and $\epsilon=\epsilon_{1}+\epsilon_{2}$, respectively. Given that this separation does not affect the temporal $e^{\nu(r)}$ or radial $e^{\lambda(r)}$ metric functions, the corresponding equations keep their original forms, depending on the total pressure and energy densities. On the other hand, one has a set of coupled TOV equations for each of the fluids. 

In our case, i.e., quark and dark matter fluids, we have these two-fluid TOV equations in its dimensionless form given by \cite{Tolos:2015qra} (see Ref. \cite{Xiang:2013xwa} for a detailed variational derivation):
\begin{eqnarray}
&&\frac{dp'_{\rm QM}}{dr'}=-\frac{(p'_{\rm QM}+\epsilon'_{\rm QM})}{2}\frac{d \nu}{dr'}, \nonumber \\
&&\frac{dm'_{\rm QM}}{dr'}=4 \pi r'^2 \epsilon'_{\rm QM}, \nonumber \\
&&\frac{dp'_{\rm DM}}{dr'}=-\frac{(p'_{\rm DM}+\epsilon'_{\rm DM})}{2}\frac{d \nu}{dr'},  \\
&&\frac{dm'_{\rm DM}}{dr'}=4 \pi r'^2 \epsilon'_{\rm DM}, \nonumber \\
&&\frac{d \nu}{dr'}=2\frac{(m'_{\rm QM}+m'_{\rm DM}) + 4 \pi r'^3(p'_{\rm QM}+p'_{\rm DM})}{r'(r'-2(m'_{\rm QM}+m'_{\rm DM}))}, \nonumber
\label{eq:2TOV}
\end{eqnarray}
where $p'$ and $\epsilon'$ are the dimensionless pressure and energy density, respectively, and $m'_{\rm QM, DM}$ are the dimensionless gravitational masses enclosed inside the dimensionless radial coordinate $r'$. 

The set of equations above is solved simultaneously by specifying the (dimensionless) EoSs for QM, i.e., $p'_{\rm QM}=p'_{\rm QM}(\epsilon'_{\rm QM})$, and DM, i.e.,  $p'_{\rm DM}=p'_{\rm DM}(\epsilon'_{\rm DM})$. As usual, the \linebreak conditions at the center should be given for QM and DM in the admixed star. The numerical integration stops when one of the pressures reaches zero, i.e., \linebreak $p'_{\rm QM/DM}(R'_{\rm QM/DM})=0$, characterizing the QM or DM core surface, allowing us to obtain the corresponding gravitational mass $m'_{\rm QM/DM}(R'_{\rm QM/DM})=M'_{\rm QM/DM}$, where, in general, $R'_{\rm QM}{~\neq~}R'_{\rm DM}$. If $R'_{\rm QM}>R'_{\rm DM}$, the admixed star has a DM core and, if $R'_{\rm QM}<R'_{\rm DM}$, it has a DM halo surrounding a QM core.\endnote{Equivalently, the total radius $R$ and total mass $M$ of the whole admixed star can be determined by the condition of the total pressure $p(R)=0$ and $M=m(R)$, where $m(r)=m_{\rm QM}(r)+m_{\rm DM}(r)$. In any case, our code for the two-fluid TOV equations matches the results of Ref. \cite{Mukhopadhyay:2015xhs} very well.}

The boundary conditions for the metric function $\nu(r)$ come from ensuring that it matches the Schwarzschild metric outside the QM or DM core in the admixed star, i.e.,:
\begin{equation}
\nu(R'_{\rm QM})=\ln\left(1-\frac{2(M'_{\rm QM}+m'_{\rm DM}(R'_{\rm QM}))}{R'_{\rm QM}}\right)
\label{eq:nu1} 
\end{equation}
or
\begin{equation}
\nu(R'_{\rm DM})=\ln\left(1-\frac{2(m'_{\rm QM}(R'_{\rm DM})+M'_{\rm DM})}{R'_{\rm DM}}\right), 
\label{eq:nu2}
\end{equation}
respectively.
  
\subsection{Pulsations of Quark and Dark Matter Cores}
  
The equations that describe the radial pulsations of one-fluid compact stars were obtained for the first time by S. Chandrasekhar \cite{Chandrasekhar:1964zza}. He found that these equations could be arranged as a Sturm--Liouville problem where the eigenvalues are the oscillation frequencies squared, $\omega^{2}$ (the eigenfunctions being the radial Lagrangian displacements). For numerical purposes, these equations can be conveniently modified to a pair of first-order differential equations for each of the Lagrangian variables with more intuitive boundary conditions \cite{Misner:1973prb,Chanmugam:1977,Glass:1983,Gondek:1997fd,Kokkotas:2000up}.

Strictly speaking, the dynamic stability of admixed stars must be studied using the full two-fluid formalism\endnote{This general-relativistic formalism was studied many years ago in different papers (see e.g. Ref. \cite{Comer:1999rs} and references therein) when investigating matter with different properties in compact-star interiors, e.g. one fluid being a proton (neutron) superconductor (superfluid) and the other being normal nuclear matter.} of Refs. \cite{Leung:2012vea,Kain:2020zjs}, which would produce unified oscillation frequencies for the entire admixed star. However, since this calculation is computationally very expensive and time consuming, we decided to solve the equivalent of the two-fluid TOV equations, realized instead in the form of oscillation equations. In order to perform that function, we used the formalism of Ref. \cite{Gondek:1997fd}, which deals with the relative radial displacement $\Delta{r'}/r'\equiv\xi'=\xi$ and the Lagrangian perturbation pressure $\Delta{p'}$, both dimensionless. 

Inspired by the previous separation for the total pressure and energy density, we separated the total Lagrangian variables as $\xi=\xi_{\rm QM}+\xi_{\rm DM}$ and $\Delta{p'}=\Delta{p'_{\rm QM}}+\Delta{p'_{\rm DM}}$ (omitting the term ${e}^{i\omega{t}}$ in both variables), obtaining the following system of equations:
	\begin{equation}
	\frac{d\xi_{\rm QM/DM}}{dr'}\equiv-\frac{1}{r'}\left(3\xi_{\rm QM/DM}+\frac{\Delta{p'_{\rm QM}}}{\Gamma{p'}}\right)-\frac{dp'}{dr'}\frac{\xi_{\rm QM/DM}}{(p'+\epsilon')} \; ,
	\label{Rad1}
	\end{equation}
	\begin{multline}
	\frac{d\Delta{p'_{\rm QM/DM}}}{dr'}\equiv\xi_{\rm QM/DM}\left\lbrace{\omega'^{2}e^{\lambda-\nu}(p'+\epsilon')r'-4\frac{dp'}{dr'}}\right\rbrace+ \\
	\xi_{\rm QM/DM}\left\lbrace\left(\frac{dp'}{dr'}\right)^{2}\frac{r'}{(p'+\epsilon')}-8\pi{e^{\lambda}}(p'+\epsilon')p'r'\right\rbrace + \\
	\Delta{p'_{\rm QM/DM}}\left\lbrace{\frac{dp'}{dr'}\frac{1}{p'+\epsilon'}-4\pi(p'+\epsilon')r'{e}^{\lambda}}\right\rbrace \; ,
	\label{Rad2}
	\end{multline}	
where $\omega'$ is the dimensionless oscillation frequency and $\Gamma$ is the adiabatic index\endnote{The physical and dimensionless definitions have the same mathematical form after being rescaled by an arbitrary factor.} \linebreak $\Gamma=(1+\epsilon'/p')(\partial{p'}/\partial{\epsilon'})$. The metric function $\lambda(r')$ is obtained from $\lambda(r')=-\ln(1-2(m'_{\rm QM}(r')+m'_{\rm DM}(r'))/r')$ with boundary conditions given by Equations (\ref{eq:nu1}) and (\ref{eq:nu2}), i.e., $\lambda(R'_{\rm QM})=-\nu(R'_{\rm QM})$ and $\lambda(R'_{\rm DM})=-\nu(R'_{\rm DM})$.

So far we have not mentioned whether $\omega'$ corresponds to the pulsation of a QM core or a DM core. Recall that these equations represent a Sturm--Liouville problem, which defines its eigenvalues in terms of the associated boundary conditions. In this case, they are 
\begin{equation}
	(\Delta{p'_{\rm QM/DM}})_{\rm center}\equiv-3(\xi_{\rm QM/DM}\Gamma{p'_{\rm QM/DM}})_{\rm center} \, ,
\label{BC1}
\end{equation}
demanding smoothness at the QM or DM stellar center, and 
\begin{equation}
	(\Delta{p'_{\rm QM/DM}})_{\rm surface}\equiv{0} \, ,
\label{BC2}
\end{equation}
since $p'_{\rm QM/DM}(R'_{\rm QM/DM})=0$, with eigenfunctions normalized to $\xi_{\rm QM/DM}(0)=1$ as usual. Thus, Equations (\ref{BC1}) and (\ref{BC2}) lead us to define $\omega'^{2}\to\omega'^{2}_{\rm QM/DM}$ if we are dealing with a QM/DM oscillating core in the admixed star. In other words, only one of the cores oscillates depending on the boundary conditions. The other fluid only affects its oscillation indirectly, by the coupling of 
 the total pressure and energy density. 

A word of caution should be added at this point. Usually, two ways of dealing with the radial oscillations of two-fluid compact stars have been explored. In the simplest one, only the radial oscillation of the whole admixed star, i.e., treated as one fluid, is studied without explicitly considering  the gravitational coupling between the QM and DM cores (see e.g., Refs. \cite{Tolos:2015qra,Panotopoulos:2017eig,Panotopoulos:2018ipq}). In the second, a consistent general-relativistic formalism to deal with the couplings between oscillation amplitudes and Lagrangian perturbations for each fluid is developed \cite{Kain:2020zjs,Kain:2021hpk}. Unfortunately, dealing with a system of highly coupled and non-linear differential equations requires very time-consuming numerical calculations which we consider unnecessary when independently solving  the oscillation equations for each DM or QM core while keeping the other fluid at rest but still coupled through the coefficients entering in the equations. In this sense, the formalism built in this work occurs more in the line of Ref. \cite{Leung:2012vea}, which considers the independent oscillations of each fluid, thus forming an Sturm--Liouville-like problem. Notice that our reasoning agrees with the fact that each of the two-fluid TOV equations can be considered an independent 'one-fluid' star only, coupled through $\nu(r')$ to the other 'one-fluid' star. Thus, radial oscillations of each 'one-fluid' star can be associated with a set of one-fluid\endnote{In the limit of one-fluid stars, our code for radial oscillations agrees very well with previous works, see e.g., Ref. \cite{Kokkotas:2000up}.} oscillation equations coupled now by the total pressures, energy densities, and polytropic indices and metric functions, $\nu(r')$ and $\lambda(r')$. For consistency, we have verified that our formalism agrees with the results of Ref. \cite{Mukhopadhyay:2015xhs} when a delay of the maximal central density is reached at higher densities for increasing amounts of DM when the zero frequencies are reached. We stress that, in this case, $f_{n=0}\to{0}$ coincides with $\partial{M}/\partial\epsilon_{c}\to{0}$, since few amounts of DM were considered, whereas in this work we explore all the available DM densities which  notoriously modifies the stability of the admixed stars, so that $\partial{M}/\partial\epsilon_{c}$ must be used with caution.

In following sections, we focus on the fundamental mode frequency, $\omega_{n=0}$. It vanishes at the maximal stable QM or DM mass configuration, marking the onset of the instability of the corresponding oscillating core which, in turn, induces the gravitational collapse of the whole admixed star.

\section{Results and Discussion}
  \label{sec:results}

The parameter space for quark stars admixed with weakly or strongly self-interacting DM is large. In this section we show only results where the effects on observables are relevant. As mentioned before, we considered dark fermion masses $m_{D}= 1, 10, 50, 100, 200, 500$ GeV in order to include all possible dark fermion candidates. 

Regarding the numerical values we chose for $\epsilon^{\rm QM/DM}_{c}$ in our calculations and showed in our plots: (i) for QM, the three values of $\epsilon^{\rm QM}_{c}$ correspond to somewhat above, twice, and nearly twice the value of the maximal central energy density of pure quark stars with $B=(\rm 145 MeV)^{4}$, i.e., $\sim{1}~{\rm GeV/fm^{3}}$. The reason for this is that higher values of $\epsilon^{\rm QM}_{c}$ are required when DM is present in the admixed star; (ii) the three values of $\epsilon^{\rm DM}_{c}$ (for strongly or weakly interacting DM) correspond to near the minimum, intermediate, and near the maximal-mass central densities for corresponding pure DM stars. In Tables \ref{tab:table1} and \ref{tab:table2} the maximal-mass values of central energy density for each $m_{D}$ are listed. This choice was made to quantify the full dependence of the stellar structure on the amounts of DM. We will show that, in some cases with a huge amount of DM, only very small objects with strangelet-like and planet-like masses are allowed. This was expected from the results of Ref. \cite{Mukhopadhyay:2015xhs}.

Although the usual criterion for static stability , $\partial{M}/\partial{\epsilon_{c}}\geq0$, consistently works for one-fluid stars, it should not be taken for granted in two-fluid stars; only the frequency analysis can decide on their stability. Our results for the oscillation frequency of the fundamental mode for QM and DM cores in the admixed stars are written in terms of the linear frequency $f^{\rm QM/DM}_{n=0}=\omega^{\rm QM/DM}_{n=0}/(2\pi)$.

\begin{table}[H]
\small
  \caption{Maximum masses $M^{\rm max}~(M_{\odot})$ and their corresponding minimum radii $R^{\rm min}$ (km) and maximum central energy densities $\epsilon^{\rm max}_{c}$ (GeV/fm$^{3}$) obtained for weakly interacting (wDM) pure dark matter stars.}
\label{tab:table1}
  
  \newcolumntype{C}{>{\centering\arraybackslash}X}
\begin{tabularx}{\textwidth}{CCCC}
      \toprule
       \boldmath{${\rm EoSs}$} & \boldmath{$\epsilon^{\rm max}_{c}({\rm wDM})$} & \boldmath{$M^{\rm max}({\rm wDM})$} & \boldmath{$R^{\rm min}({\rm wDM})$}\\
      \midrule
 ${\rm DM}~(m_{D}/{\rm GeV}=1)$ & $\sim$3 & $\sim$0.63 & $\sim$8.1 \\
 $10$ 	& $\sim$$3\times{10^{4}}$ & $\sim$$6.27\times{10^{-3}}$ & $\sim$$7.8\times{10^{-2}}$\\
 $50$ 	& $\sim$$1.97\times{10^{7}}$ & $\sim$$2.50\times{10^{-4}}$ & $\sim$$3.2\times{10^{-3}}$\\
 $100$ 	& $\sim$$3\times{10^{8}}$ & $\sim$$6.27\times{10^{-5}}$ & $\sim$$8.1\times{10^{-4}}$\\
 $200$ 	& $\sim$$4.99\times{10^{9}}$ & $\sim$$1.56\times{10^{-5}}$ & $\sim$$2\times{10^{-4}}$\\
 $500$  & $\sim$$1.97\times{10^{11}}$ & $\sim$$2.50\times{10^{-6}}$ & $\sim$$3\times{10^{-5}}$\\\bottomrule
        \end{tabularx}

\end{table}
  
%
\begin{table}[H]
\small
\caption{Same notation as in Table \ref{tab:table1} but now for strongly interacting (sDM) pure dark matter stars.}
\label{tab:table2}

  \newcolumntype{C}{>{\centering\arraybackslash}X}
\begin{tabularx}{\textwidth}{CCCC}
    \toprule
       \boldmath{${\rm EoSs}$} & \boldmath{$\epsilon^{\rm max}_{c} ({\rm sDM})$} & \boldmath{$M^{\rm max}({\rm sDM})$} & \boldmath{$R^{\rm min}({\rm sDM})$}\\
      \midrule
 ${\rm DM}~(m_{D}/{\rm GeV}=1)$ & $\sim$$4.9\times{10^{-5}}$ & $\sim$$2.67\times{10^{2}}$ & $\sim$$1.87\times{10^{3}}$\\
 $10$ 	& $\sim$0.6 & $\sim$2.67 & $\sim$18.5\\
 $50$ 	& $\sim$$4\times{10^{2}}$ & $\sim$$1.07\times{10^{-1}}$ & $\sim 7.4\times{10^{-1}}$\\
 $100$ 	& $ \sim$$4.9\times{10^{3}}$ & $\sim$$2.67\times{10^{-2}}$ & $\sim$$18.7\times{10^{-2}}$\\
 $200$ 	& $\sim$$8\times{10^{4}}$ & $\sim$$6.68\times{10^{-3}}$ & $\sim$$4.7\times{10^{-2}}$\\
 $500$  & $\sim$$4\times 10^{6}$ & $\sim$$1.07\times{10^{-3}}$ & $\sim$$7.5\times{10^{-3}}$\\\bottomrule
        \end{tabularx}

\end{table}
%


\subsection{Admixtures of Quark Matter and Weakly $(y=0.1)$ Interacting Dark Matter}

\subsubsection{Solving the Two-Fluid TOV Equations}

In Figure \ref{fig:MRdens01QM}, we display  the results obtained from solving the two-fluid TOV \linebreak Equation~(\ref{eq:2TOV}) with the condition $p_{\rm QM}(R_{\rm QM})=0$ for different central energy densities of weakly interacting DM. One can easily see that only the solutions for $m_{D}=1500$ GeV display sizable modifications on the QM stellar masses and radii. In particular, the case of $m_{D}=500$~GeV suffers a marked reduction of $~1.2~M_{\odot}$ due to the very high DM central energy densities (${\sim}10^{11}~{\rm GeV/fm^{3}}$). Additionally, the central QM densities increased considerably, by a factor $\sim$7. Normally such QM densities would generate unstable pure quark stars with central energy densities at most $\sim${1}${\rm GeV/fm^{3}}$ without the DM component .

\begin{figure}[H]

\begin{adjustwidth}{-\extralength}{0cm}
\centering 

{\includegraphics[scale=0.37]{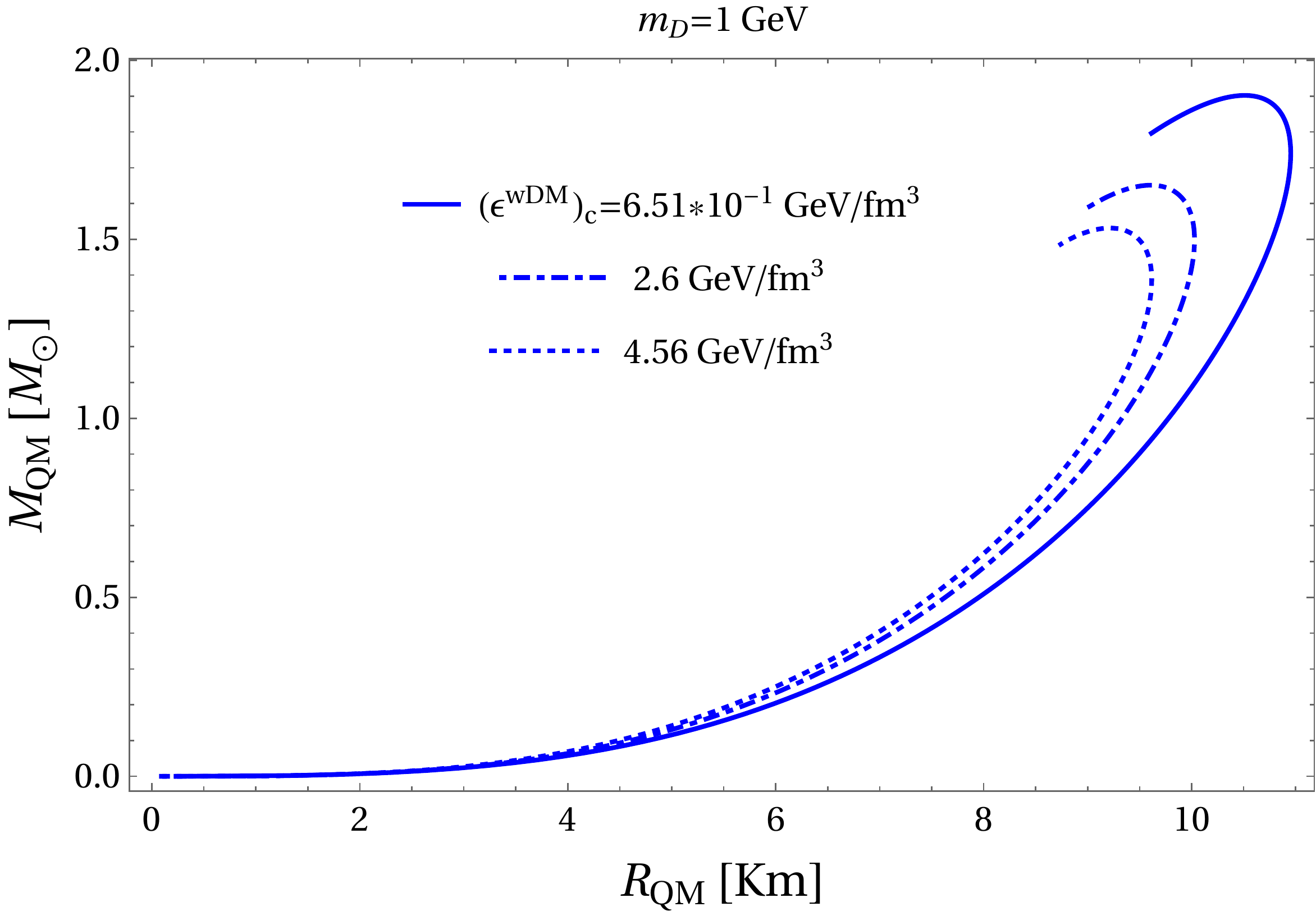}\hspace{3pt}
	  \includegraphics[scale=0.37]{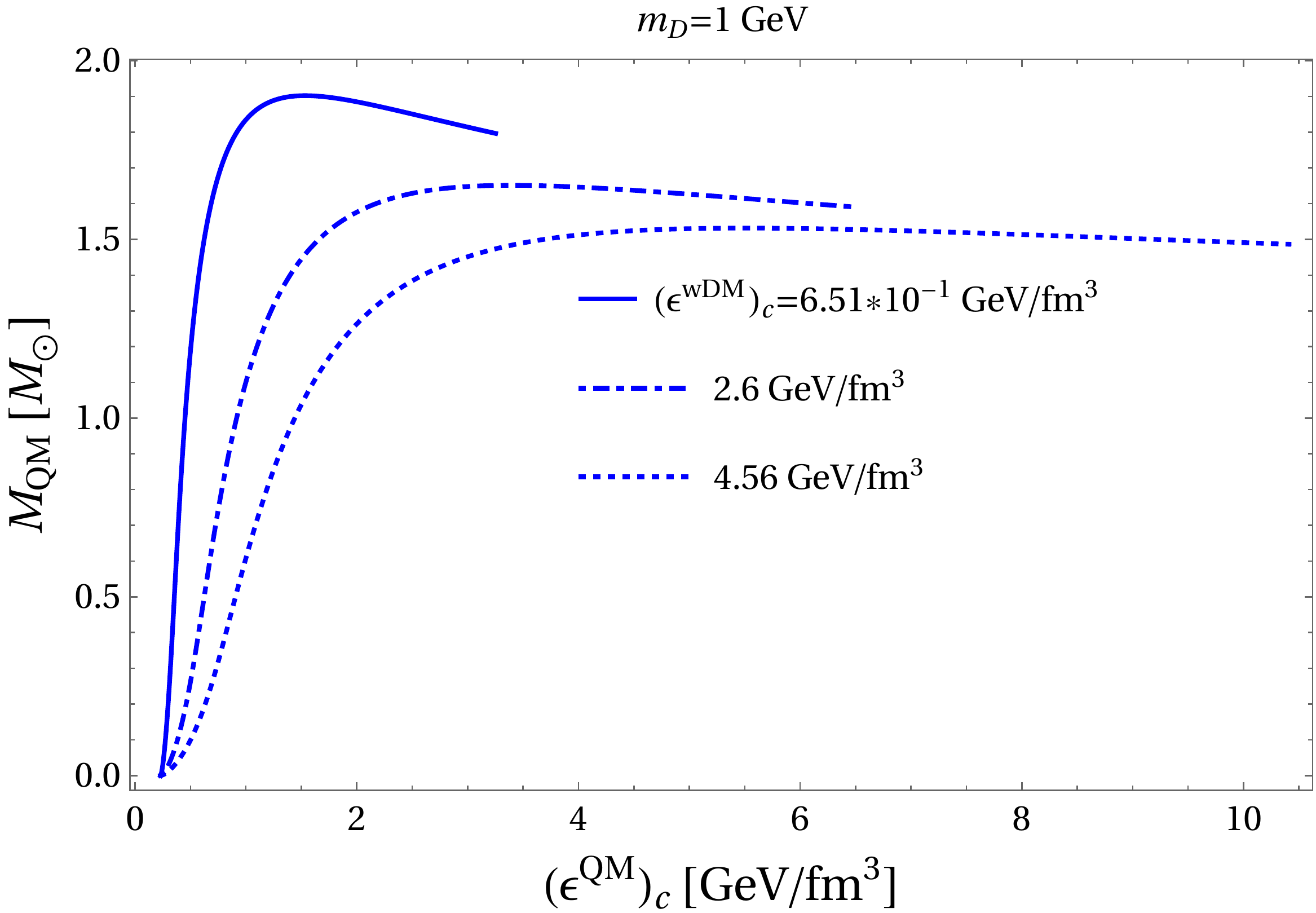}}\\
{\includegraphics[scale=0.38]{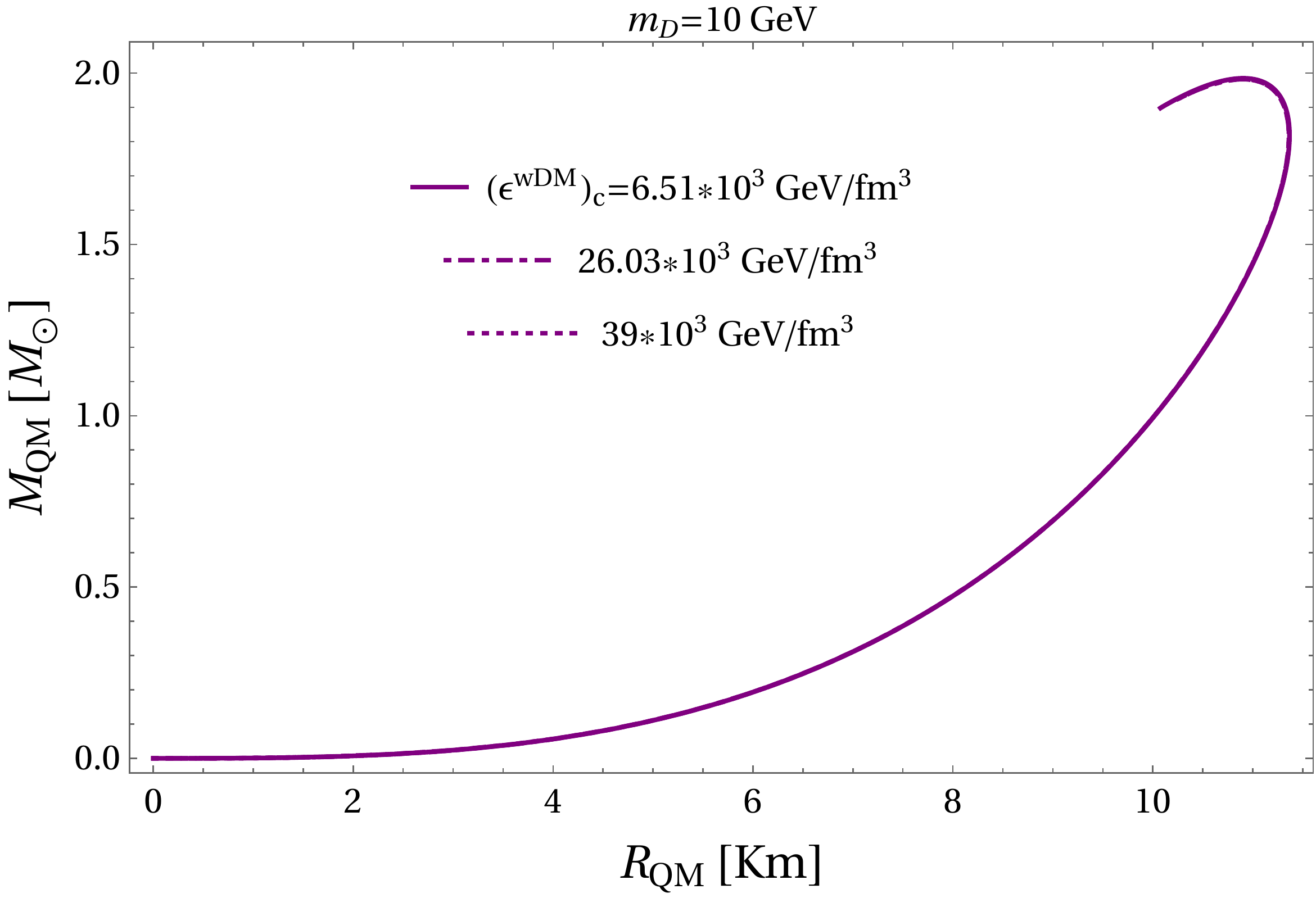}\hspace{3pt}
	  \includegraphics[scale=0.395]{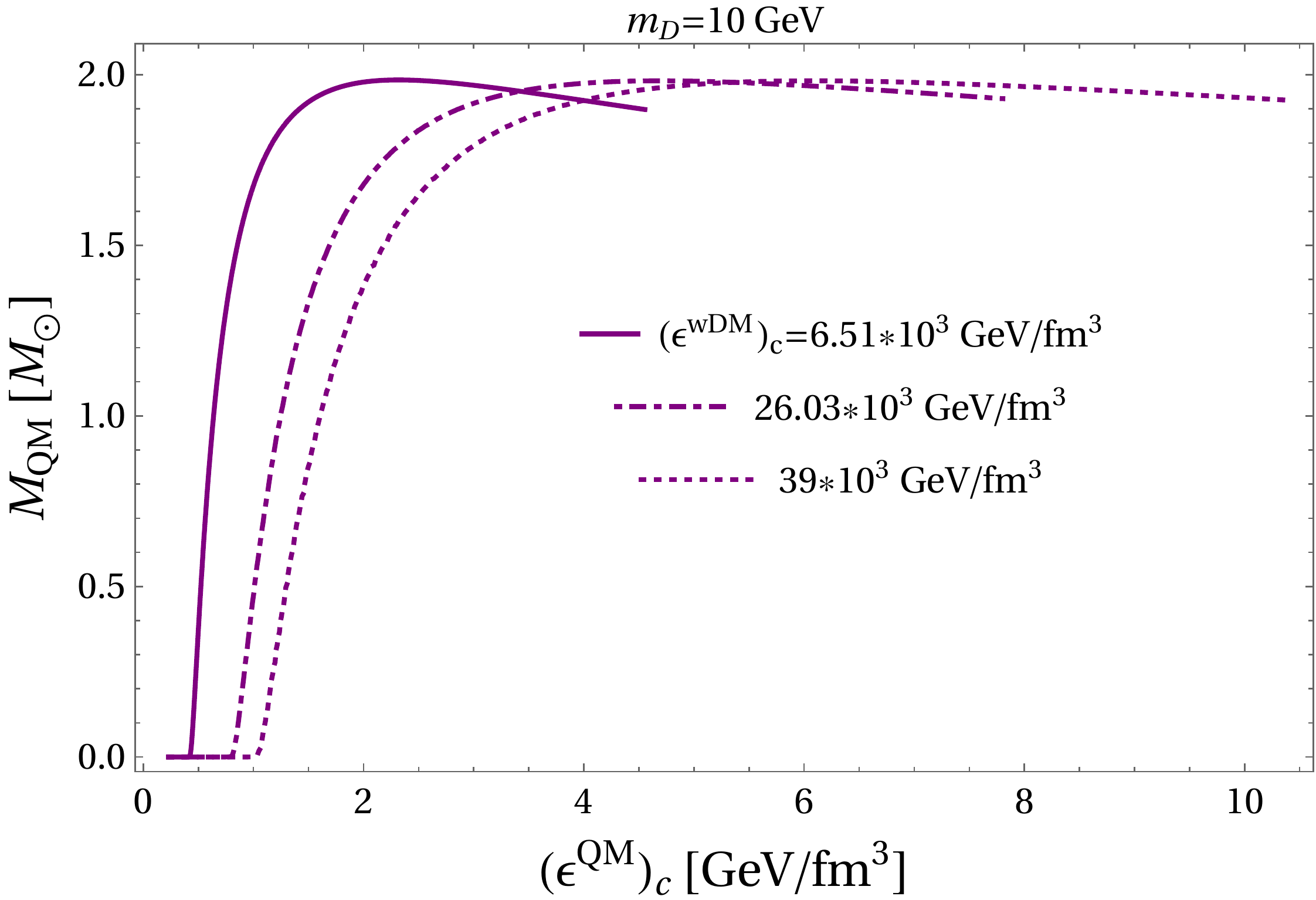}}\\
\end{adjustwidth}
\caption{\emph{Cont.}}
\label{fig:MRdens01QM}

\end{figure}

\begin{figure}[H]\ContinuedFloat

\begin{adjustwidth}{-\extralength}{0cm}
\centering 

{\includegraphics[scale=0.37]{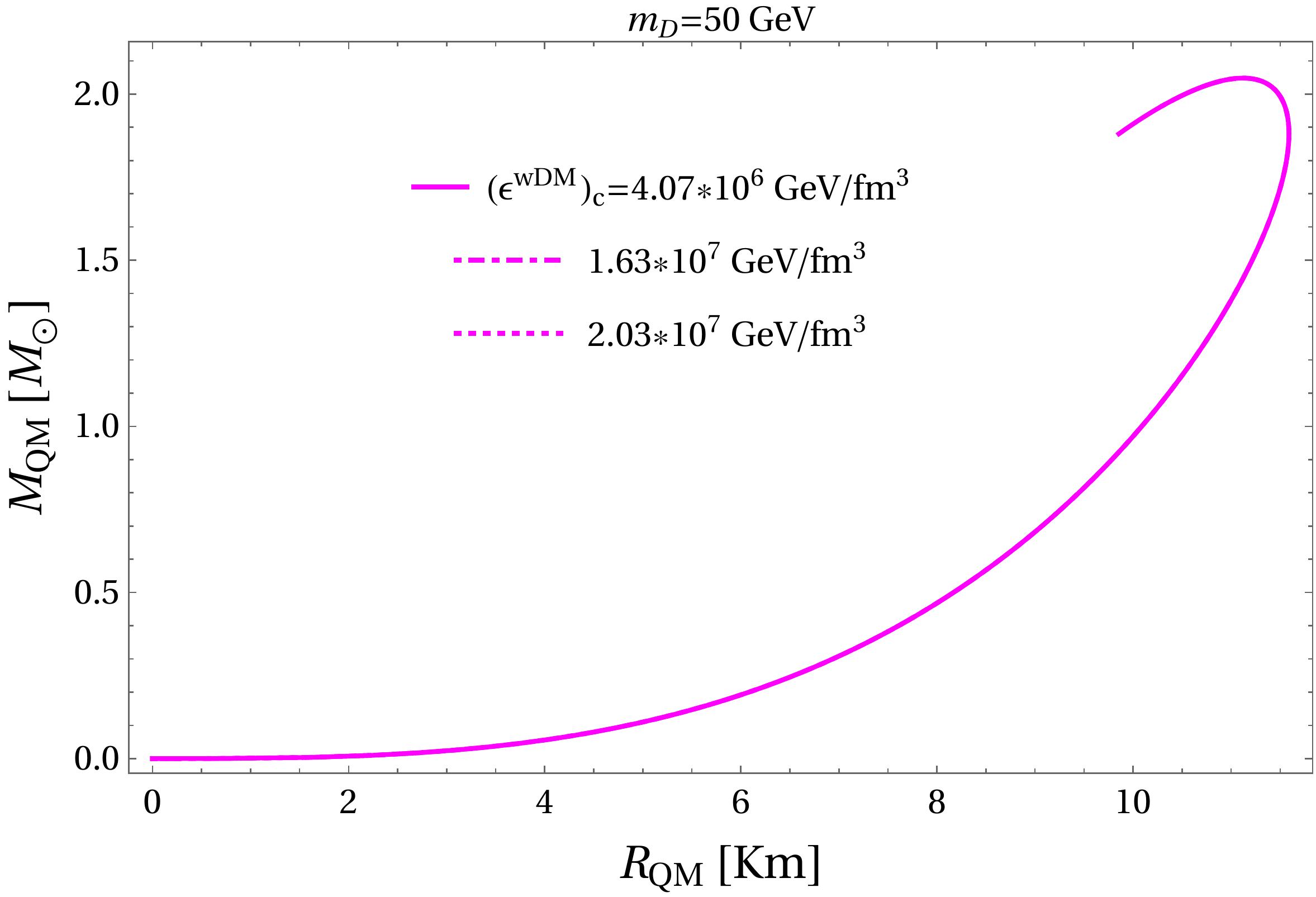}
	  \includegraphics[scale=0.39]{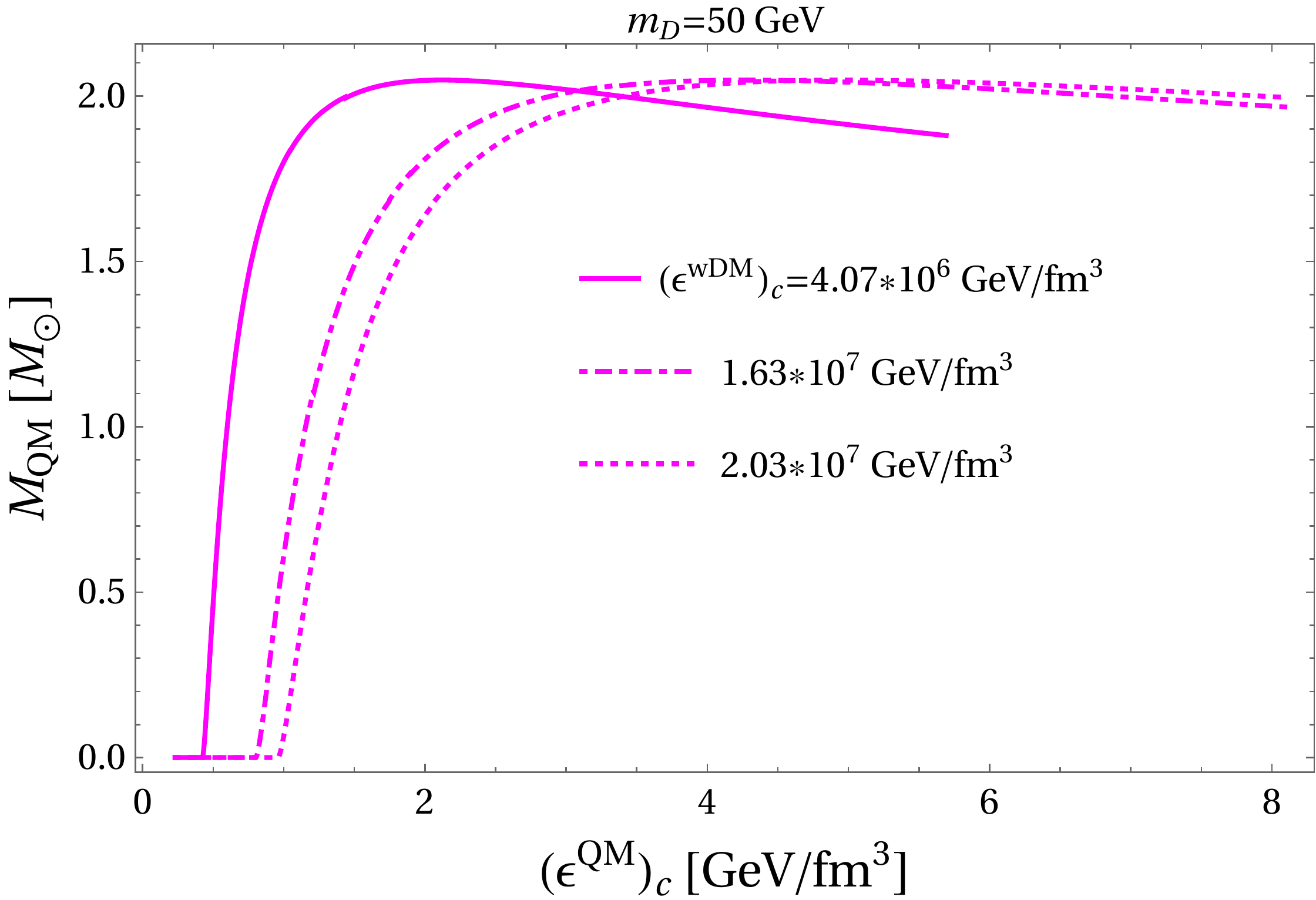}}\\
{\includegraphics[scale=0.36]{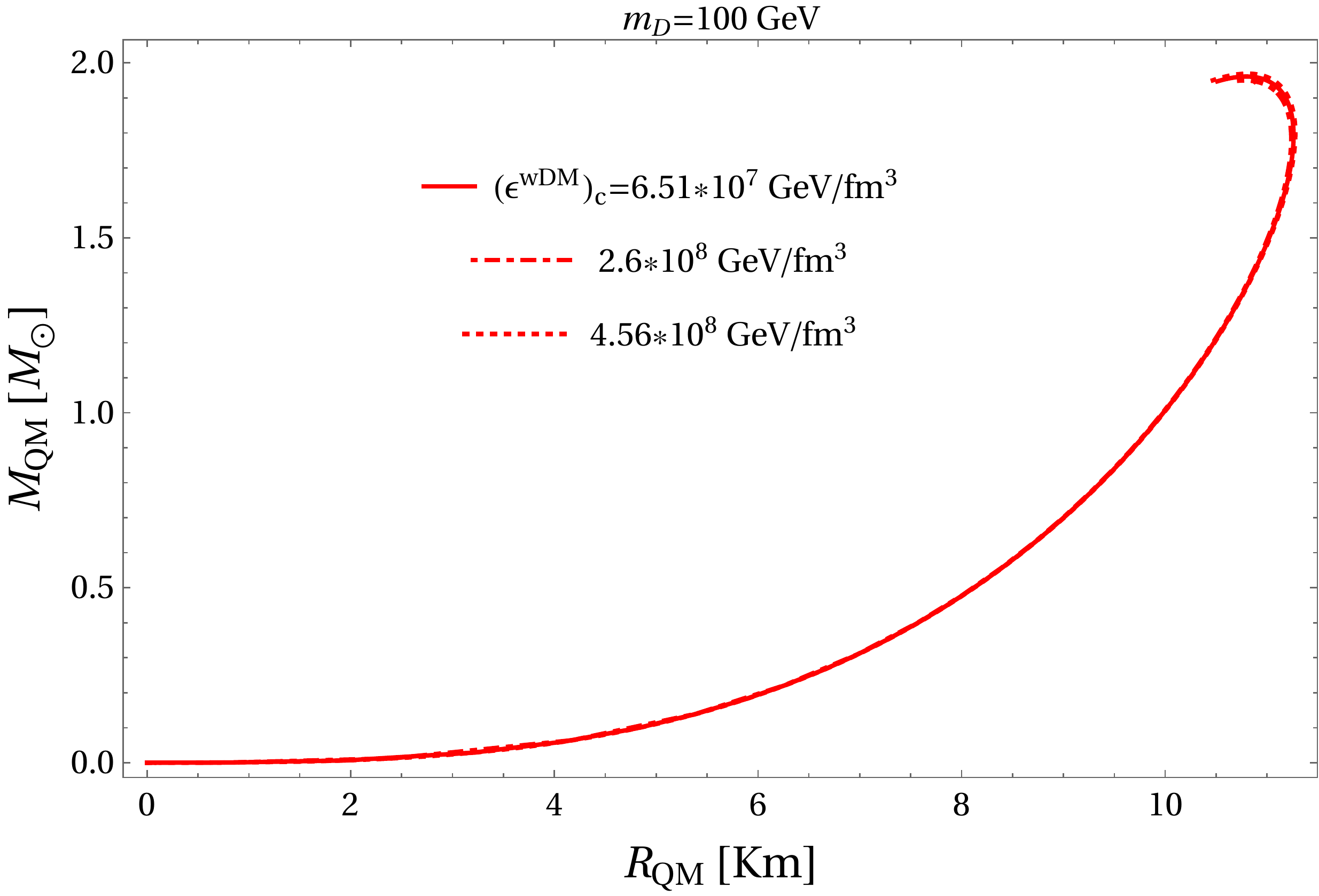}
	  \includegraphics[scale=0.325]{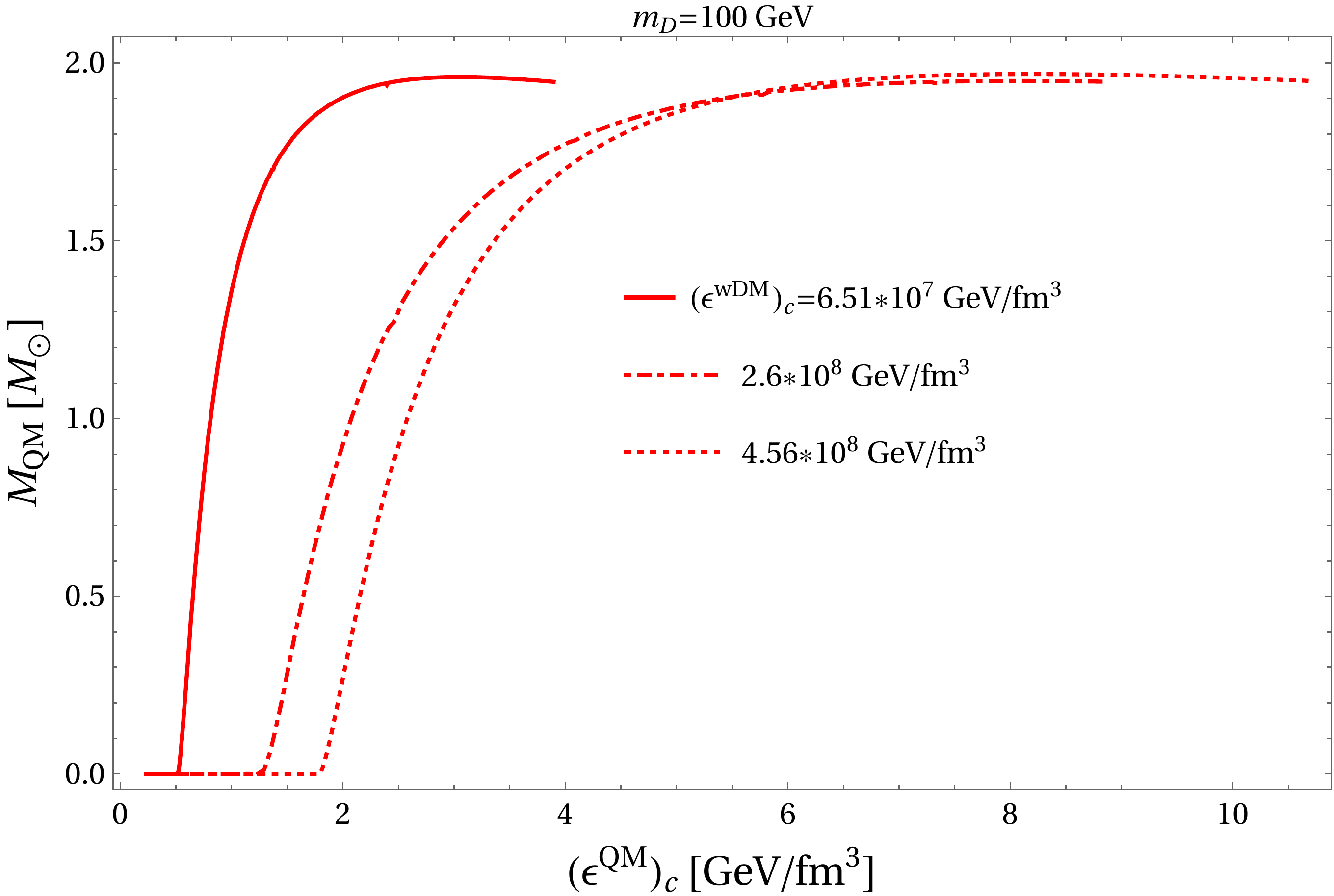}}\\
{\includegraphics[scale=0.39]{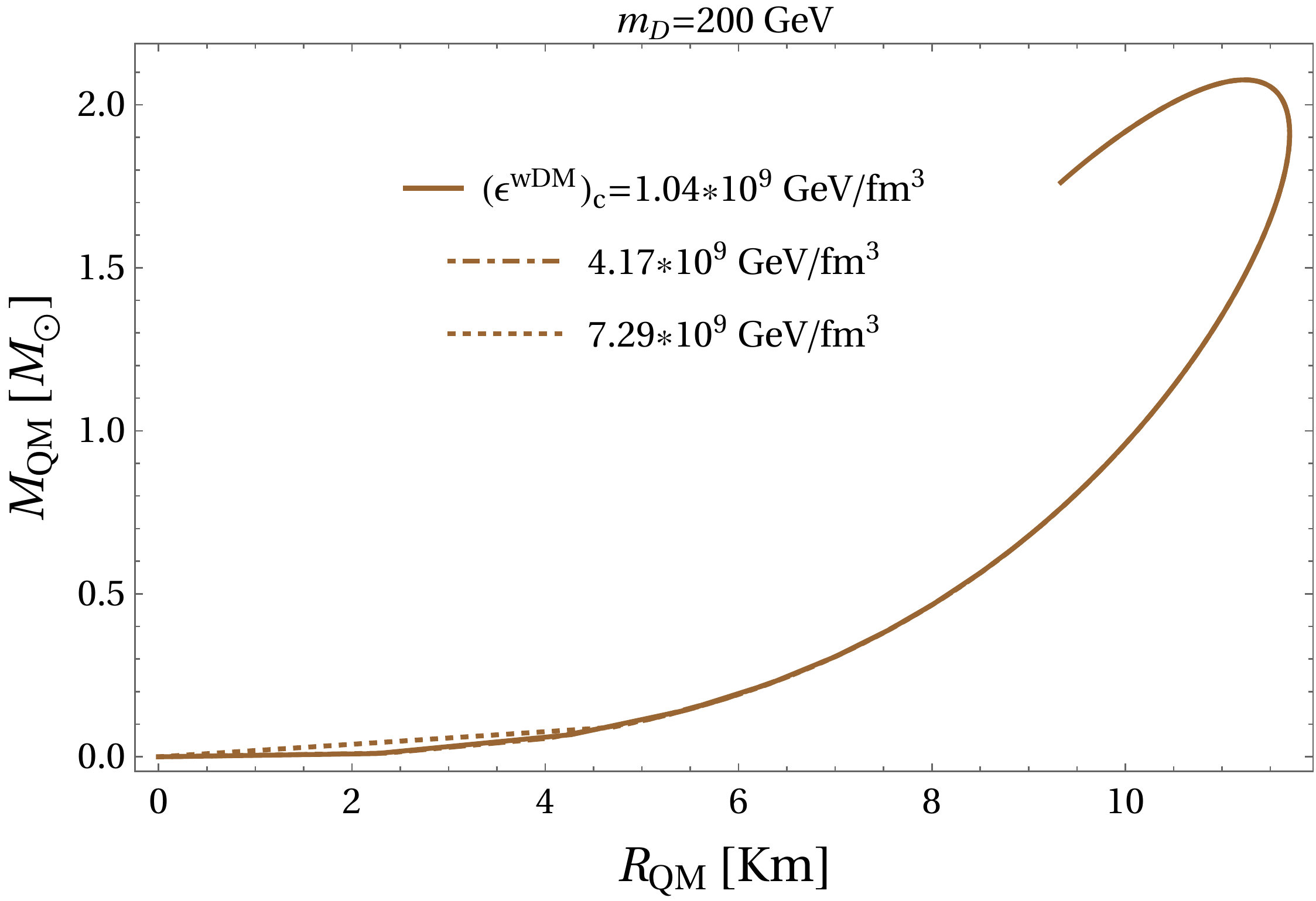}
	  \includegraphics[scale=0.405]{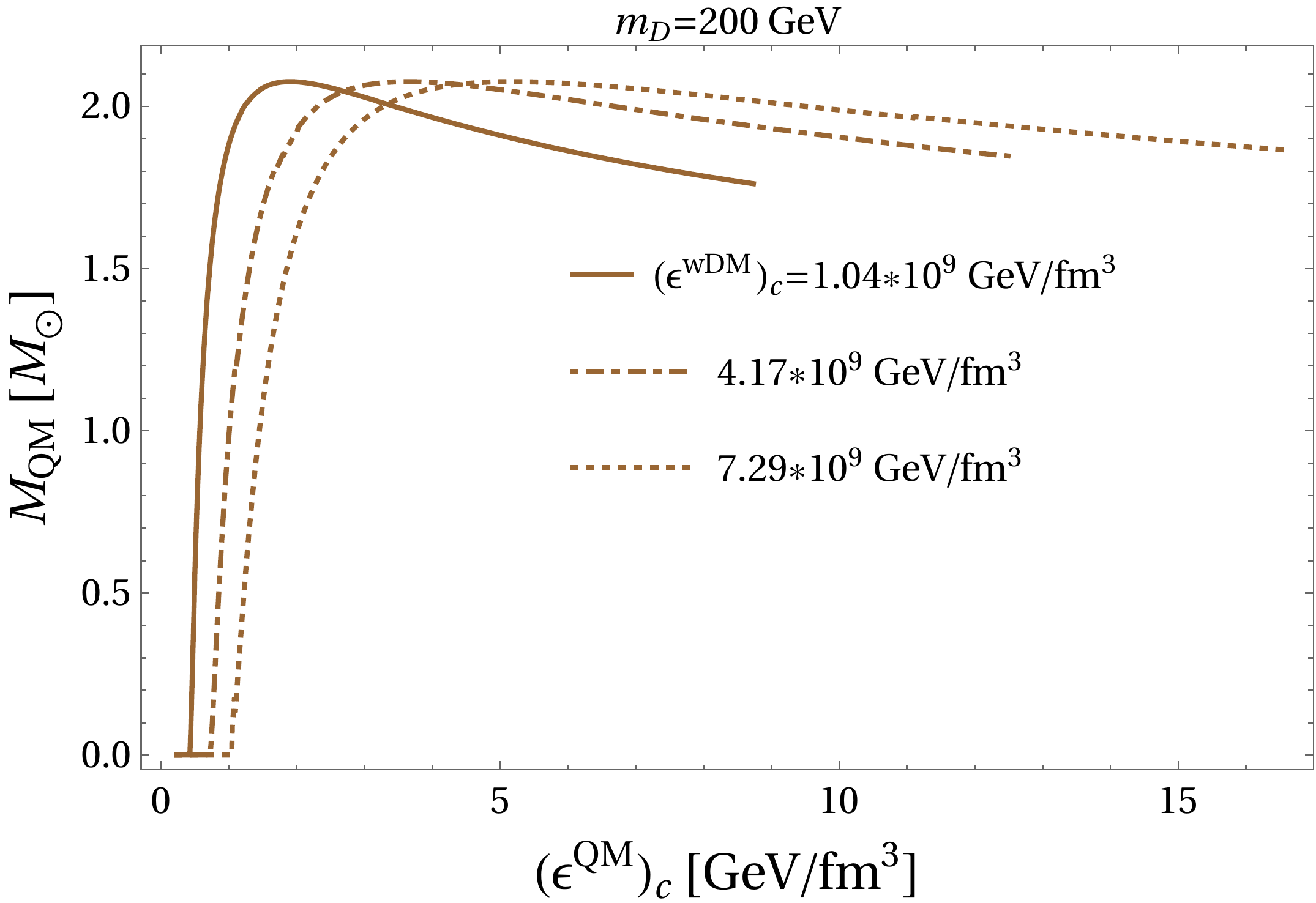}}\\
\end{adjustwidth}
\caption{\emph{Cont.}}
\label{fig:MRdens01QM}

\end{figure}

\begin{figure}[H]\ContinuedFloat

\begin{adjustwidth}{-\extralength}{0cm}
\centering 

{\includegraphics[scale=0.375]{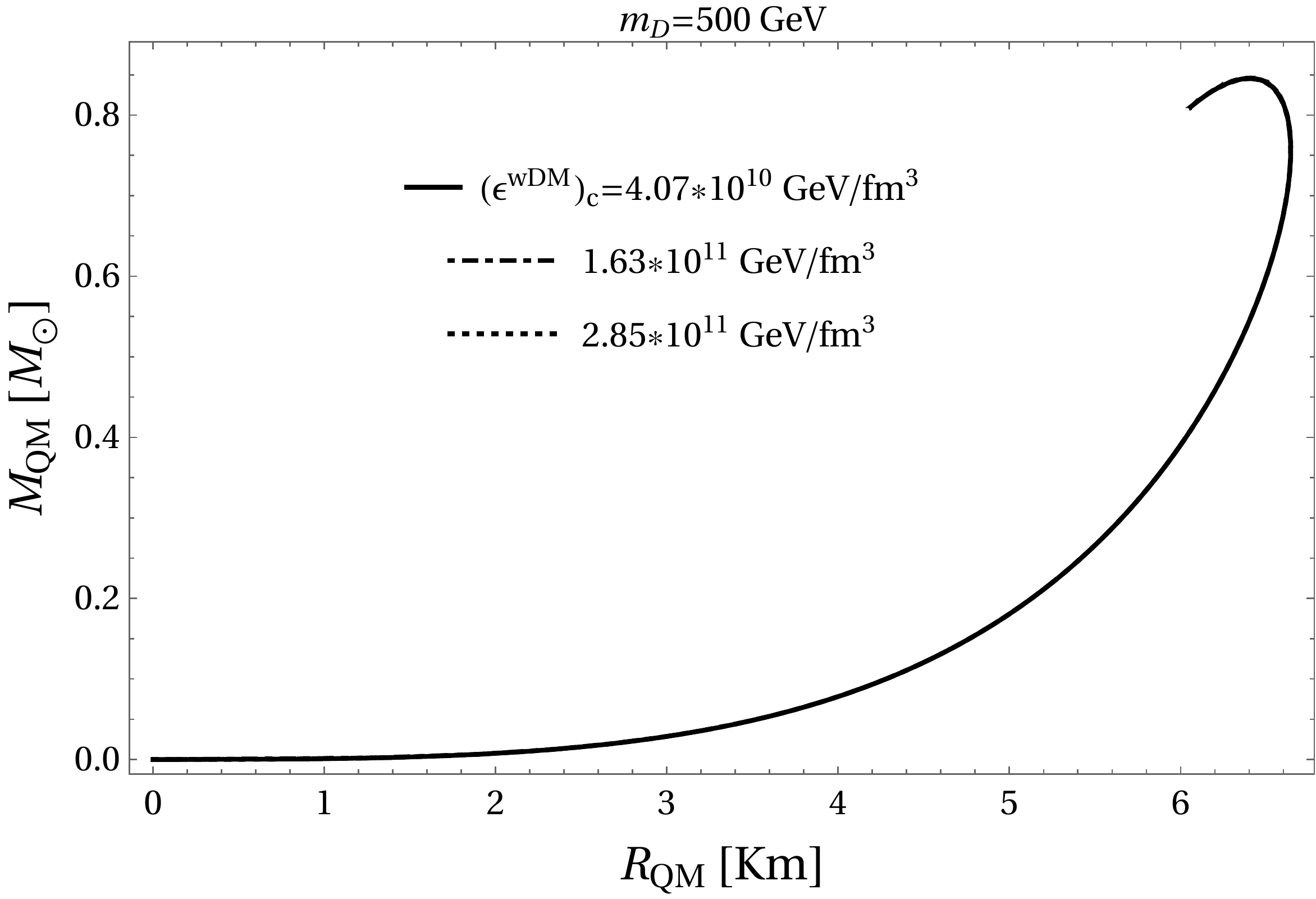}\hspace{5pt}
	  \includegraphics[scale=0.37]{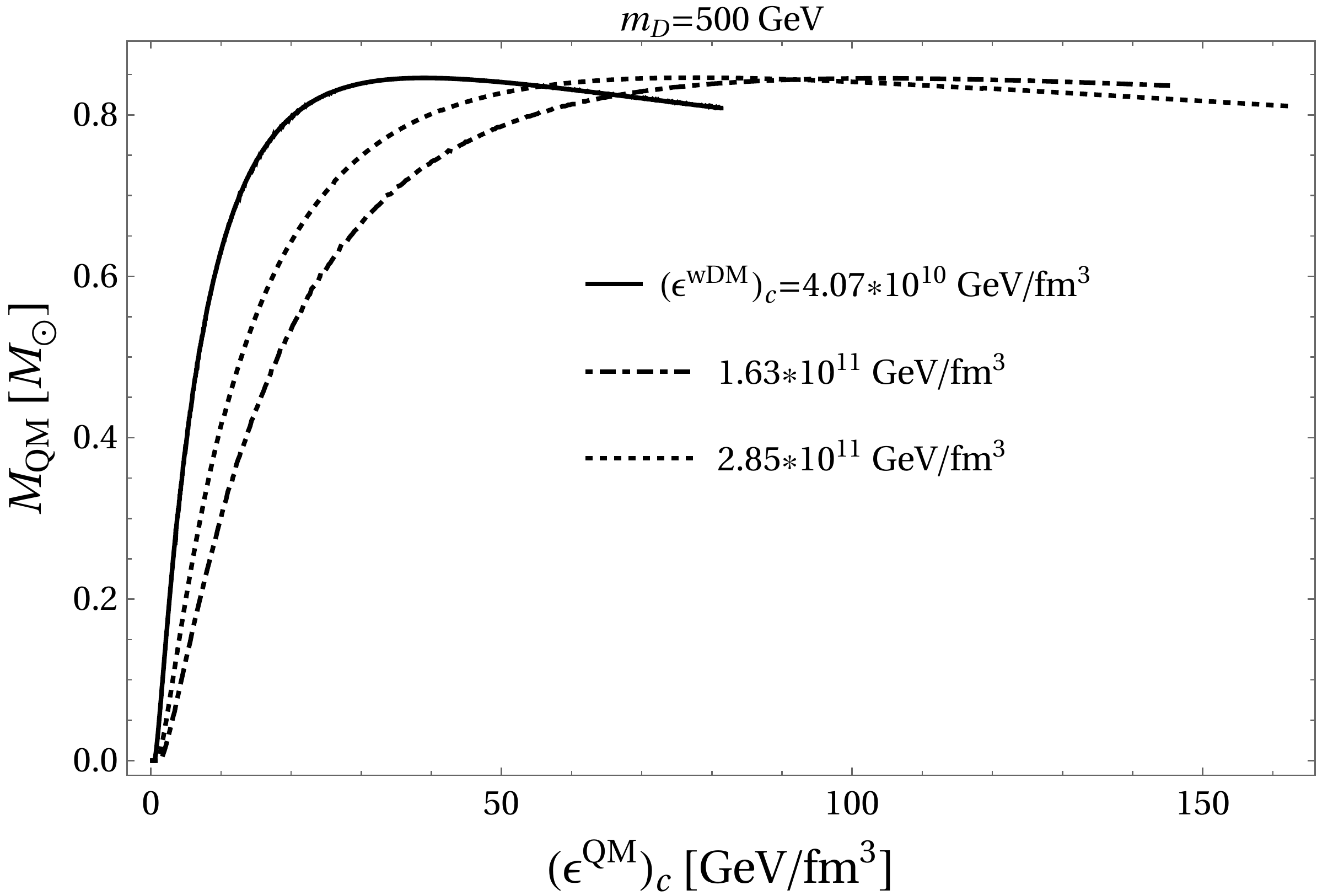}}
\end{adjustwidth}
\caption{Each pair of panels with the same color for the plots displays the mass--radius relation and the mass as a function of the energy density for quark matter cores, i.e., $p_{\rm QM}(R_{\rm QM})=0$, with different amounts of weakly ($y=0.1$) interacting dark matter ($w$DM) for dark fermion masses of $m_{D}=1,10,50,100,200,500$ GeV.}
\label{fig:MRdens01QM}

\end{figure}
  

On the other hand, solutions for other values of the dark fermion mass show negligible effects. Again, these QM stars require higher central energy densities in order to compensate for the extra gravitational pull from the DM. Furthermore, solutions for all the dark fermion masses, except $m_{D}=1$ and $500$ GeV, develop a plateau at low QM stellar masses in the mass vs. central energy density plots, which became wider for higher DM central energy densities. Our calculations show that these QM cores have masses between $10^{-18}$ to $10^{-4}~M_{\odot}$ with radii between $10^{-4}$ and $10^{-2}$ km, depending on the value of the dark fermion mass. For example, $m_{D}\sim 10$ GeV mostly commonly produces  stellar masses around $10^{-5}~M_{\odot}$ with radii of $10^{-3}$ km. As we increase the mass of the dark fermion, the values of $M_{\rm QM}$ and $R_{\rm QM}$  are reduced by many orders of magnitude. We note that all these stars satisfy the criterion $\partial{M}_{\rm QM}/\partial{\epsilon^{\rm QM}_{c}}>0$ and can be tentatively considered stable objects, "dark strange planets" in analogy to the results of Ref. \cite{Tolos:2015qra}, and "dark strangelets".

Figure \ref{fig:MRdens01DM} shows our results from the two-fluid TOV Equations (\ref{eq:2TOV}) with the condition $p_{\rm DM}(R_{\rm DM})=0$ for different values of QM central energy densities. The DM stars that are most affected by the presence of QM are the ones with $m_{D}=1$ and $10$ GeV. For stars with $m_{D}= 1$ GeV, the central QM energy densities are high enough to convert the usual behavior of pure DM stars in the mass--radius diagram into a self-bound-like behavior, making them more compact. To see this more quantitatively, Table \ref{tab:table1} shows values of masses and radii for pure $y=0.1$ DM stars for the whole range of dark fermion masses considered. One can see that the stellar masses and radii are slightly affected by the presence of QM near the maximum mass, but the radii of less massive DM stars are significantly modified. The same is true in the case of $m_{D}=10$ GeV. For $m_{D}=50, 100, 200, 500$ GeV, the DM high central energy densities completely dominate the QM contribution.

\subsubsection{Solving the Coupled Radial Oscillation Equations}

The solutions to the coupled radial pulsation Equations (\ref{Rad1}) and (\ref{Rad2}), assuming an oscillating QM core with fixed DM central energy density with boundary conditions (\ref{BC1}) and (\ref{BC2}), are shown in Figure \ref{fig:f0Mdens01QM}. We show the zero-mode frequency as a function of central energy density and stellar mass. 

\begin{figure} [H]

\begin{adjustwidth}{-\extralength}{0cm}
\centering 

{\includegraphics[scale=0.355]{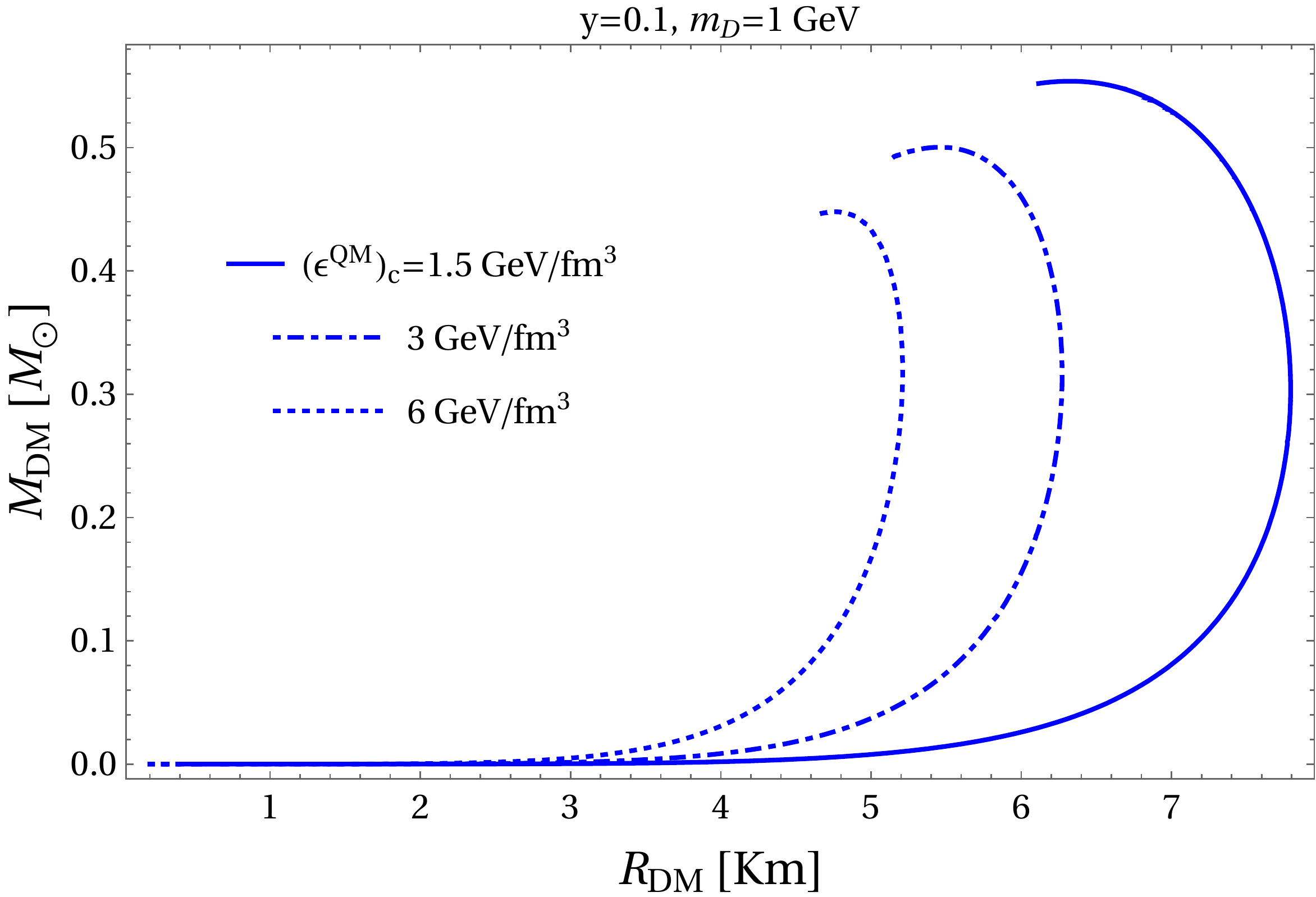}\vspace{3pt}
	  \includegraphics[scale=0.355]{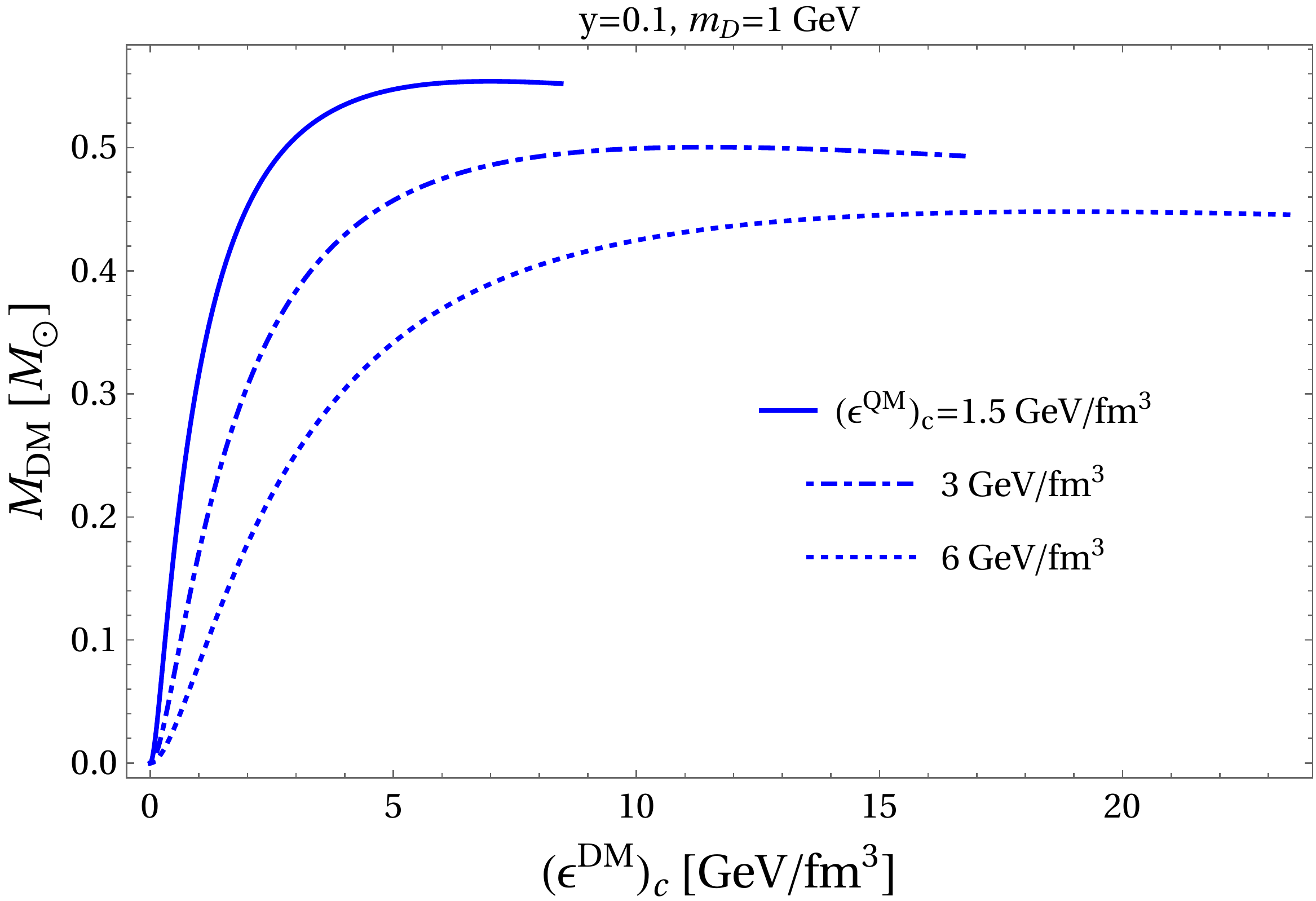}}\\\vspace{5pt}
{\includegraphics[scale=0.355]{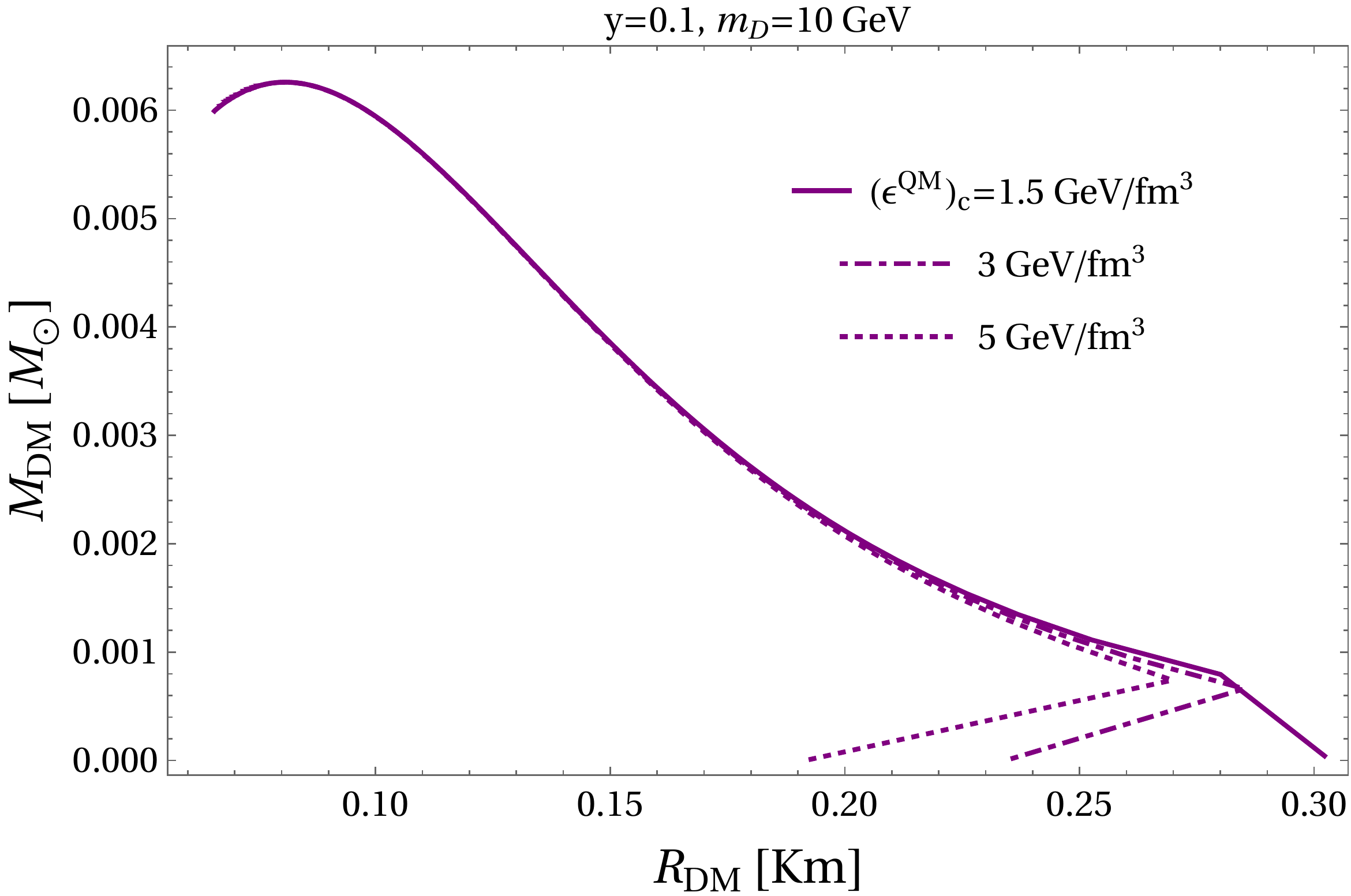}\vspace{5pt}
	  \includegraphics[scale=0.37]{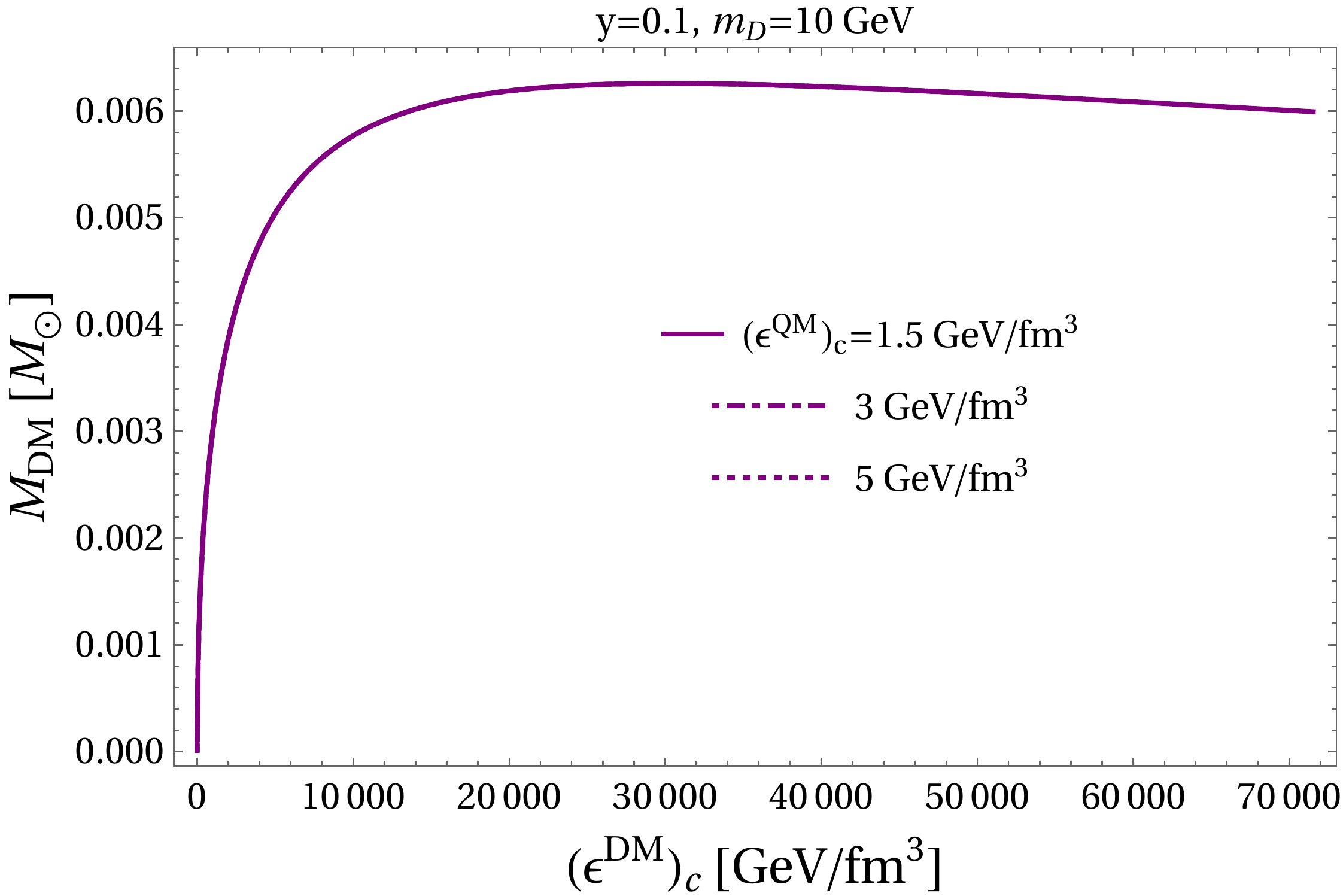}}\\\vspace{5pt}
{\includegraphics[scale=0.35]{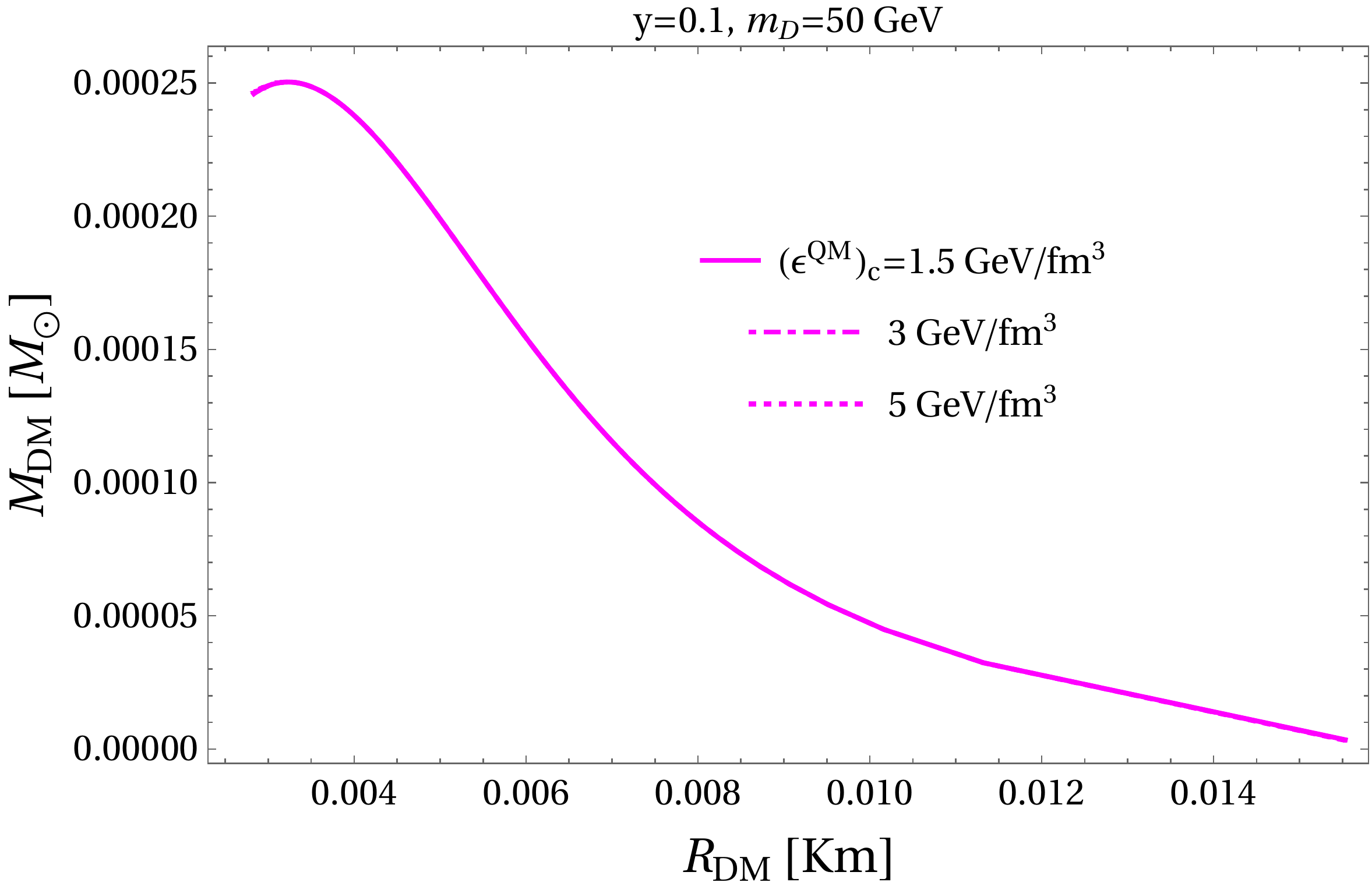}\vspace{3pt}
	  \includegraphics[scale=0.365]{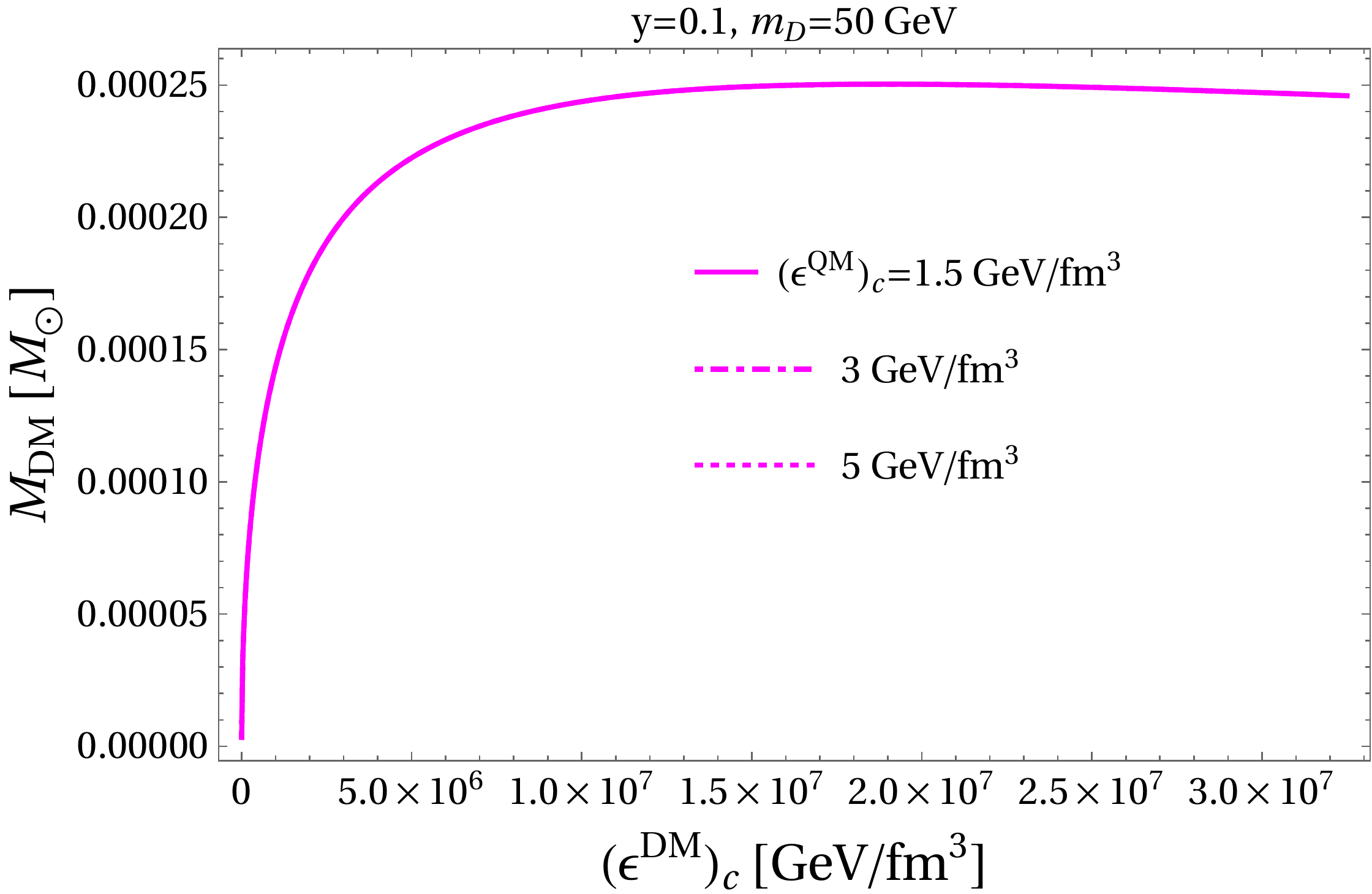}}\\\vspace{5pt}
{\includegraphics[scale=0.33]{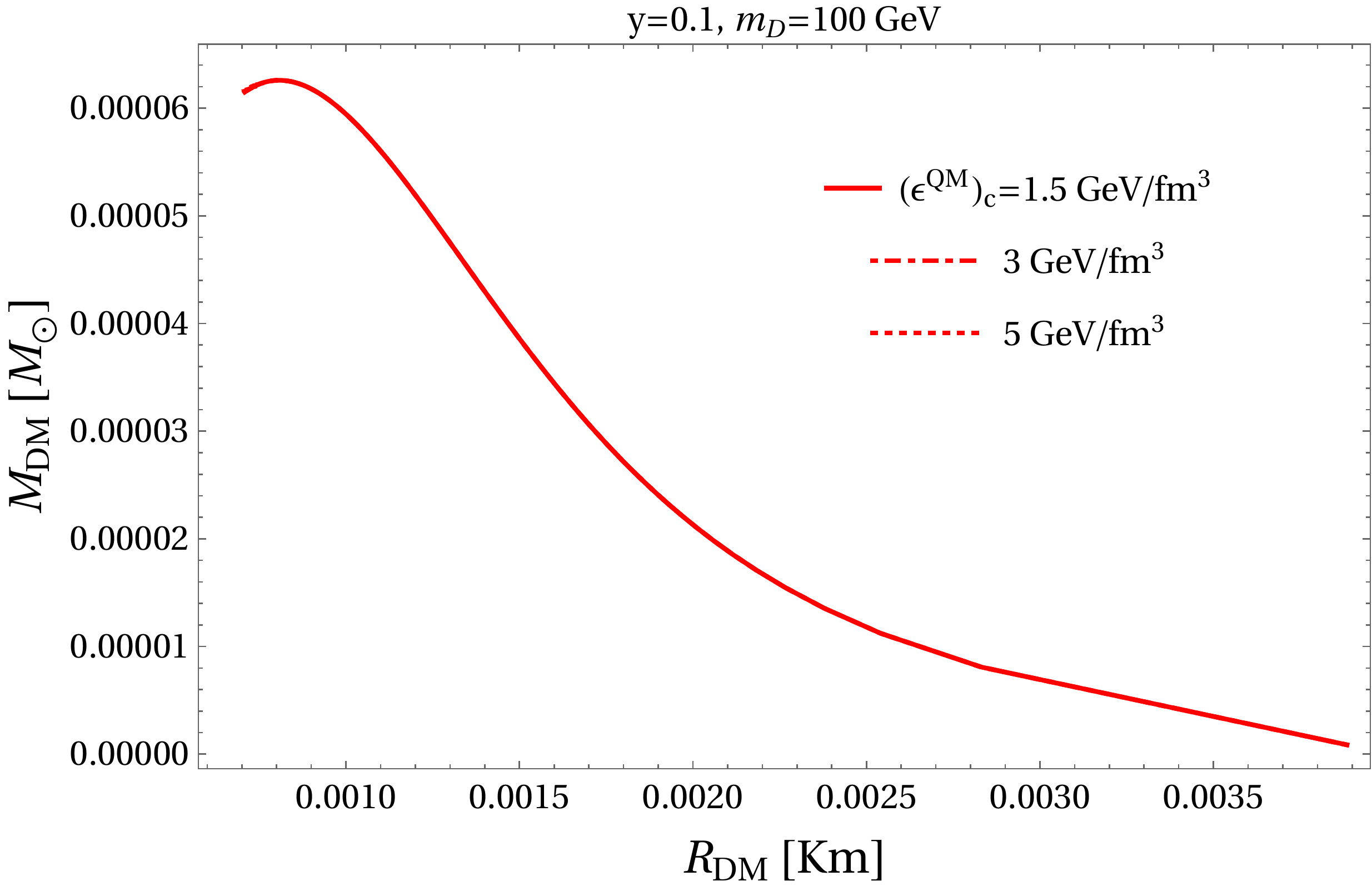}\vspace{3pt}
	  \includegraphics[scale=0.30]{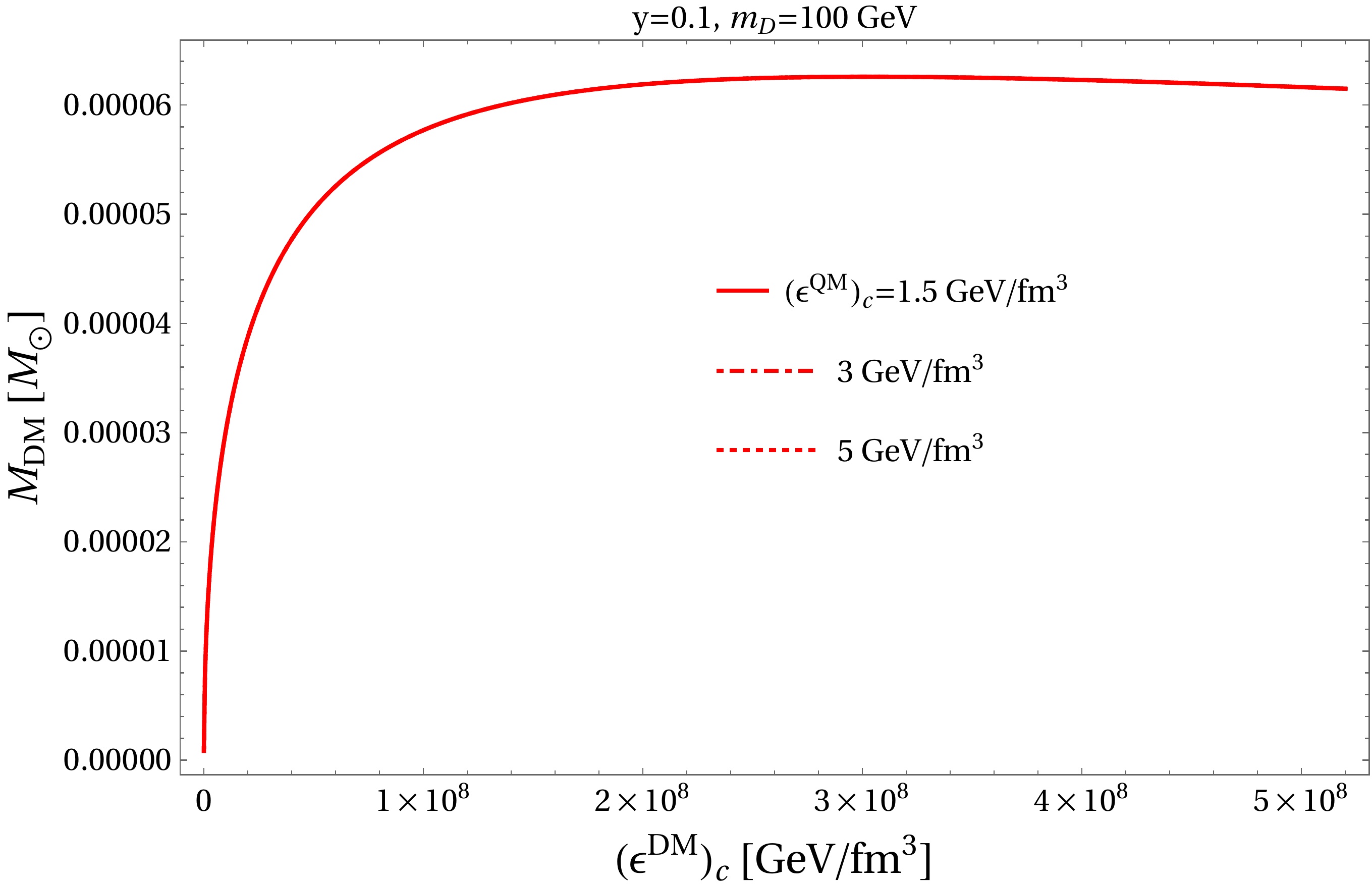}}\\
\end{adjustwidth}
\caption{\emph{Cont.}}
\label{fig:MRdens01DM}

\end{figure}

\begin{figure} [H]\ContinuedFloat

\begin{adjustwidth}{-\extralength}{0cm}
\centering 

{\includegraphics[scale=0.385]{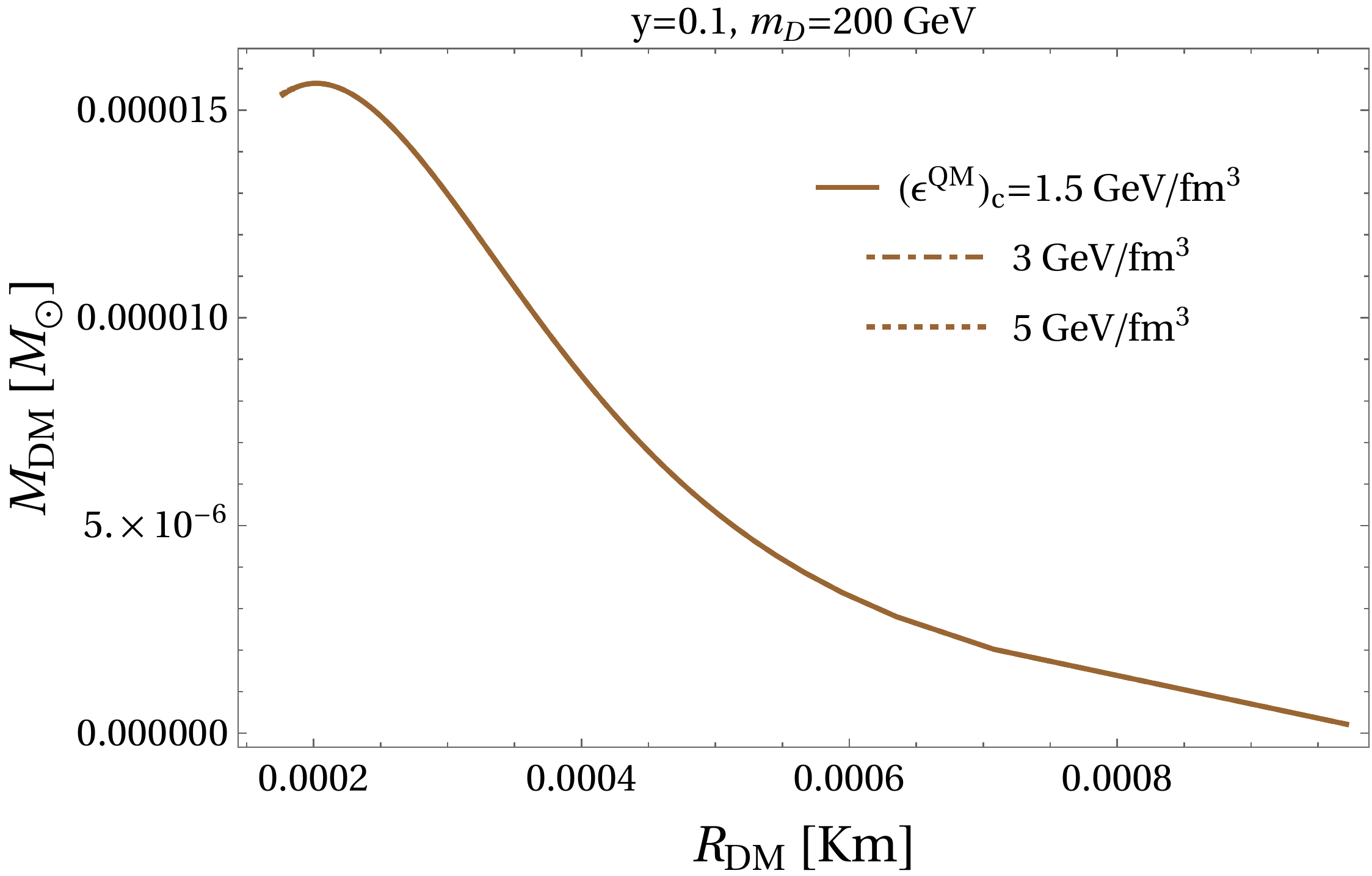}\vspace{3pt}
	  \includegraphics[scale=0.40]{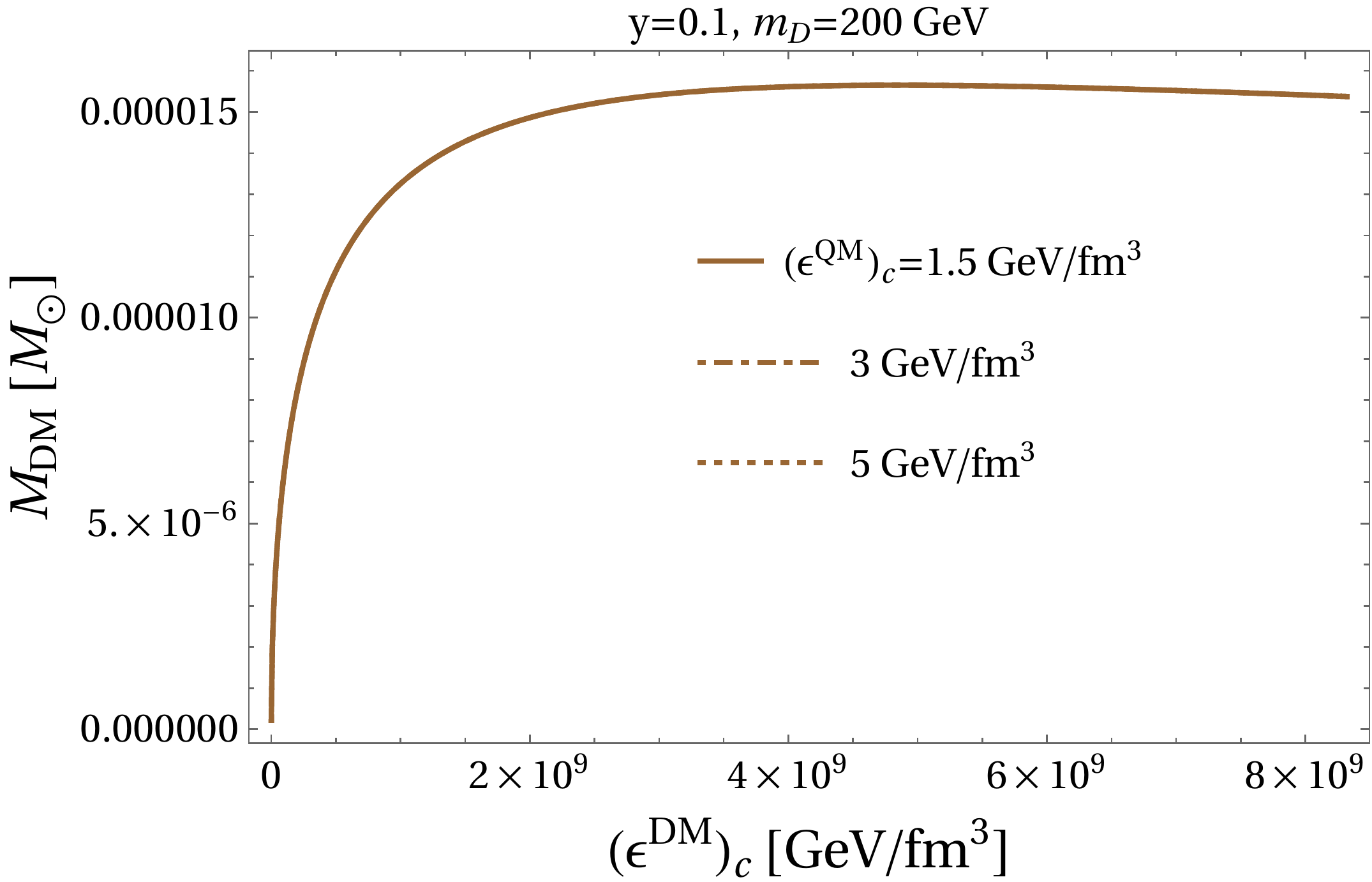}}\\\vspace{5pt}
{\includegraphics[scale=0.38]{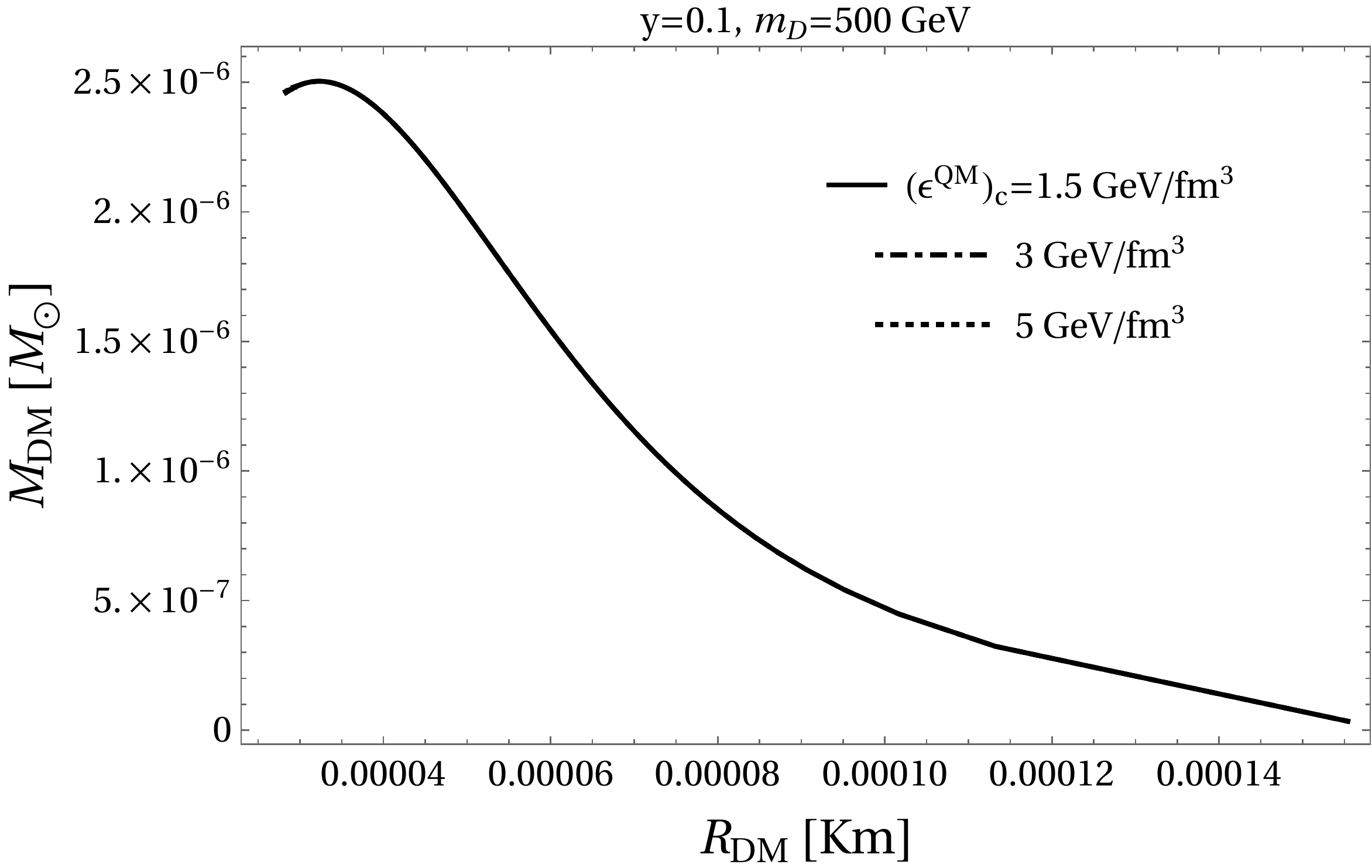}\vspace{3pt}
	  \includegraphics[scale=0.38]{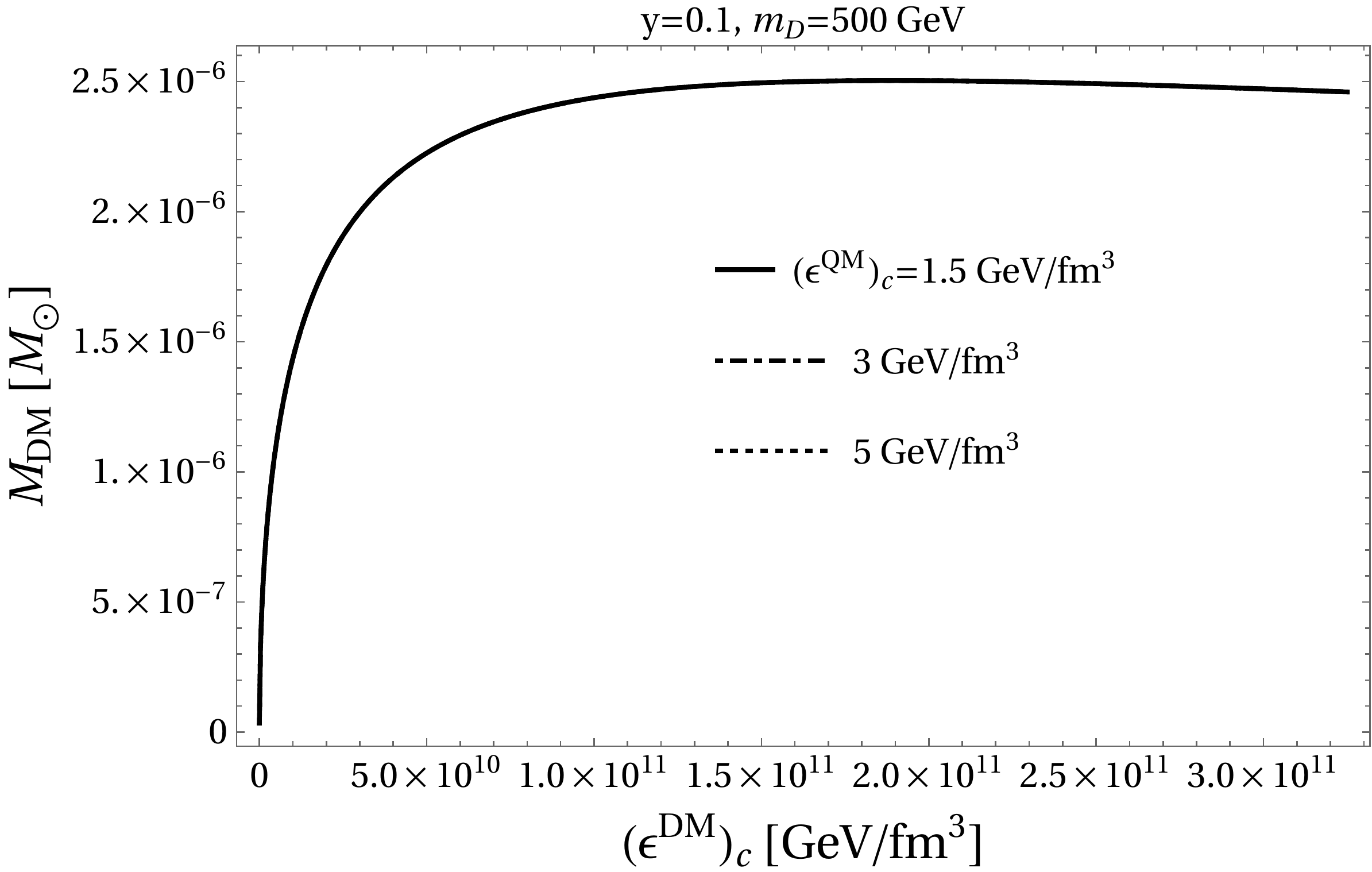}}
\end{adjustwidth}
\caption{Same notation as in Figure \ref{fig:MRdens01QM} but now for dark matter cores satisfying $p_{\rm DM}(R_{\rm DM})=0$.}
\label{fig:MRdens01DM}

\end{figure}


\begin{figure}[H]

\begin{adjustwidth}{-\extralength}{0cm}
\centering 

{\includegraphics[scale=0.36]{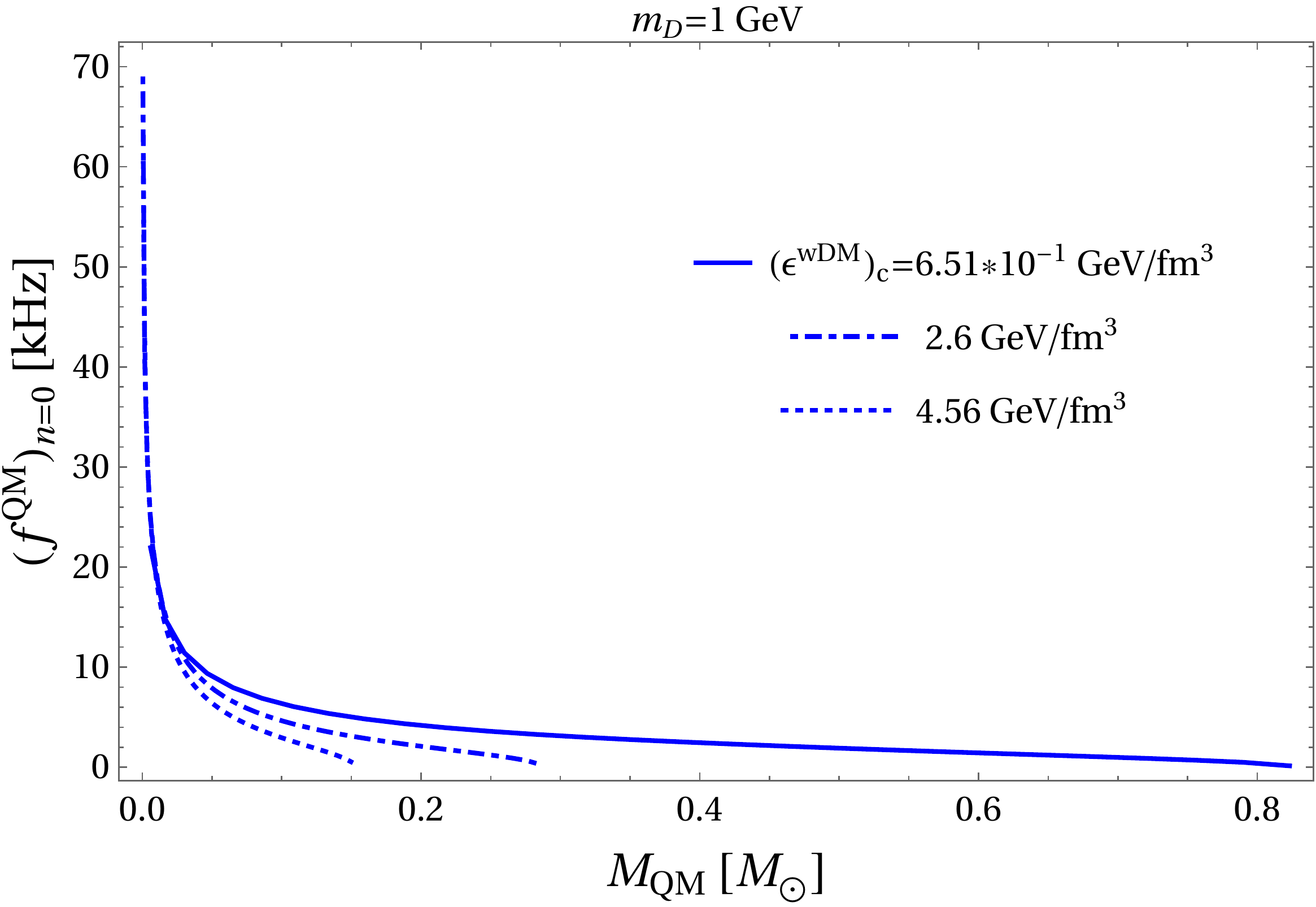}\vspace{3pt}
	  \includegraphics[scale=0.38]{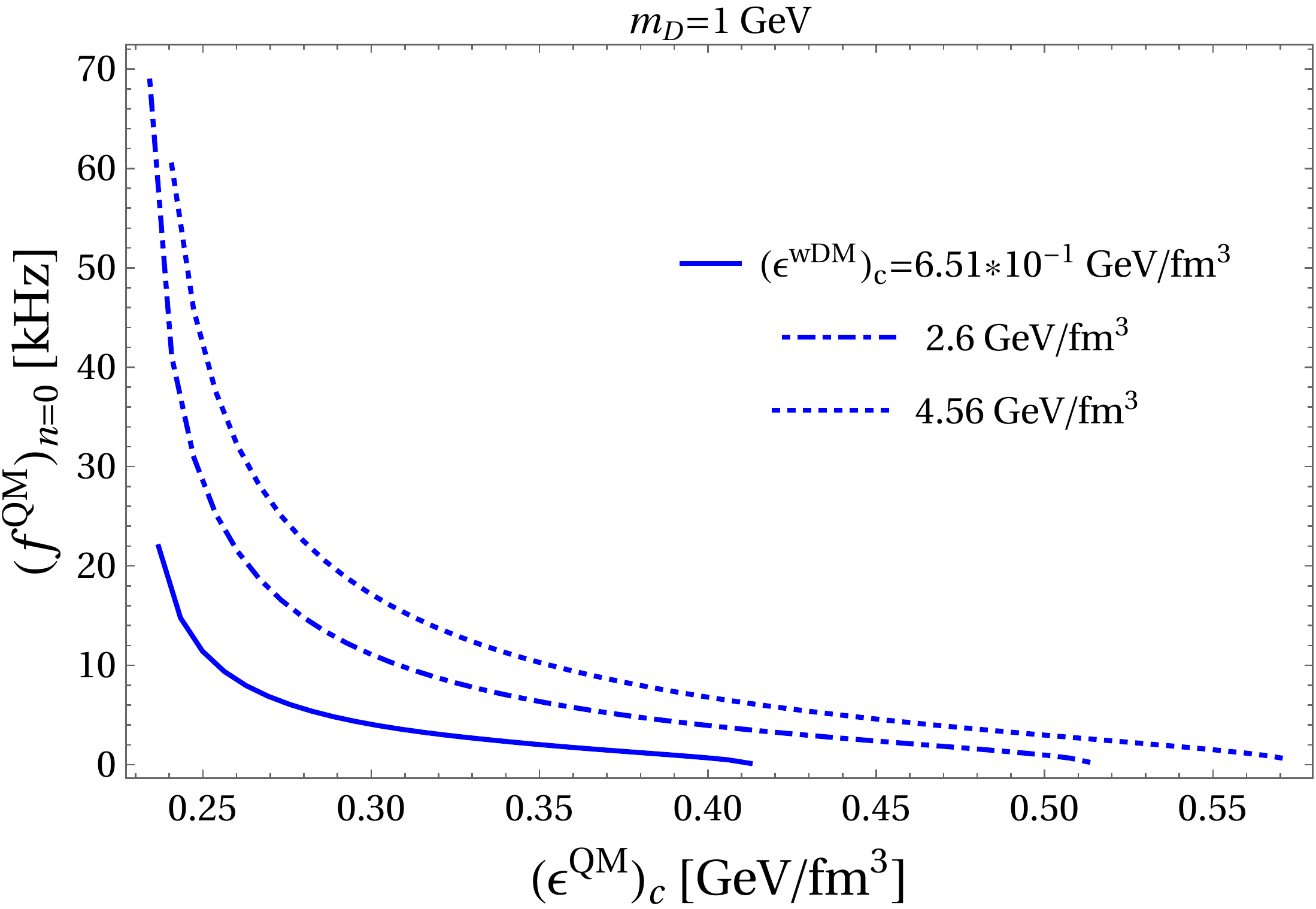}}\\\vspace{5pt}
{\includegraphics[scale=0.335]{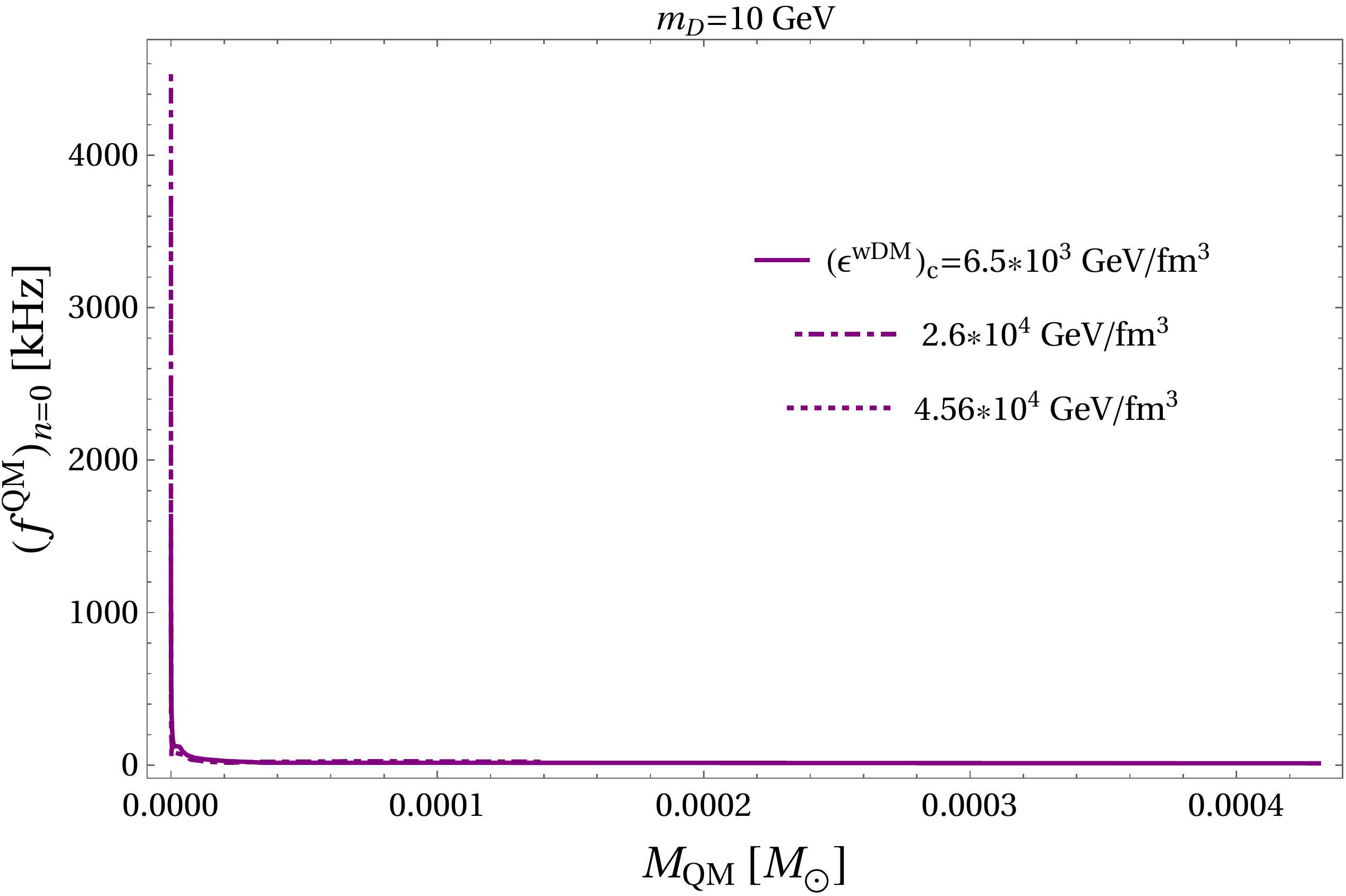}\vspace{3pt}
	  \includegraphics[scale=0.38]{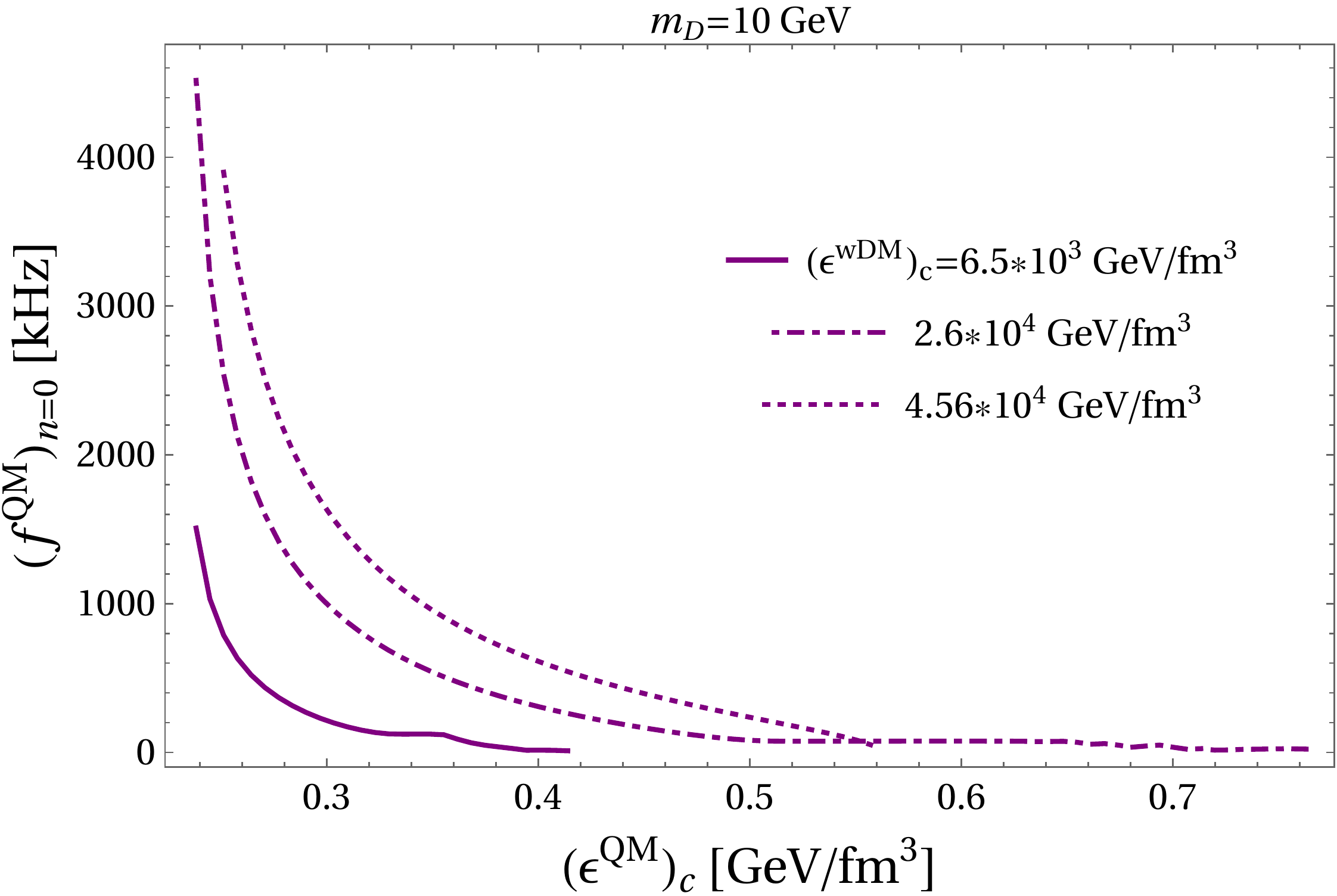}}\\
\end{adjustwidth}
\caption{\emph{Cont.}}
\label{fig:f0Mdens01QM}

\end{figure}

\begin{figure}[H]\ContinuedFloat

\begin{adjustwidth}{-\extralength}{0cm}
\centering 

{\includegraphics[scale=0.335]{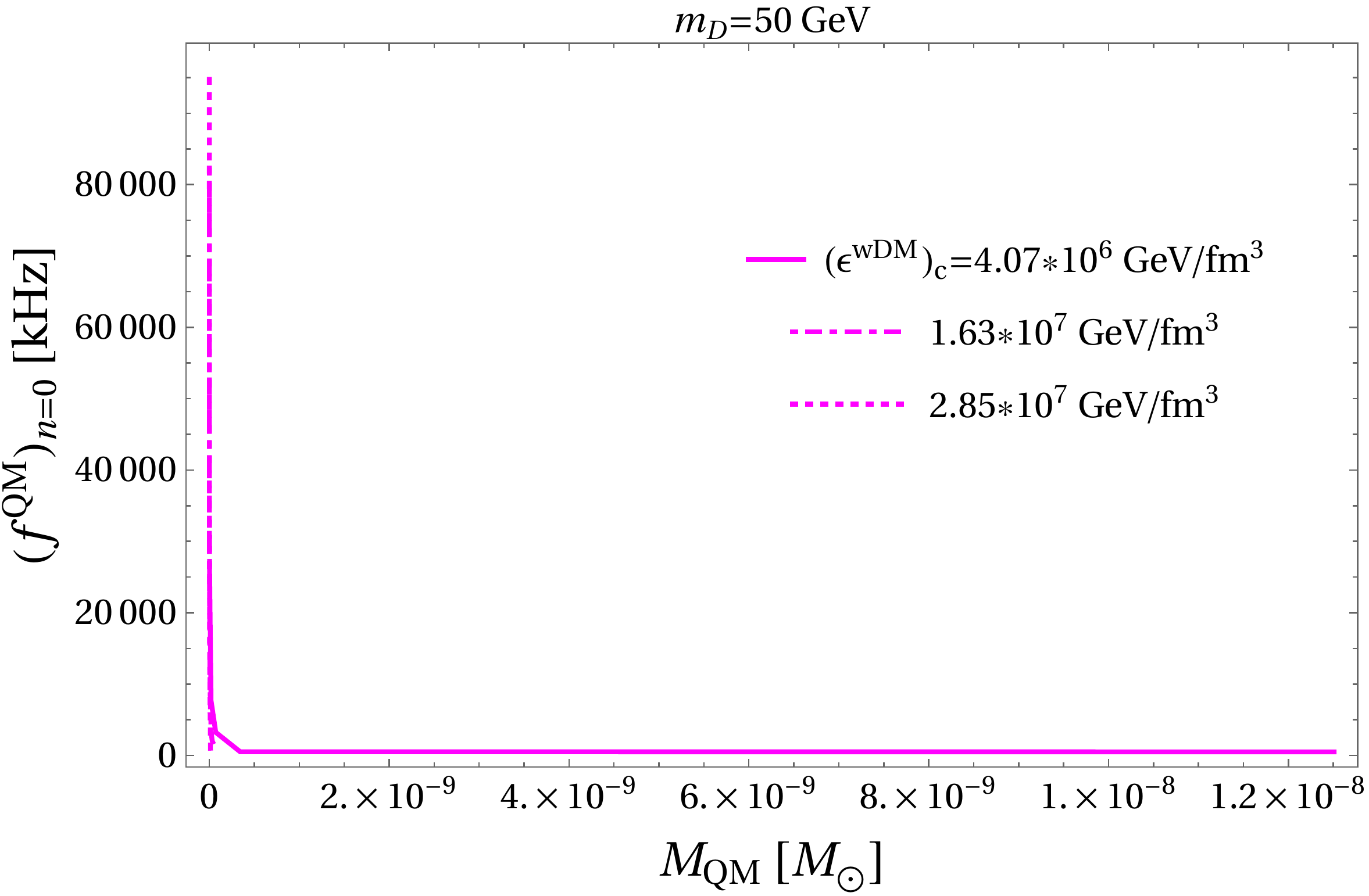}\vspace{3pt}
	  \includegraphics[scale=0.35]{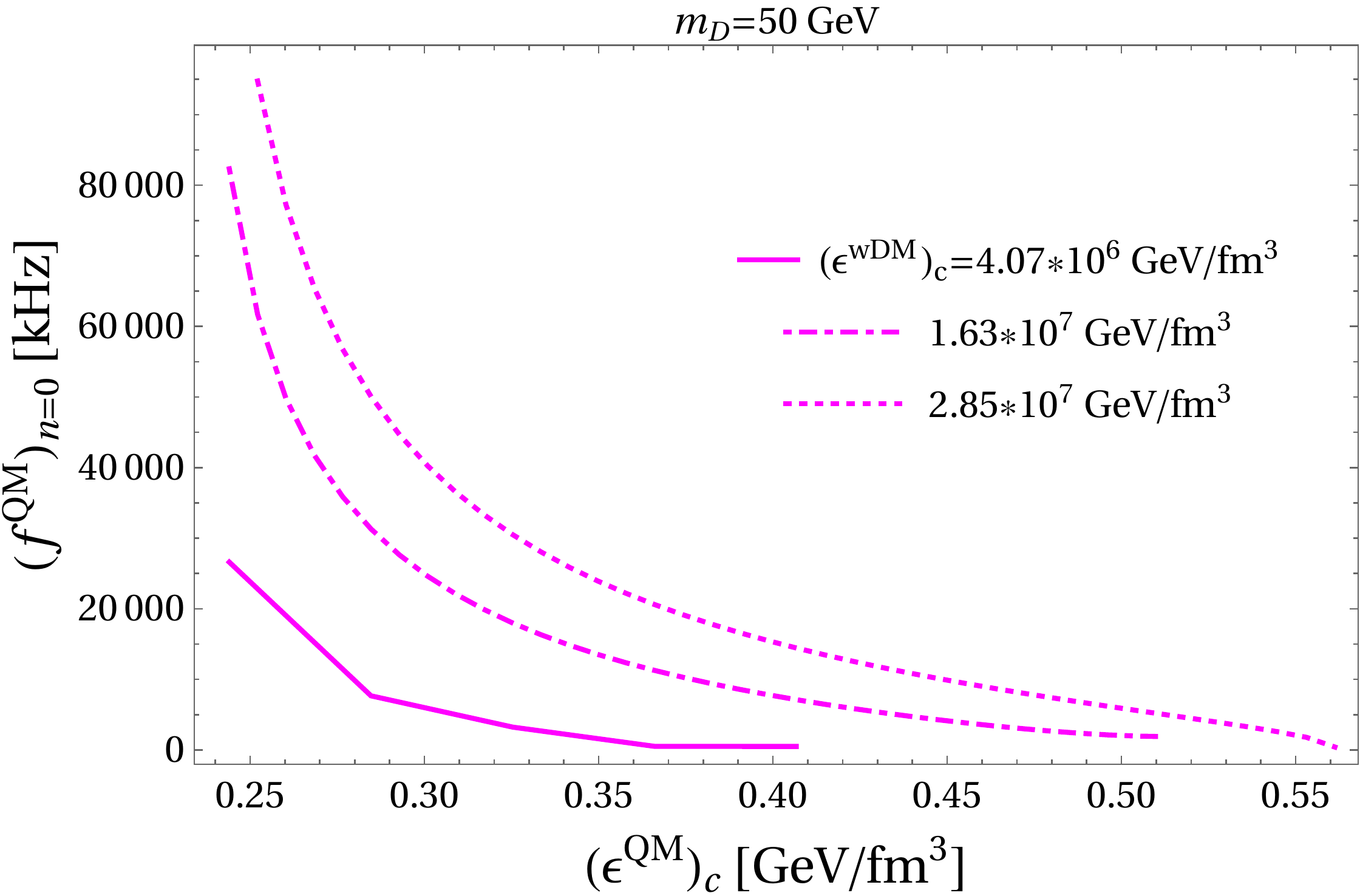}}\\\vspace{5pt}
{\includegraphics[scale=0.33]{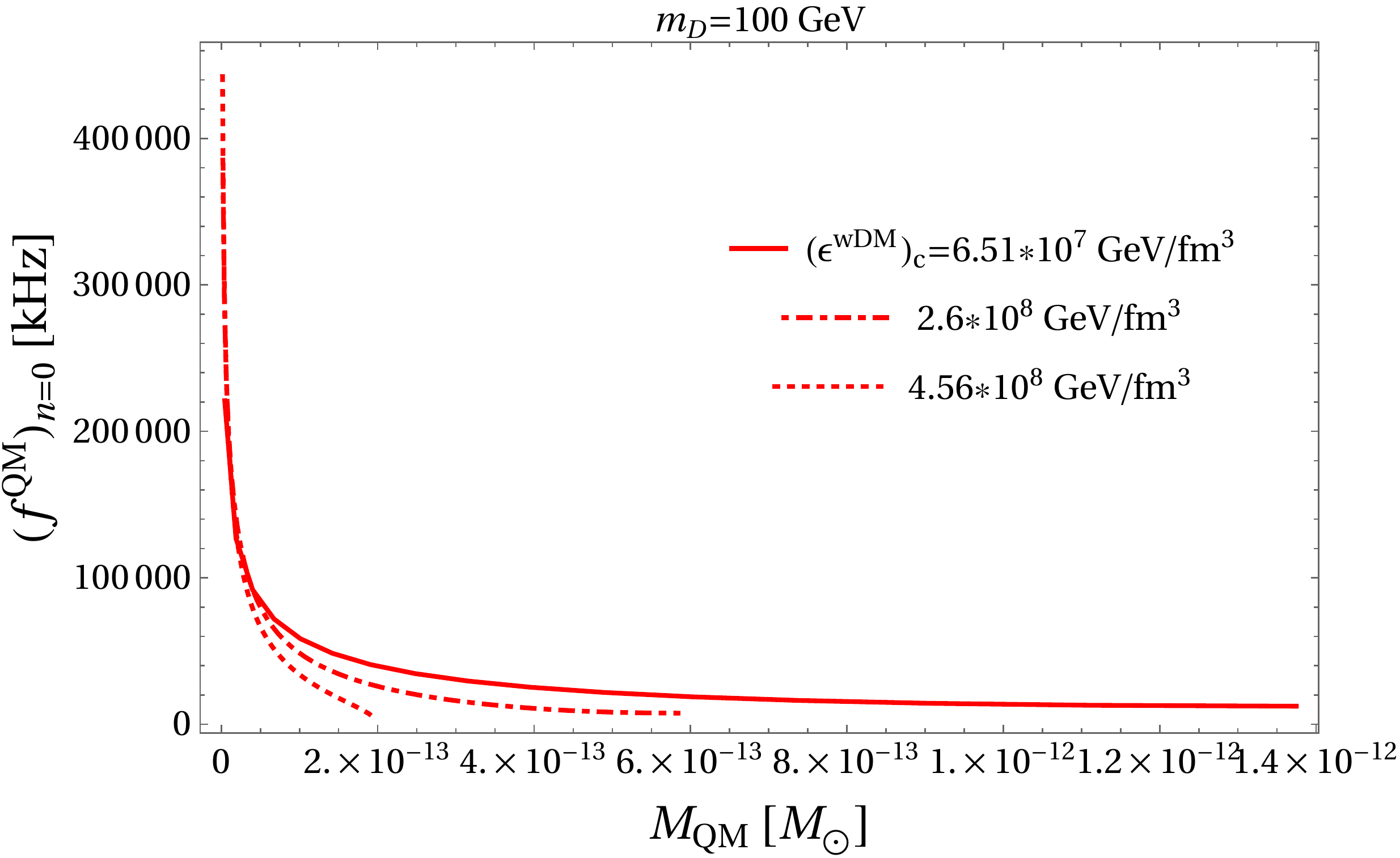}\vspace{1pt}
	  \includegraphics[scale=0.325]{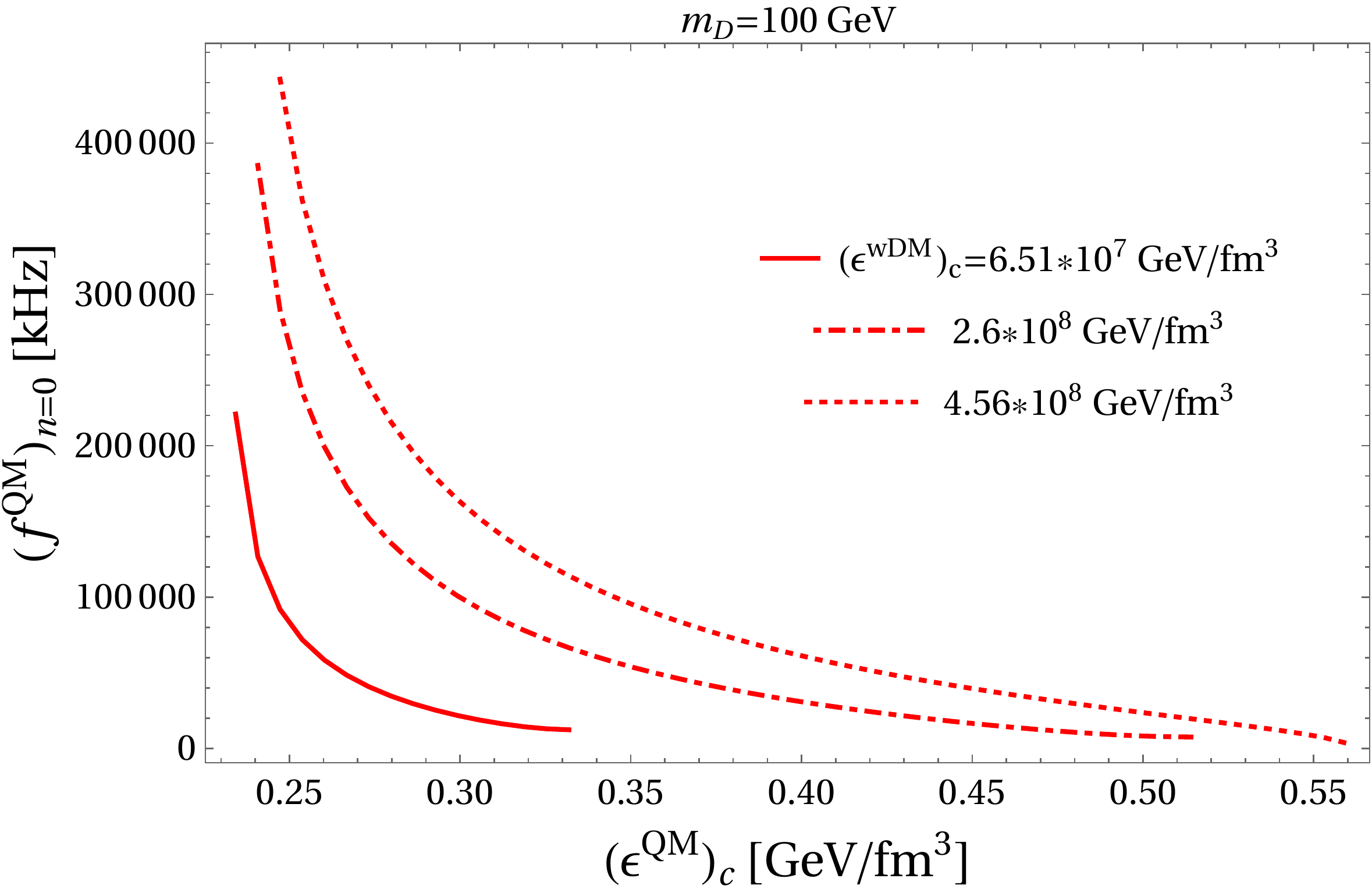}}\\\vspace{5pt}
{\includegraphics[scale=0.36]{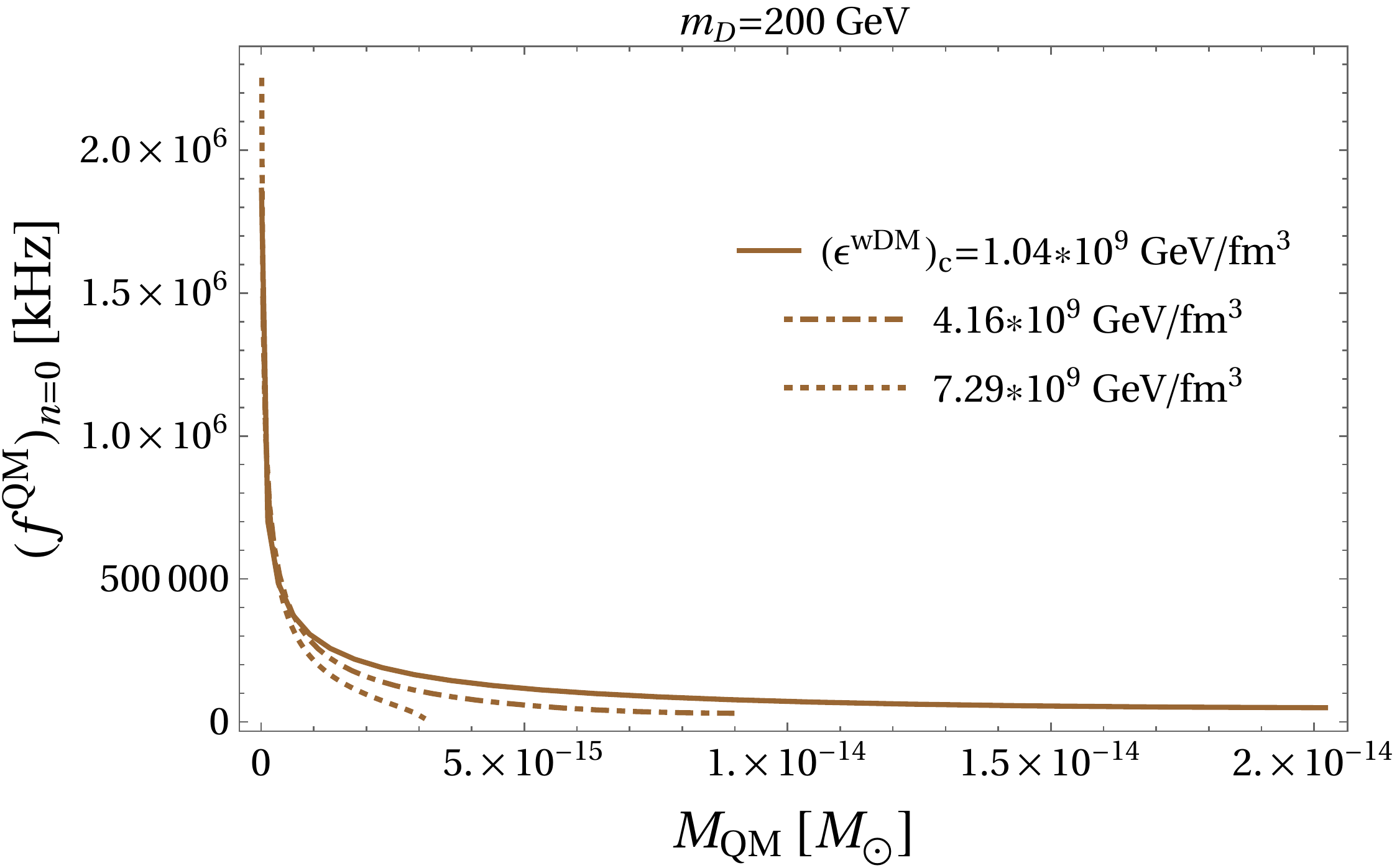}\vspace{3pt}
	  \includegraphics[scale=0.365]{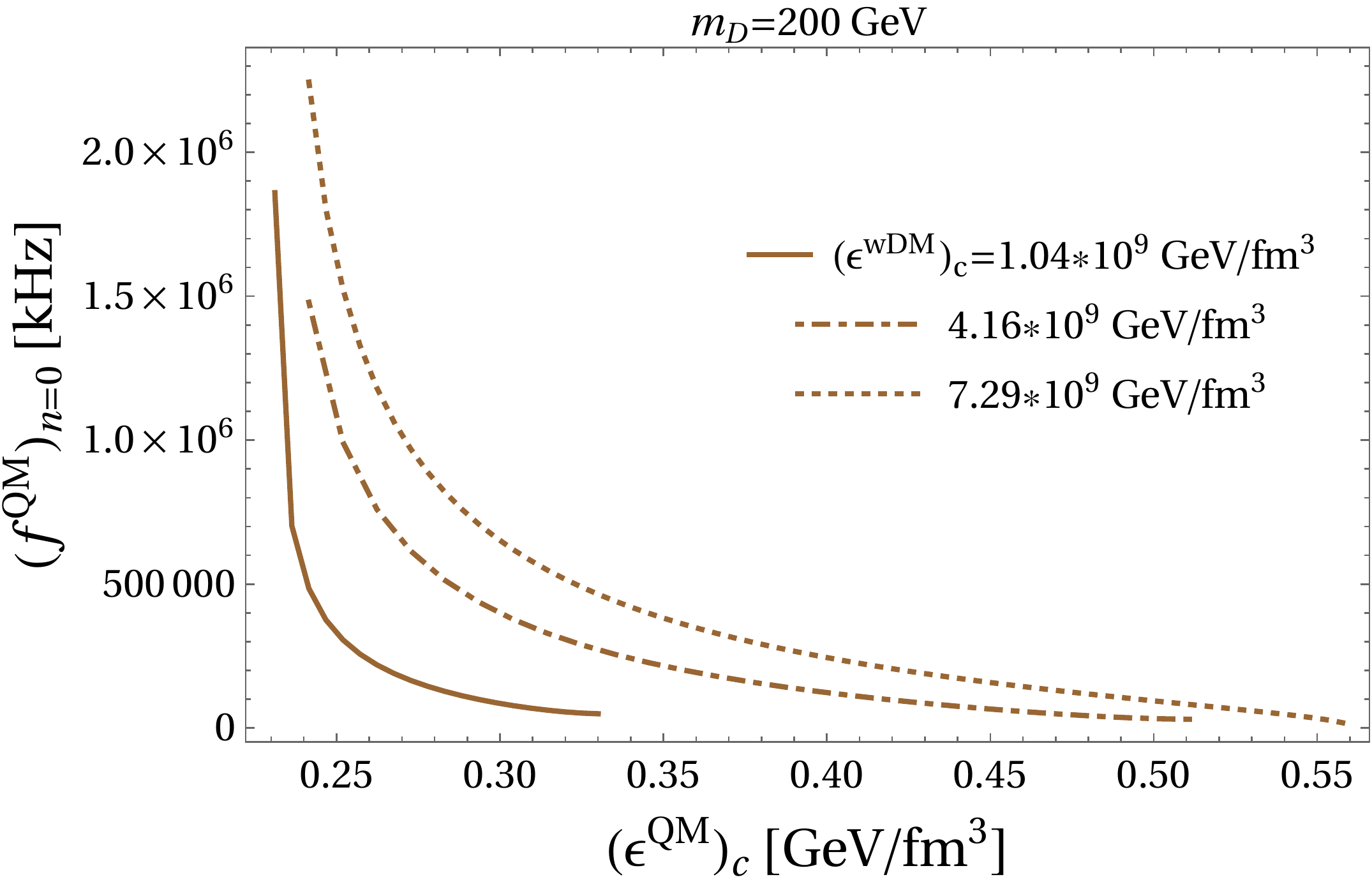}}\\\vspace{5pt}
{\includegraphics[scale=0.34]{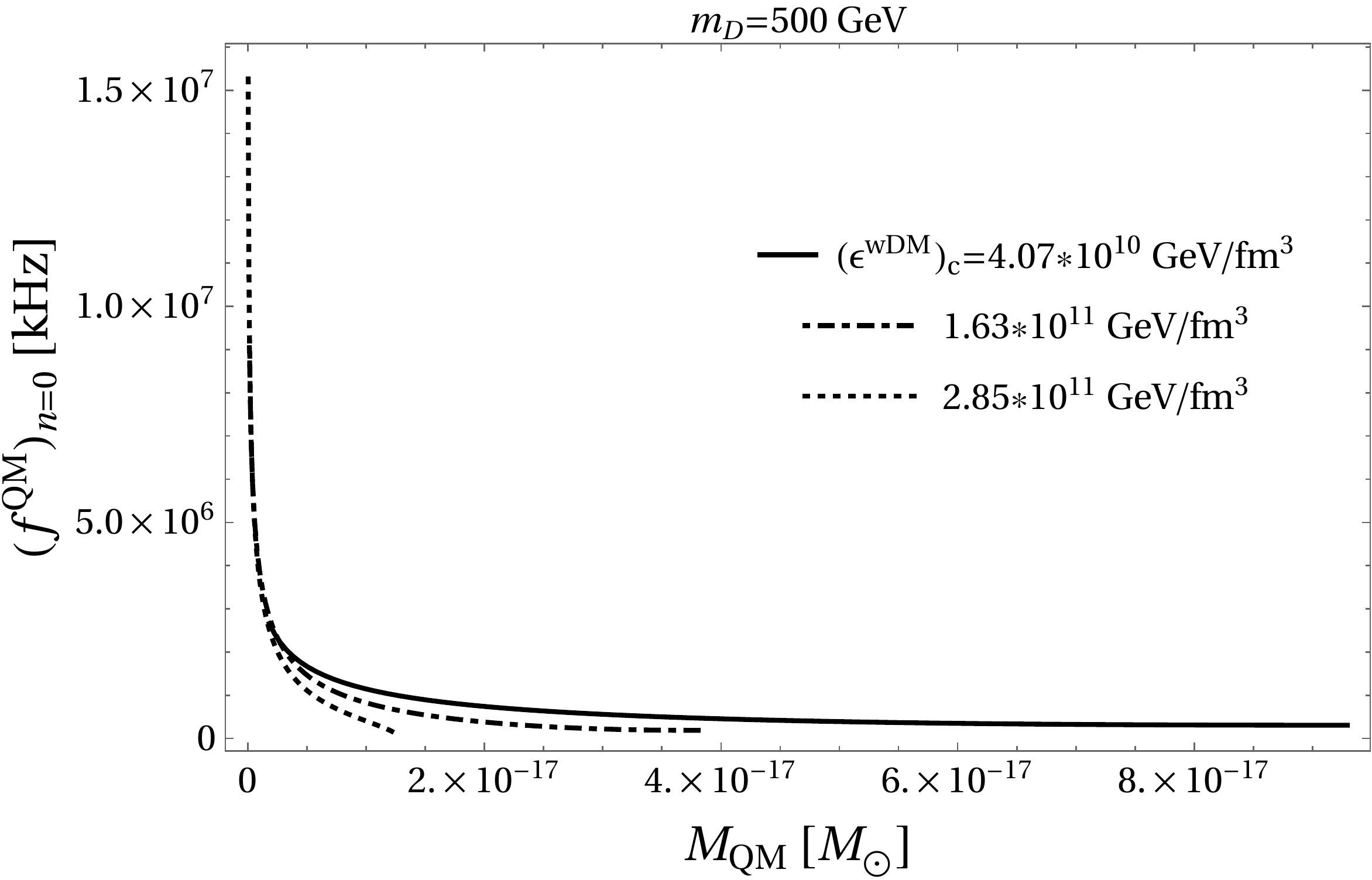}\vspace{3pt}
	  \includegraphics[scale=0.335]{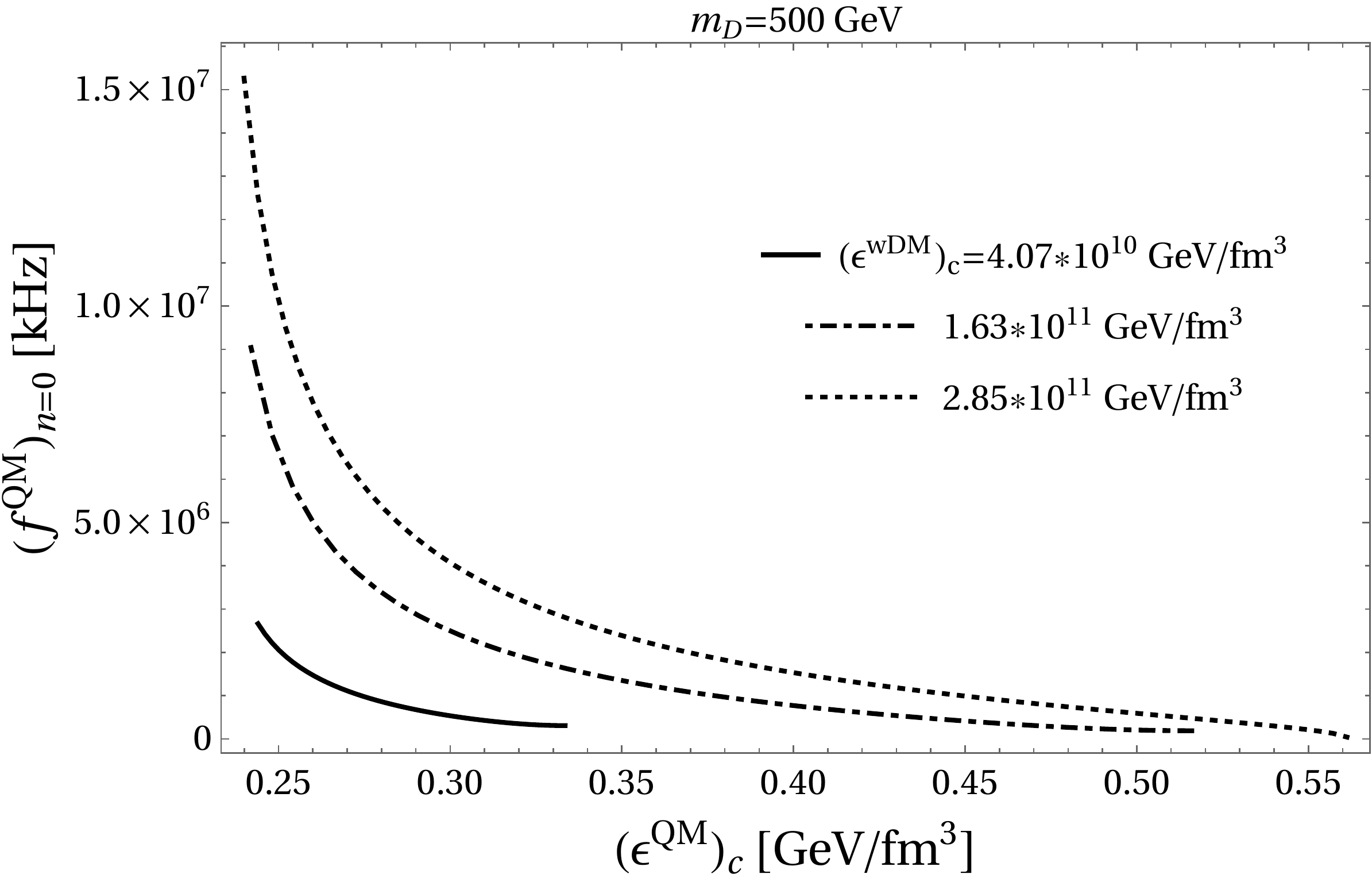}}
\end{adjustwidth}
\caption{Fundamental-mode frequencies, $f_{n=0}$, versus gravitational masses, $M$, and central energy densities, $\epsilon_{c}$, all for the oscillating QM cores with different values of central wDM and dark fermion masses $m_{D}$. Panels with the same color belong to the same class of admixed stars. Notice that the behavior in the planes $f_{n=0}(M)$ is highly dependent on the value of $\epsilon^{\rm wDM}_{c}$, especially for the low-mass QM cores. In a marked contrast, the changes are more modest in the $f_{n=0}(\epsilon_{c})$ planes.}
\label{fig:f0Mdens01QM}

\end{figure}


One can see in Figure \ref{fig:f0Mdens01QM} that the increments in the DM central energy density tend to delay the onset  of radial instability (except in the case of $m_{D}=10$ GeV), which happens when $f^{\rm QM}_{n=0}=0$. At the same time, this results in the maximum QM stellar masses in the admixed star becoming smaller (in some cases by a factor of 10). This opens a new stability window of ultra-low QM masses (when surrounded by DM) in the range between $10^{-18}$ and $10^{-4}$ $M_{\odot}$, depending on $m_{D}$, which correspond to the dark strange planets and strangelets discussed~above.


In the same way, Figure \ref{fig:f0Mdens01DM} shows the results for the coupled radial pulsation Equations (\ref{Rad1}) and (\ref{Rad2}) assuming an oscillating DM core with boundary conditions (\ref{BC1}) and (\ref{BC2}) for different fixed central energy densities of QM. Clearly, the general behavior is qualitatively different, resembling the behavior of nucleonic stars. However, frequencies are very large, reaching $\sim$$3\times{10^{5}}$ kHz, in contrast with the few and tens of kHz for hadronic and quark stars, respectively \cite{Jimenez:2019iuc}. In almost all cases, the DM core is essentially unaffected by QM due to its very large central density, the exception being the case with $m_{D}=1$ GeV. On the other hand, non-trivial effects show up in the $f^{\rm DM}_{n=0}$ vs. $M$ diagrams: the maximum stable masses do not correspond to the the ones in the mass--radius diagram of Figure \ref{fig:MRdens01DM}; they are smaller.

\begin{figure}[H]

\begin{adjustwidth}{-\extralength}{0cm}
\centering 

{\includegraphics[scale=0.38]{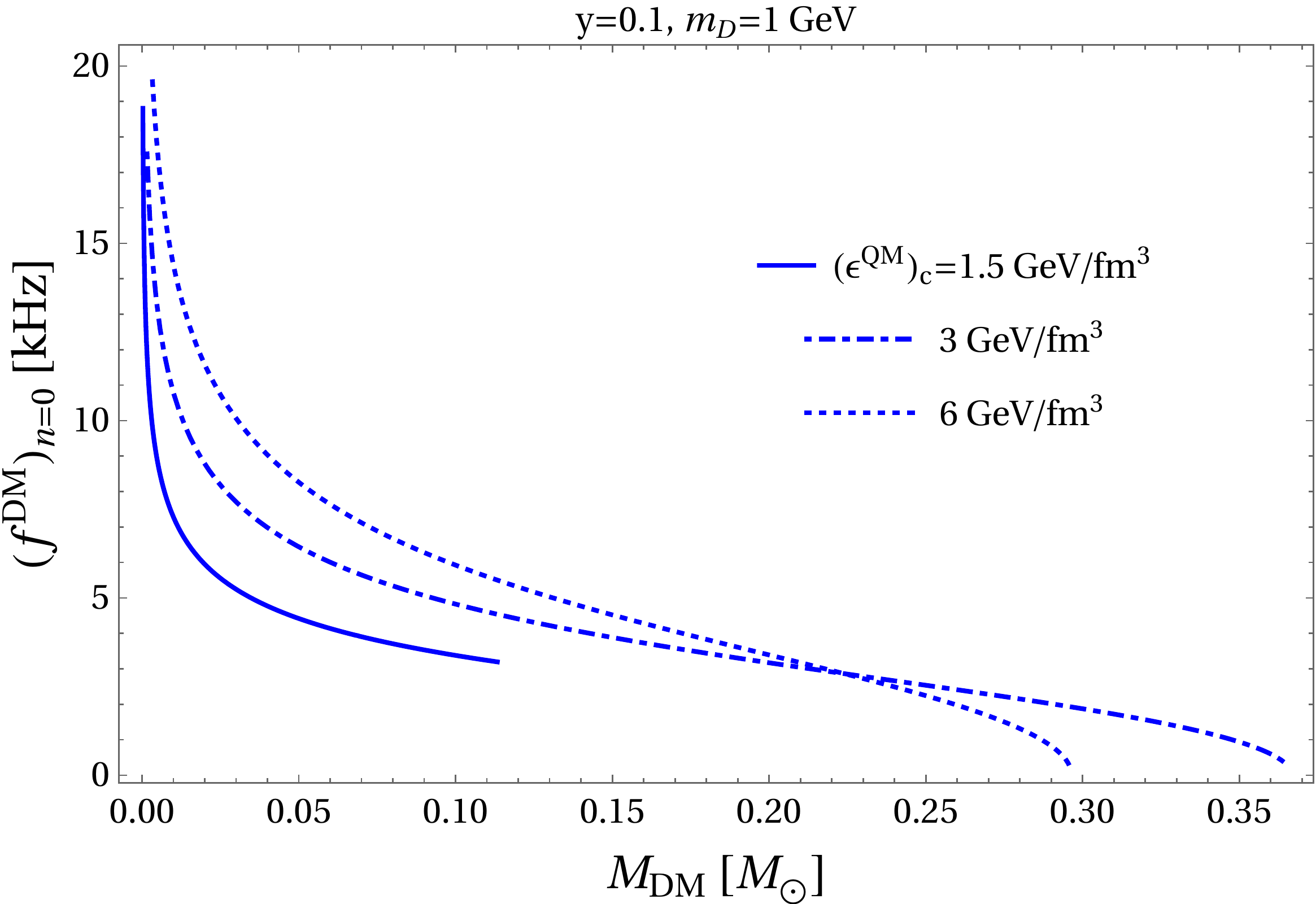}\vspace{3pt}
	  \includegraphics[scale=0.405]{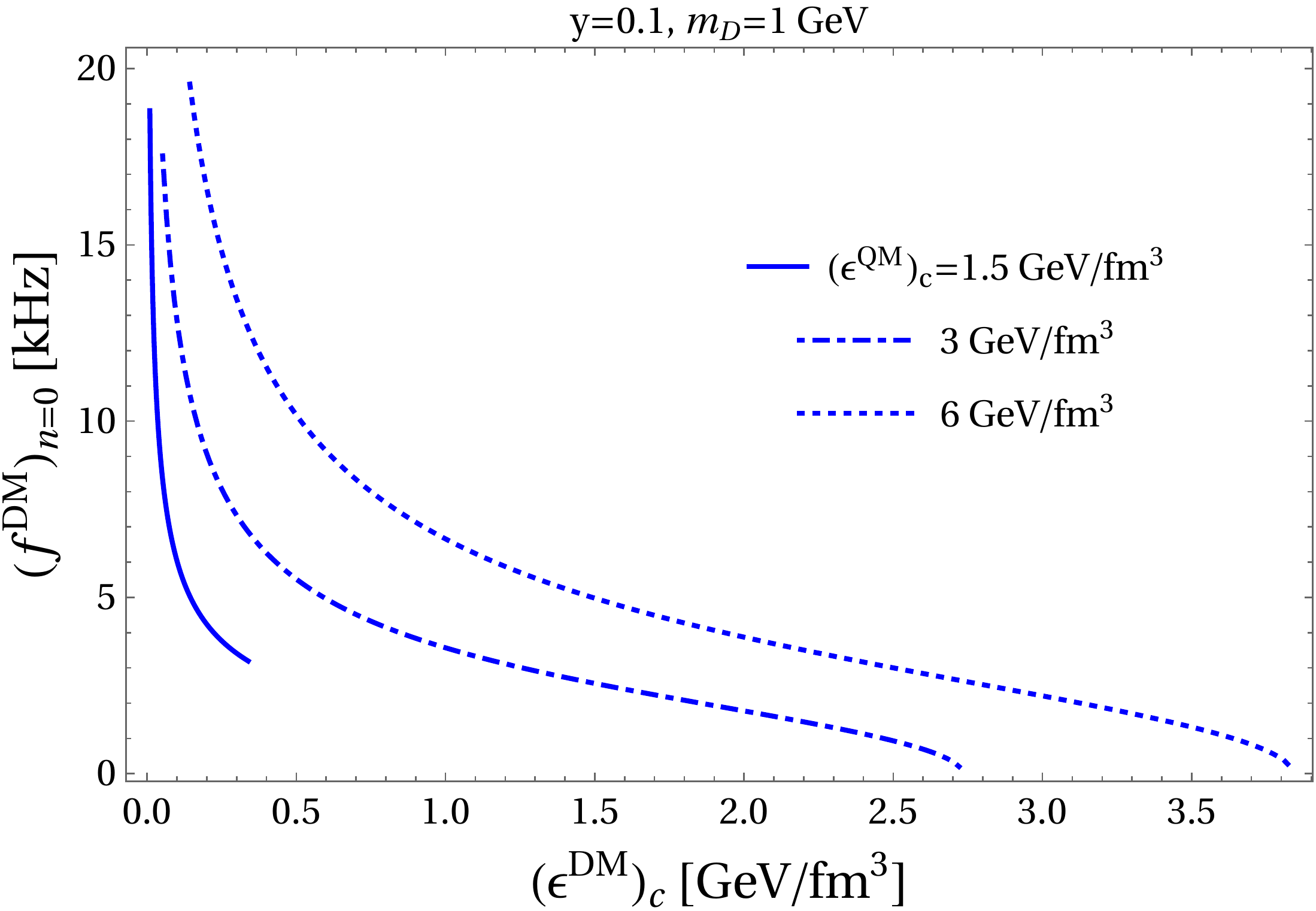}}\\\vspace{5pt}
{\includegraphics[scale=0.38]{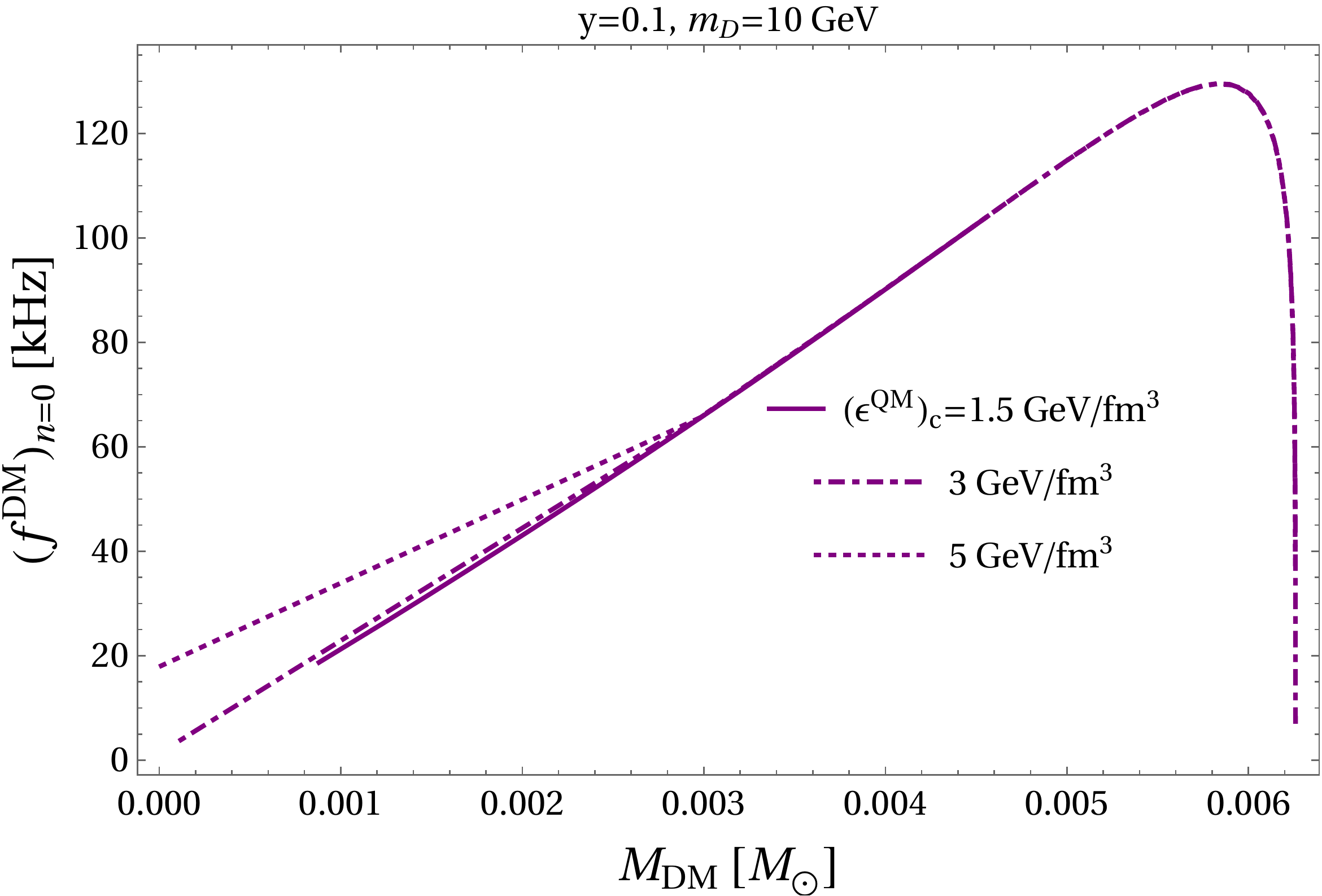}\vspace{3pt}
	  \includegraphics[scale=0.4]{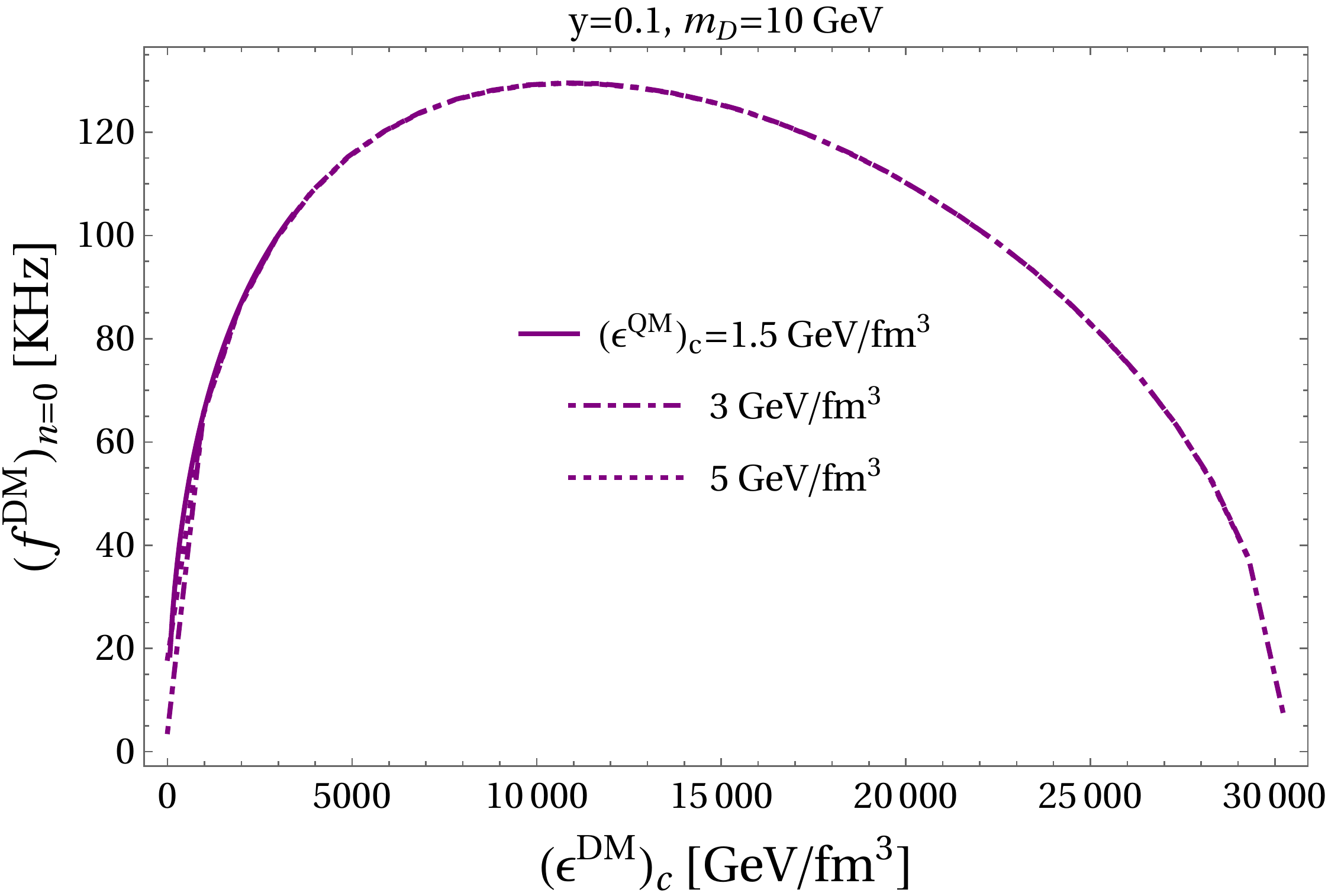}}\\
\end{adjustwidth}
\caption{\emph{Cont.}}
\label{fig:f0Mdens01DM}

\end{figure}

\begin{figure}[H]\ContinuedFloat

\begin{adjustwidth}{-\extralength}{0cm}
\centering 

{\includegraphics[scale=0.33]{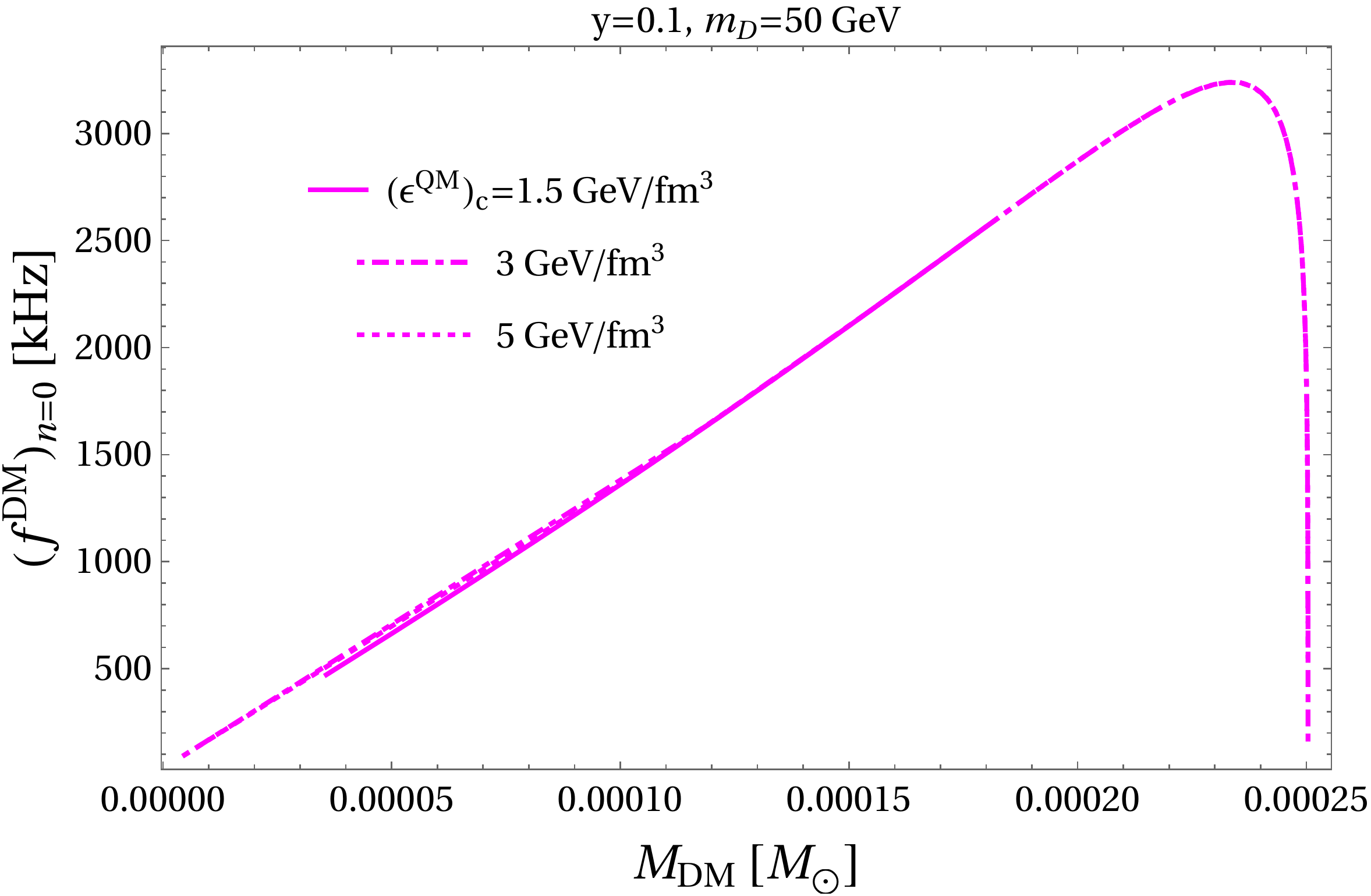}\vspace{1pt}
	  \includegraphics[scale=0.335]{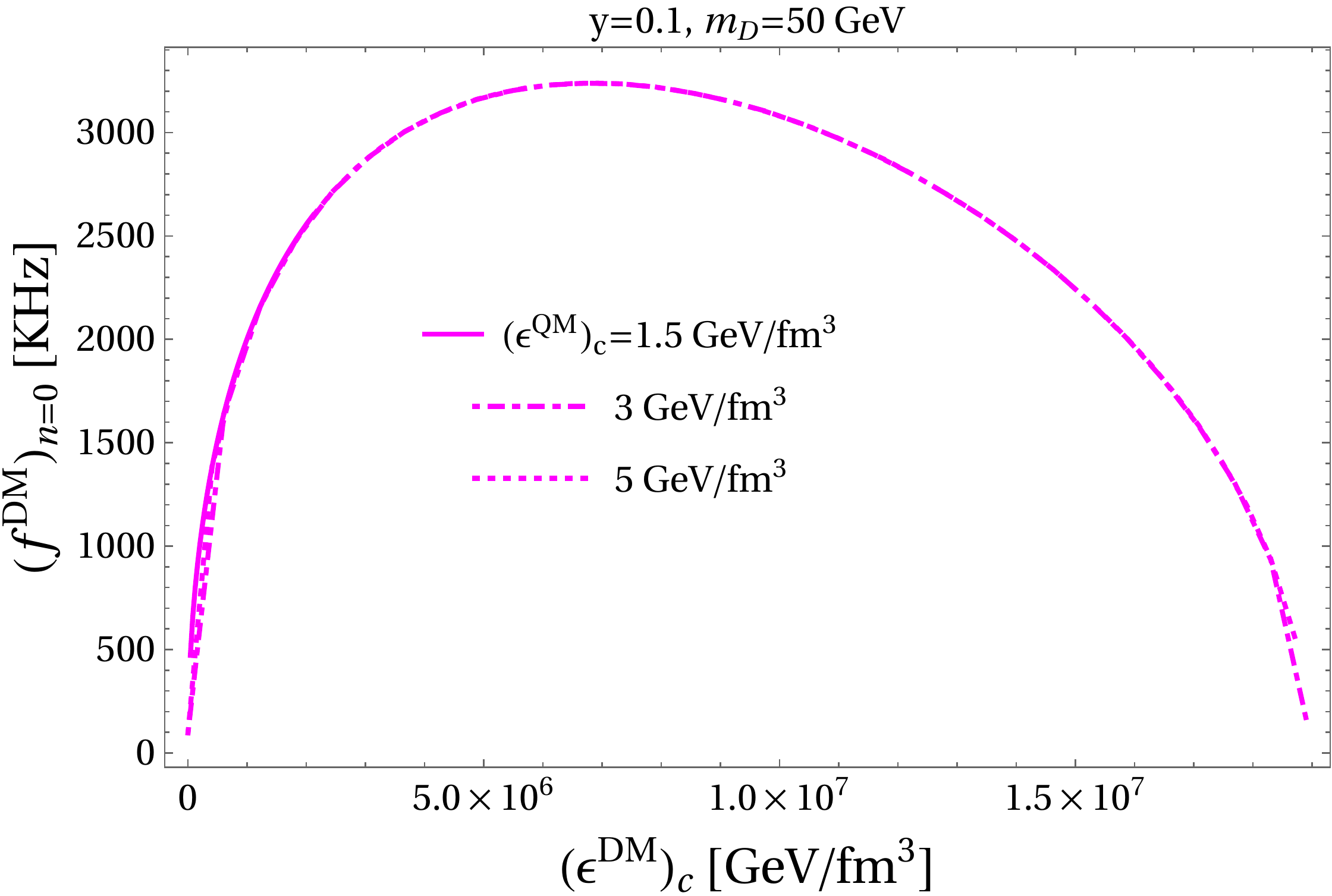}}\\\vspace{5pt}
{\includegraphics[scale=0.315]{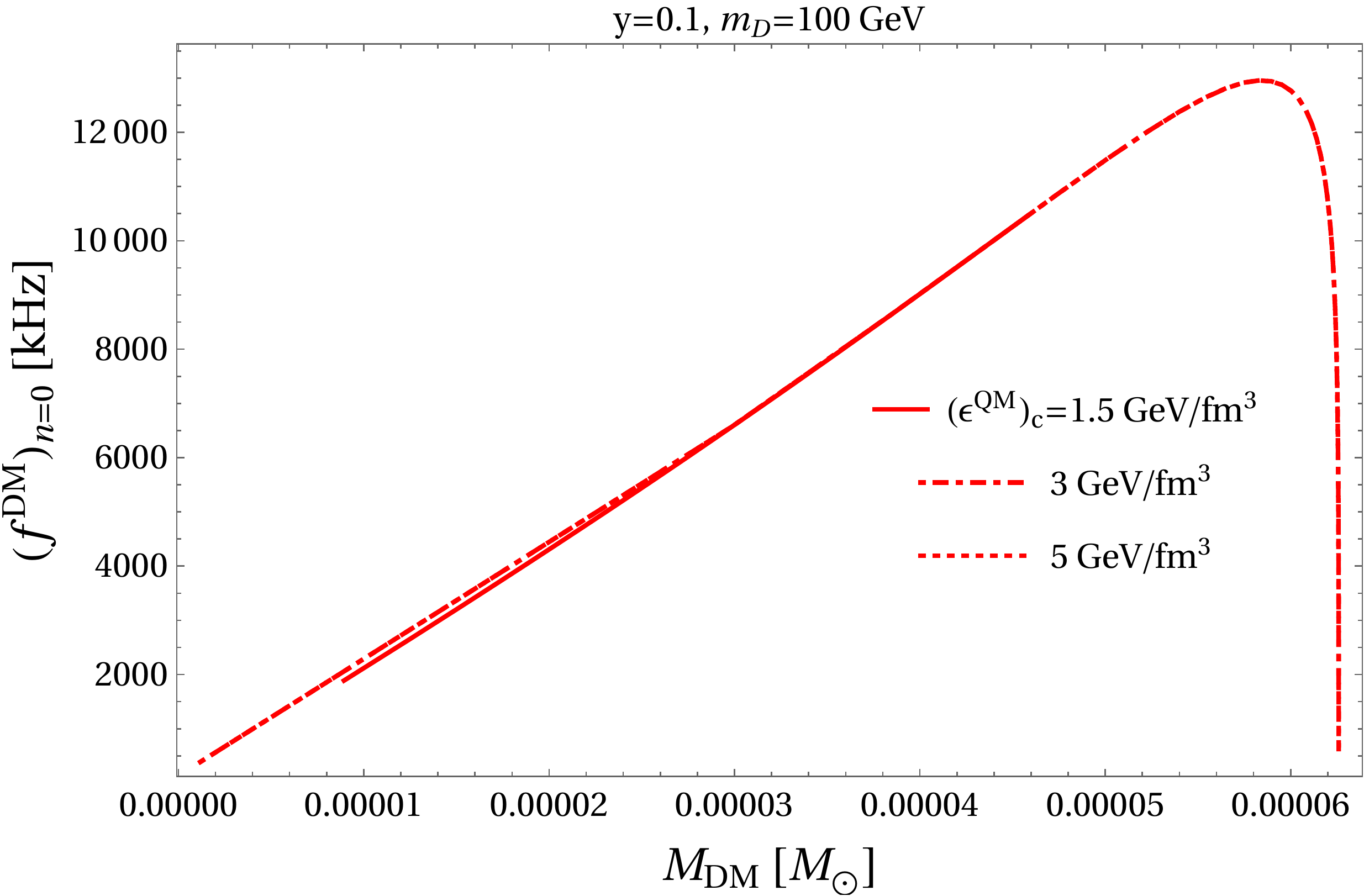}\vspace{3pt}
	  \includegraphics[scale=0.335]{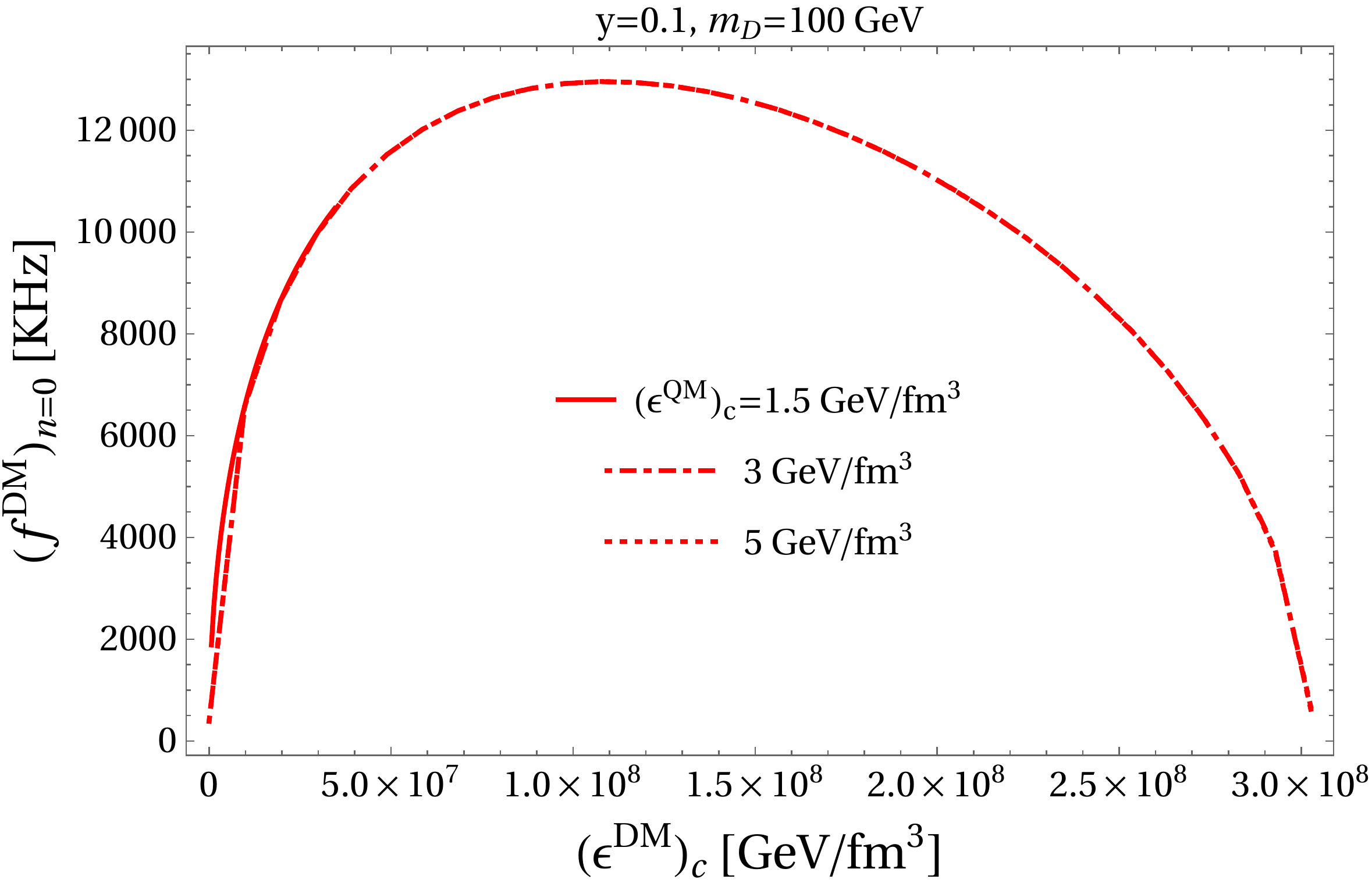}}\\\vspace{5pt}
{\includegraphics[scale=0.34]{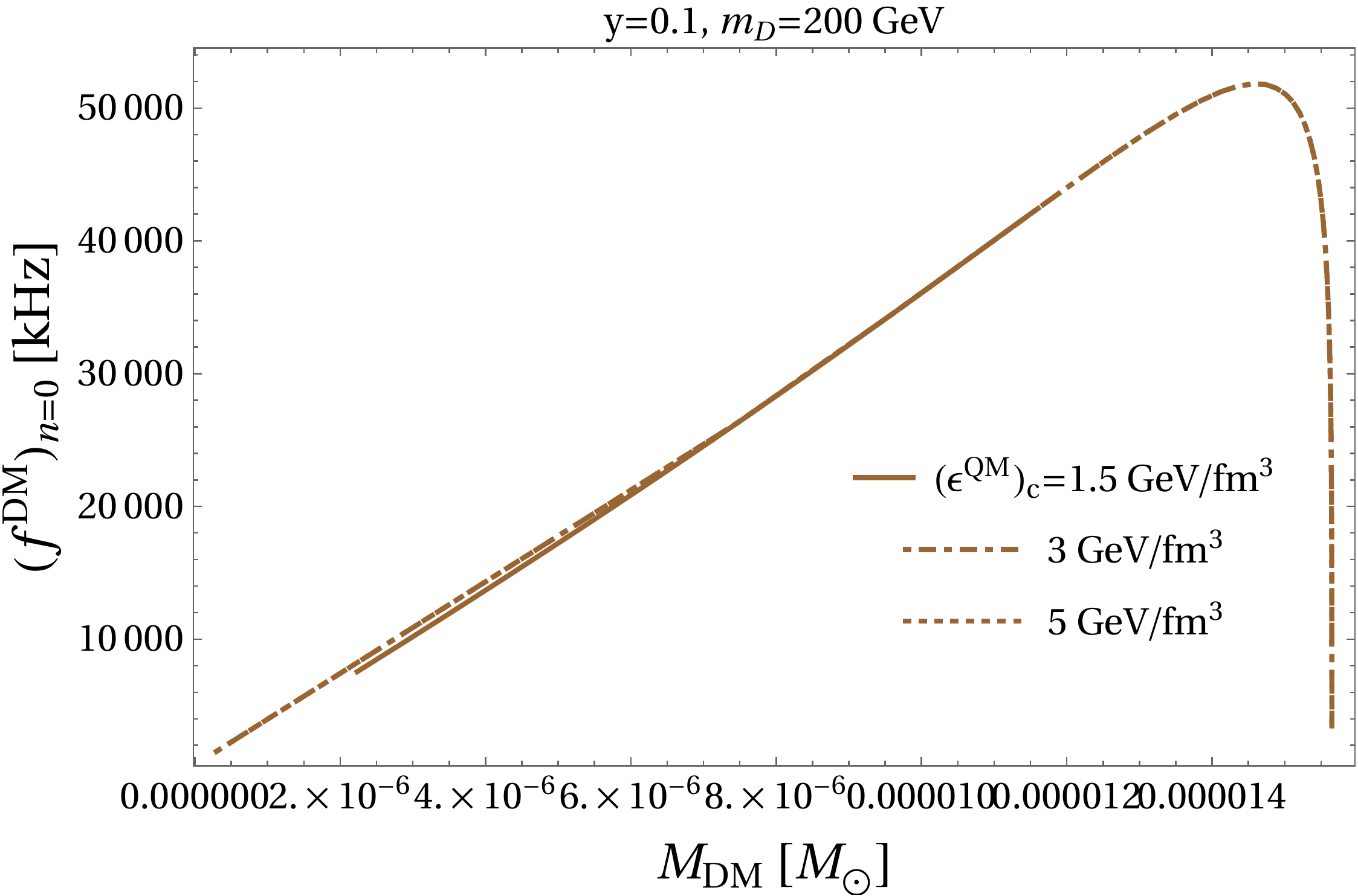}\vspace{3pt}
	  \includegraphics[scale=0.355]{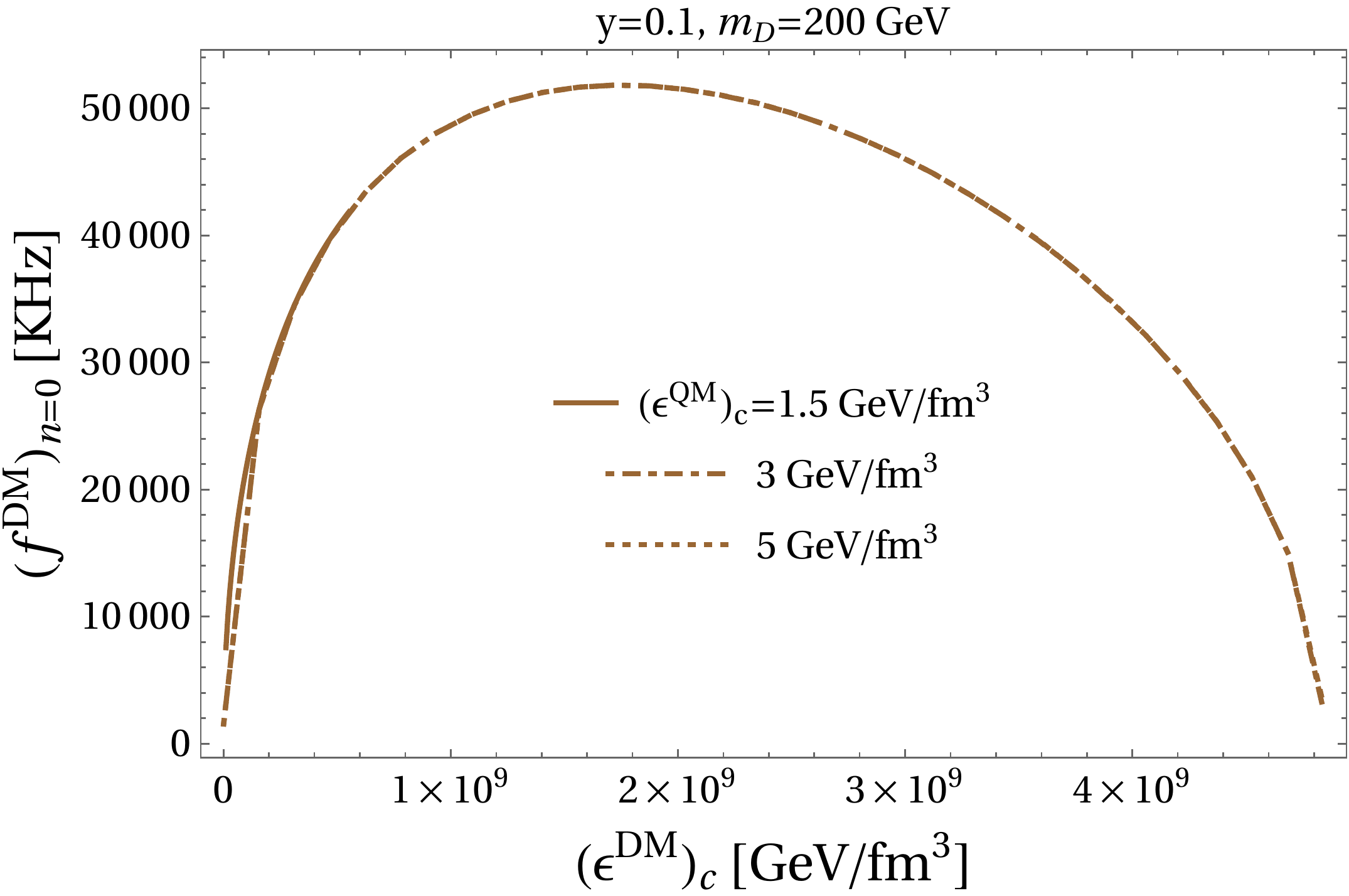}}\\\vspace{5pt}
{\includegraphics[scale=0.34]{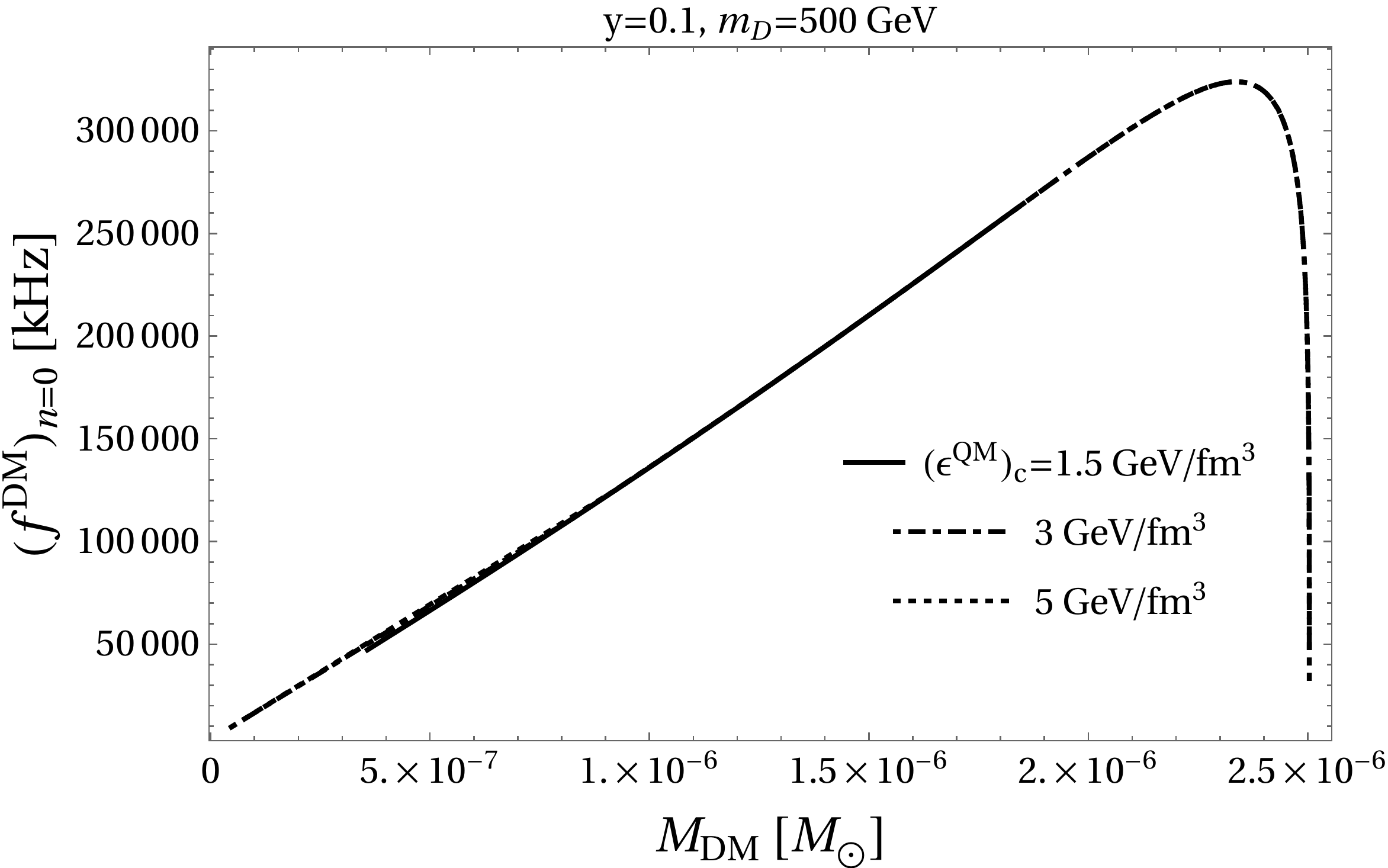}\vspace{1pt}
	  \includegraphics[scale=0.32]{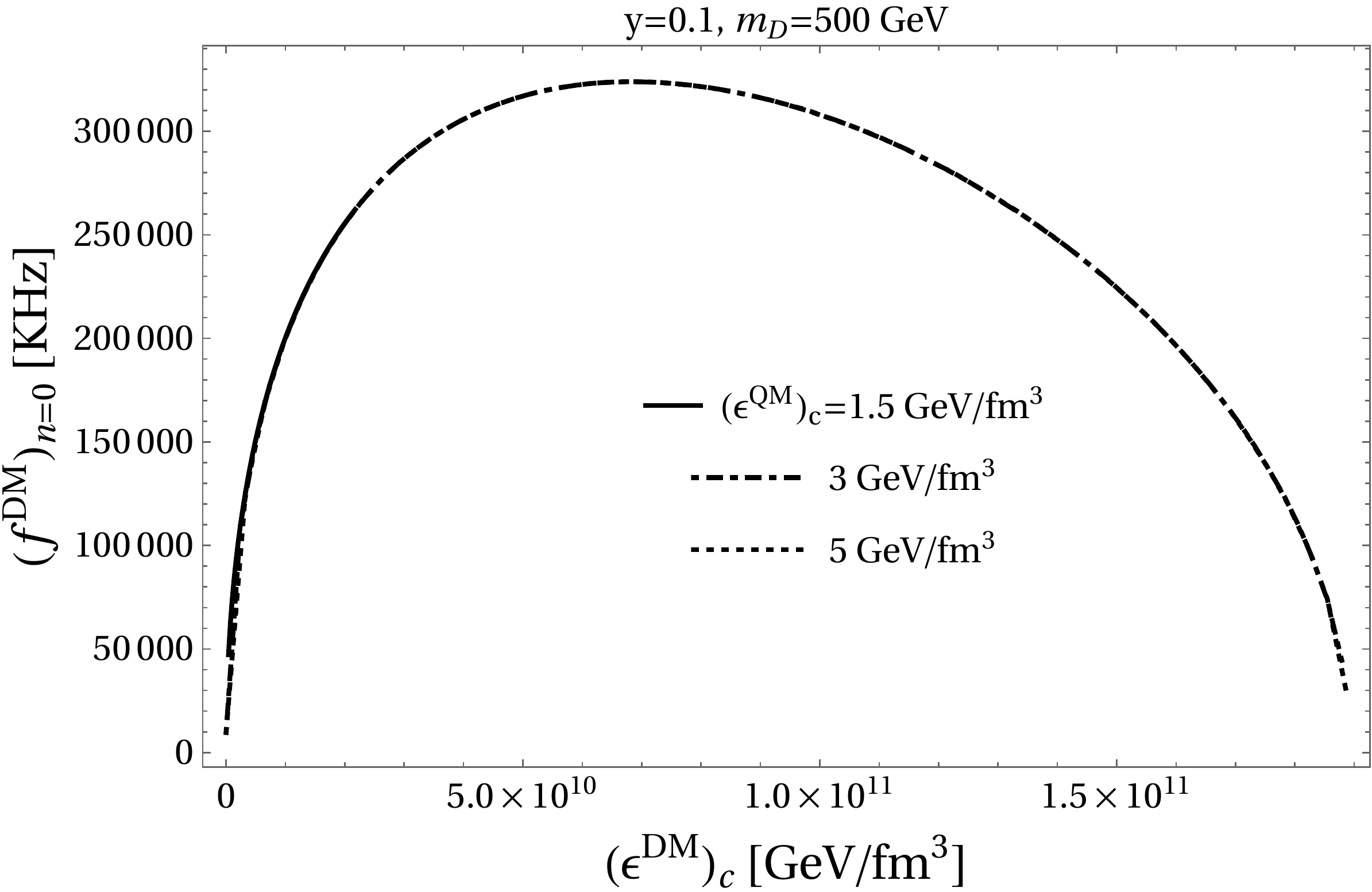}}
\end{adjustwidth}
\caption{Same notation as in Figure \ref{fig:f0Mdens01QM} but now for oscillating wDM cores and different central QM energy densities. Notice that, although the frequencies still reach high values, e.g., $\sim$$3\times{10}^{5}$ KHz for $m_{D}=1$ GeV, the qualitative behavior in the $f_{n=0}(M)$ and $f_{n=0}(\epsilon_{c})$ planes is markedly different and characteristic of dominating wDM in the admixed star for any amount of central QM.}
\label{fig:f0Mdens01DM}

\end{figure}


Furthermore, in Figure \ref{fig:f0Mdens01DM} one sees that, for the case of $\epsilon^{\rm QM}_{c}=1.5~{\rm GeV/fm^{3}}$ with $m_{D}=1$ GeV, zero frequencies are not reached in the corresponding DM stars. After some point, the solutions become mechanically unstable in the sense of having negative QM pressure profiles inside the DM star. A similar phenomenon occurs for larger $m_{D}$, though it is much less visible. Systems exhibiting negative pressures, e.g., dark energy inducing an accelerated expansion of the universe, are not strictly prohibited but should be carefully interpreted. For instance, fluids develop negative-pressure states when stresses are applied for long periods. In the case of hybrid neutron stars with a first-order hadron--quark transition, oscillations to negative-pressure states may accelerate the nucleation of bubbles around the transition region, which, in the limit of large amplitudes, induce mechanical instabilities \cite{Kapusta:2006pm}. In our case of oscillating QM and DM cores with small amplitudes having negative-pressure profiles, the two-fluid TOV equations allows for their existence as hydrostatically-equilibrated configurations that are potentially unstable when disturbed by radial perturbations leading to the automatic collapse of the whole admixed star. In other words, negative-pressure interiors lead immediately to complex oscillation frequencies. Only by increasing $\epsilon^{\rm QM/DM}_{c}$ do the instabilities disappear and one is able to find only real frequencies.
 
\subsection{Admixtures of Quark Matter and Strongly $(y=10^{3})$ Interacting Dark Matter}

\subsubsection{Solving the Two-Fluid TOV Equations}

Similar to the case of weakly self-interacting DM, we solved the two-fluid TOV Equation (\ref{eq:2TOV}) with the condition $p_{\rm QM}(R_{\rm QM})=0$ for different central energy densities of strongly self-interacting DM. We present our results in Figure \ref{fig:MRdens1000QM}. As in the weak limit for DM, in most of the cases stellar masses, radii and central energy densities of the QM core are not appreciably affected. However, for increasing DM central energy densities, some relevant variations occur. In particular, when $m_{D}=1, 100, 200$ GeV, the maximum QM central energy densities are increased by a factor of $\sim$$20$. The cases with $m_{D}=10, 50$ GeV show sizable variations of the masses and radii, especially near the maximum mass, where the presence of DM reduces the QM core masses down to $\sim$${0.4}~M_{\odot}$. Interestingly, in the case of $m_{D}=10$ GeV, the central QM energy density is almost unaffected by DM, whereas, for $m_{D}=50$ GeV, it is dramatically increased. In the case with $m_{D}=500$ GeV, the QM masses and radii increase by any amount of DM, and the QM central energy density is augmented by a factor of $10$. Analogously to what we have seen before, in the cases with $m_{D}=50$ to $500$ GeV, there are plateaus whose widths increase with $m_{D}$. 

\begin{figure}[H]

\begin{adjustwidth}{-\extralength}{0cm}
\centering 

{\includegraphics[scale=0.405]{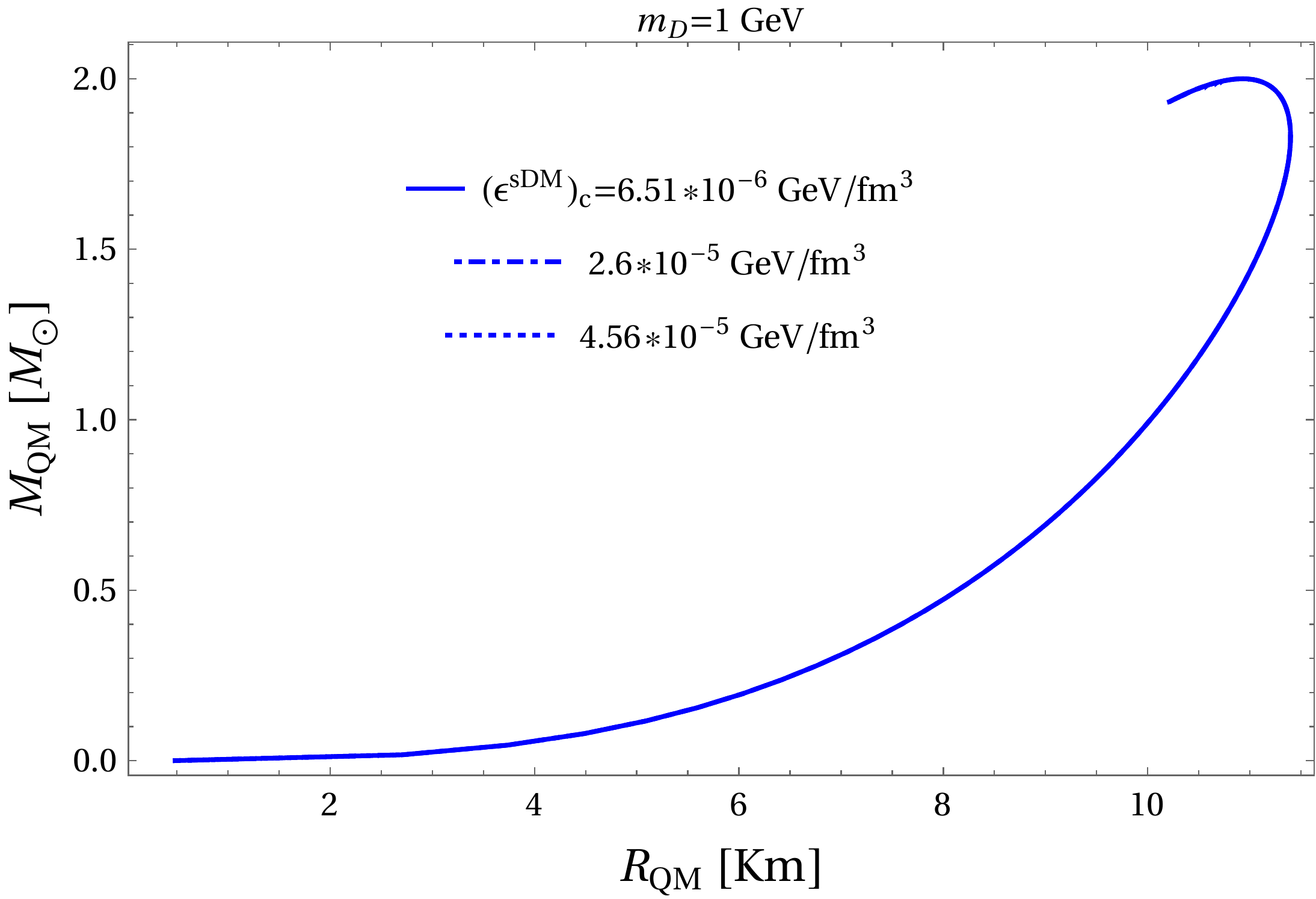}\vspace{3pt}
	  \includegraphics[scale=0.405]{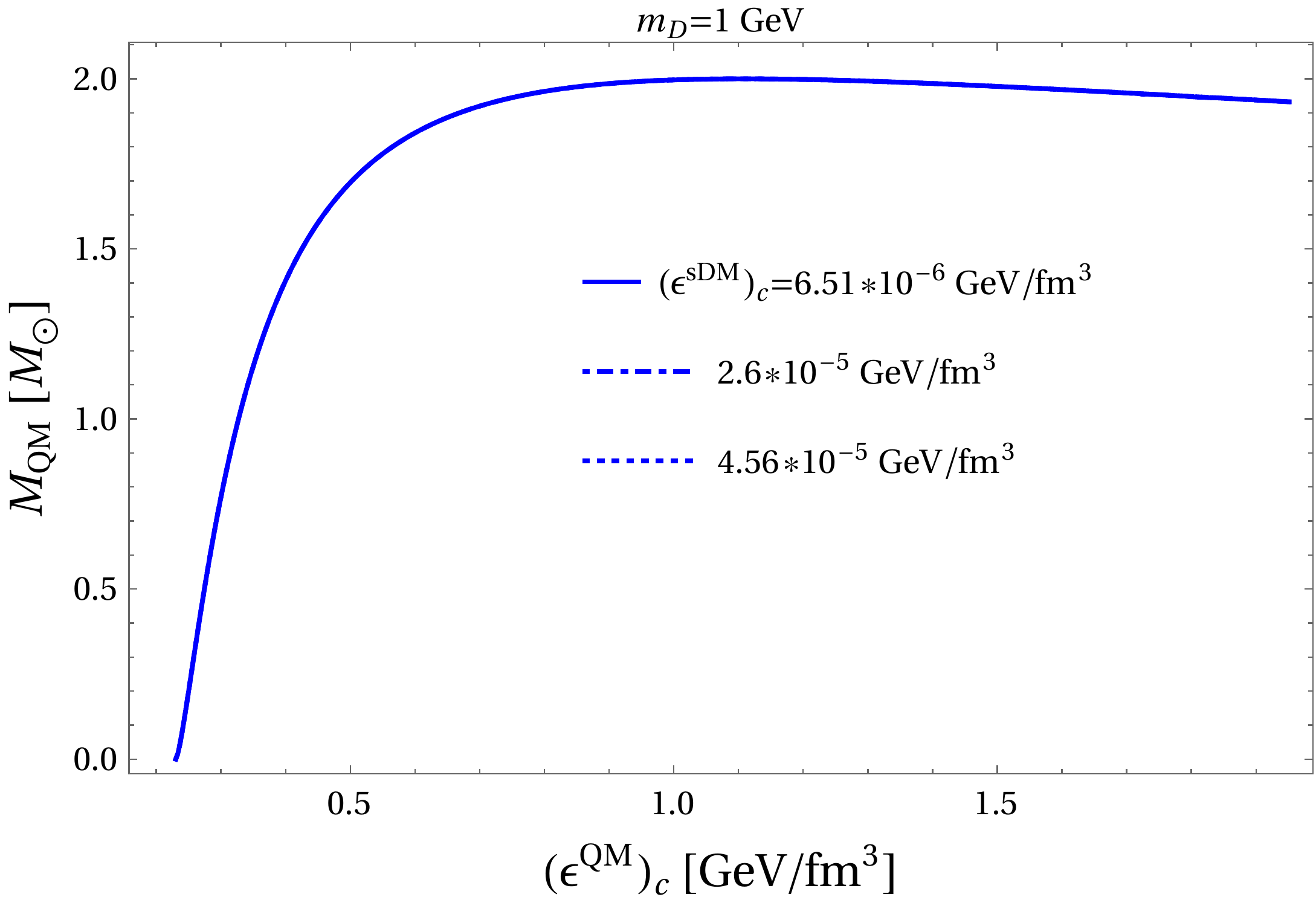}}\\
\end{adjustwidth}
\caption{\emph{Cont.}}
\label{fig:MRdens1000QM}

\end{figure}
\begin{figure}[H]\ContinuedFloat

\begin{adjustwidth}{-\extralength}{0cm}
\centering 

{\includegraphics[width=0.58\textwidth]{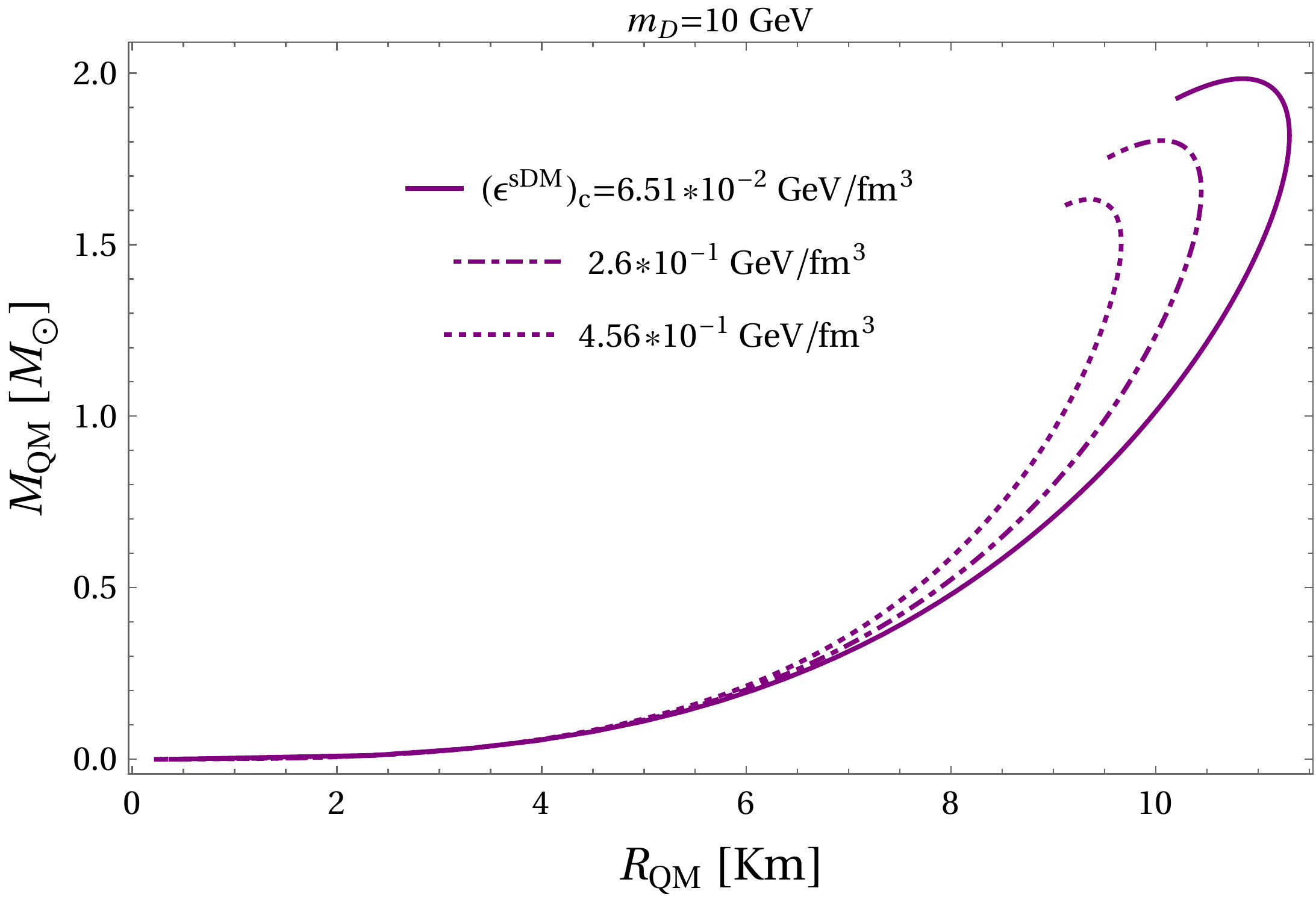}\vspace{3pt}
	  \includegraphics[scale=0.38]{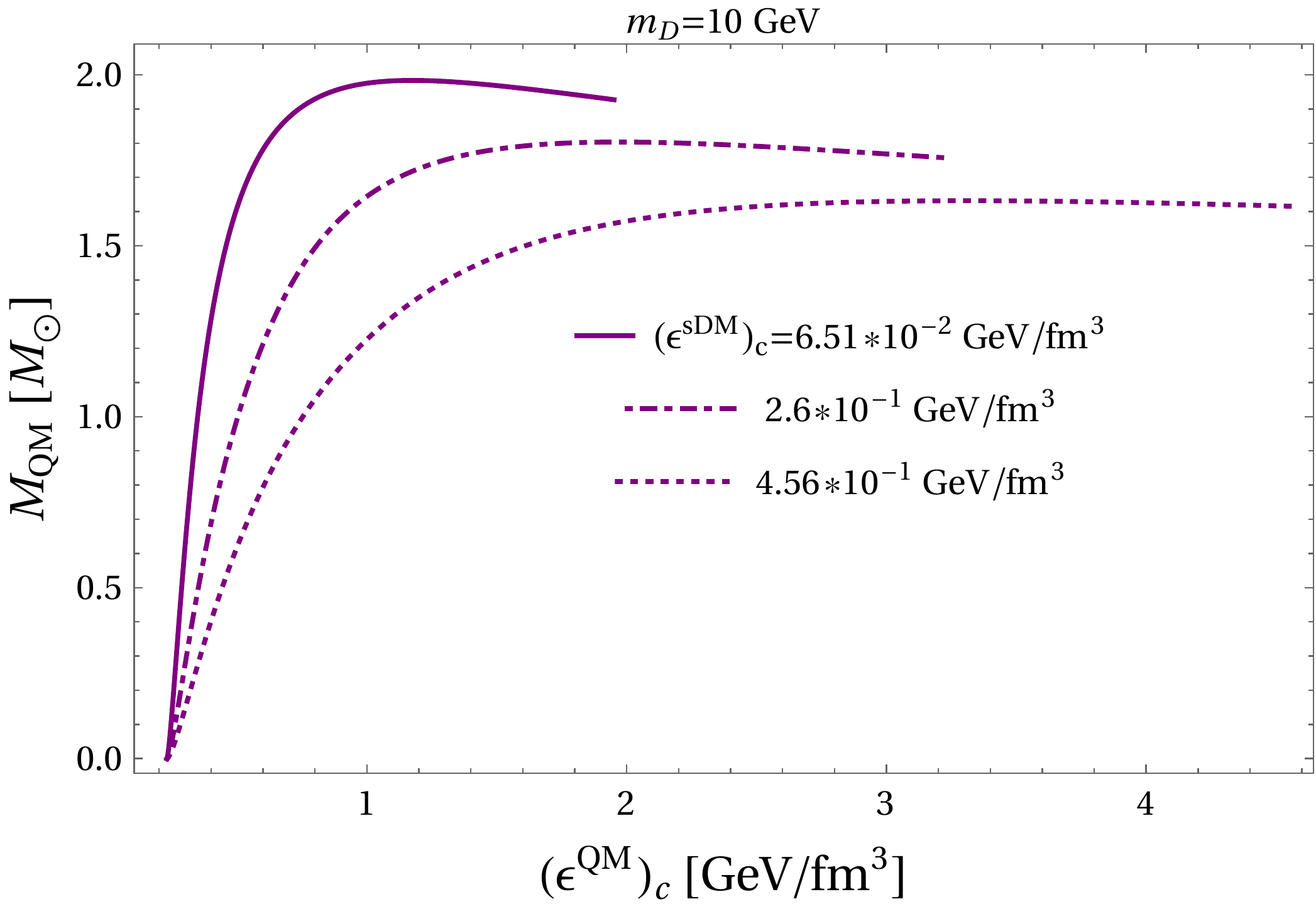}}\\\vspace{2pt}
{\includegraphics[width=0.58\textwidth]{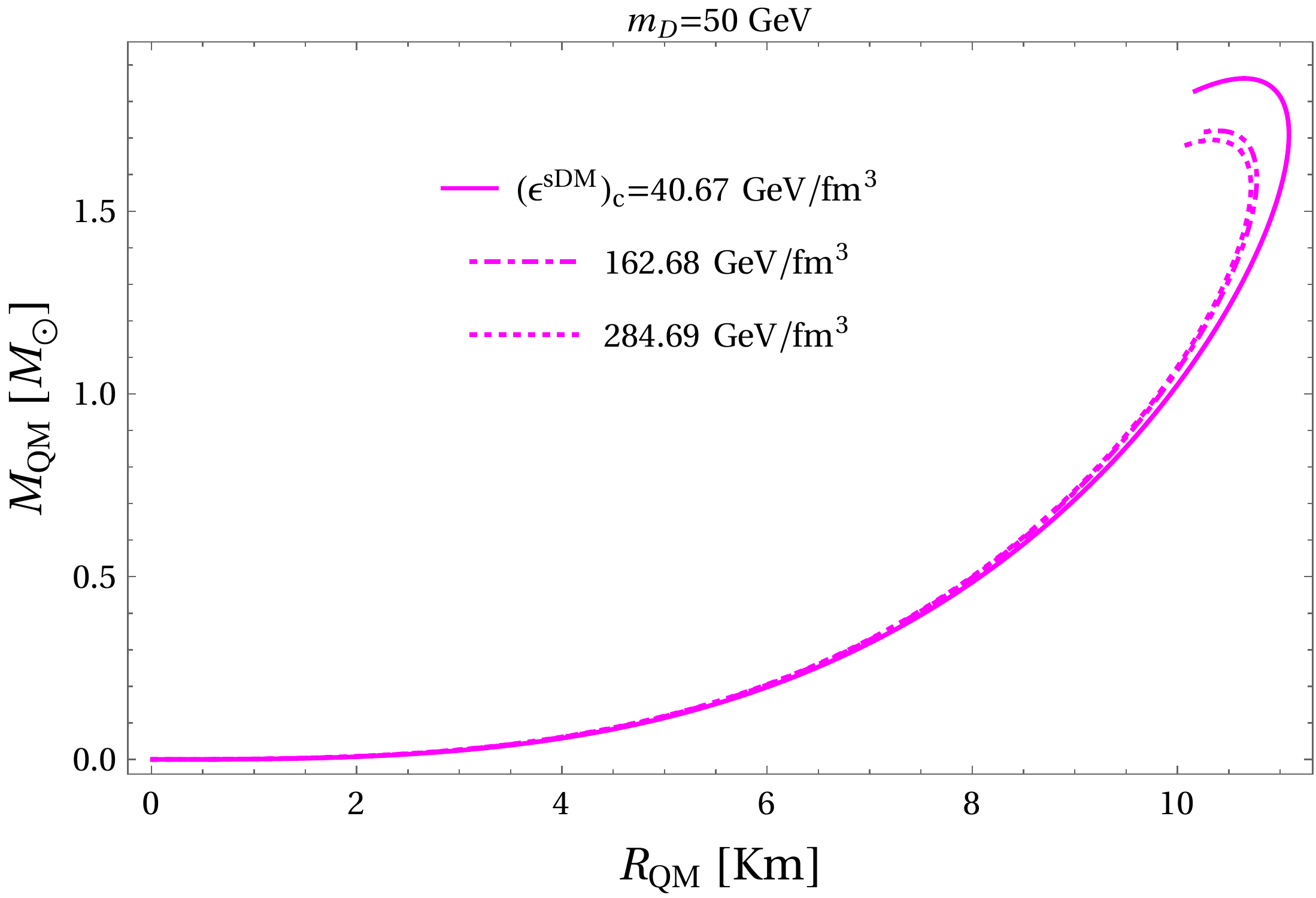}\vspace{3pt}
	  \includegraphics[scale=0.38]{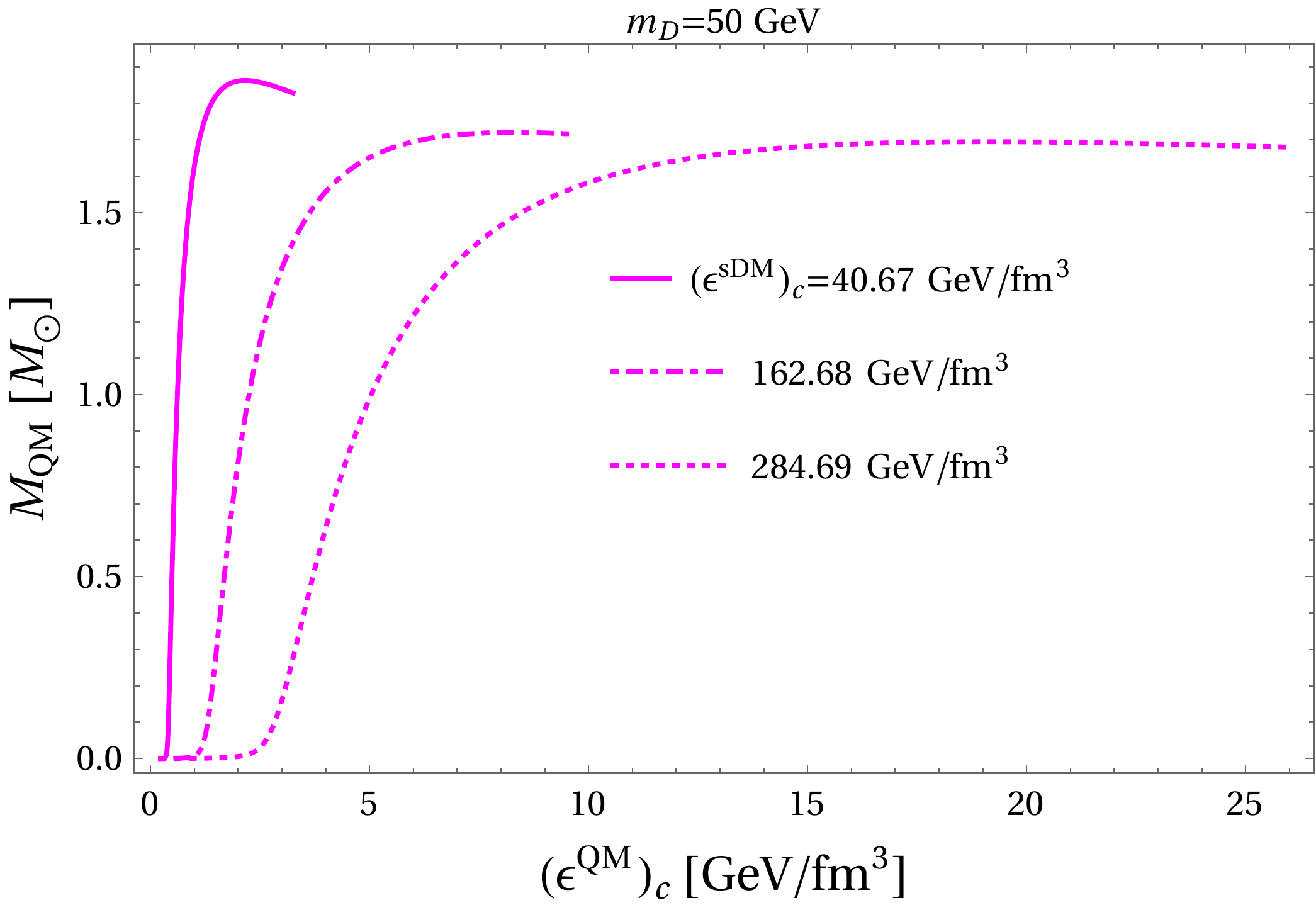}}\\\vspace{2pt}
{\includegraphics[scale=0.345]{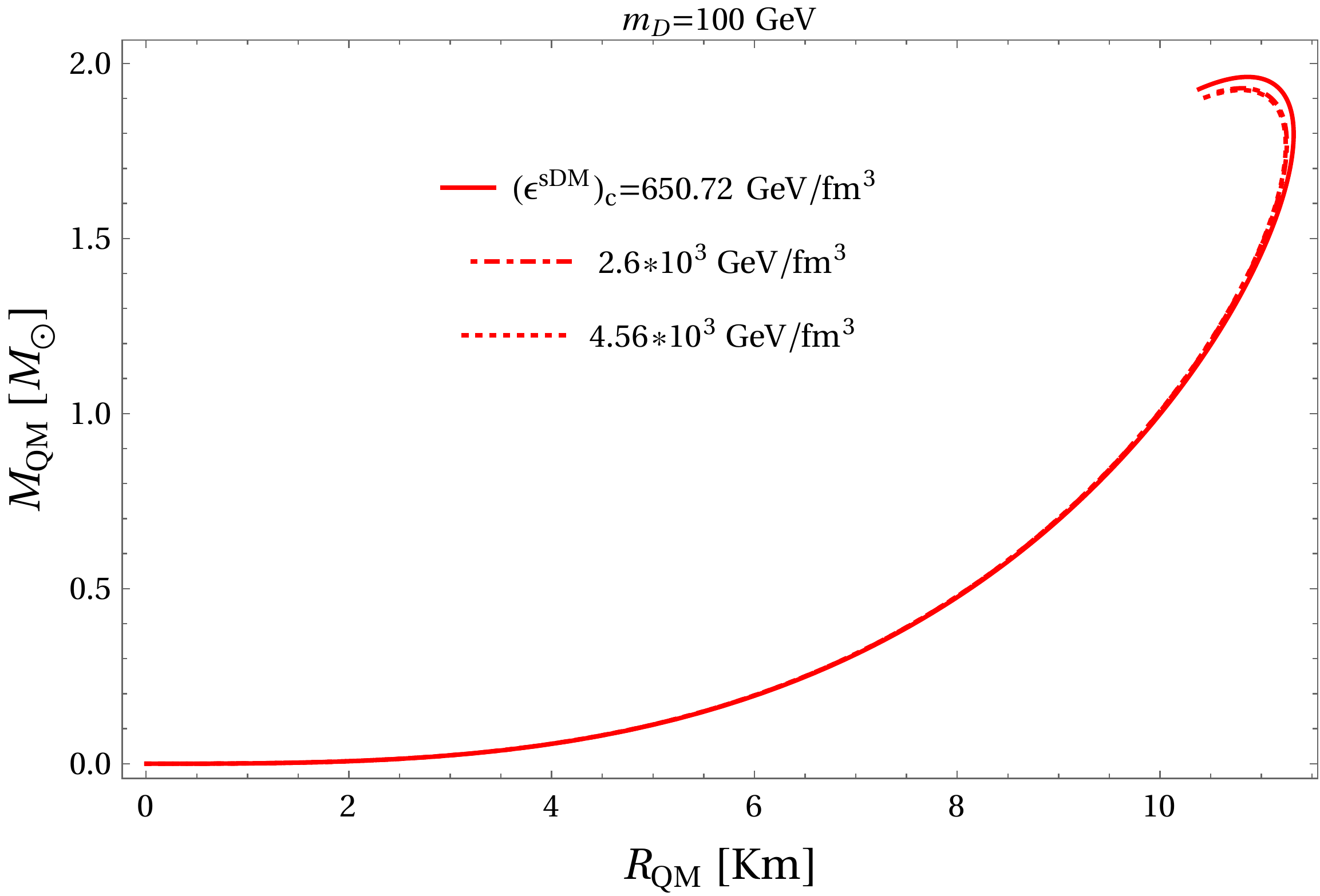}\vspace{3pt}
	  \includegraphics[scale=0.31]{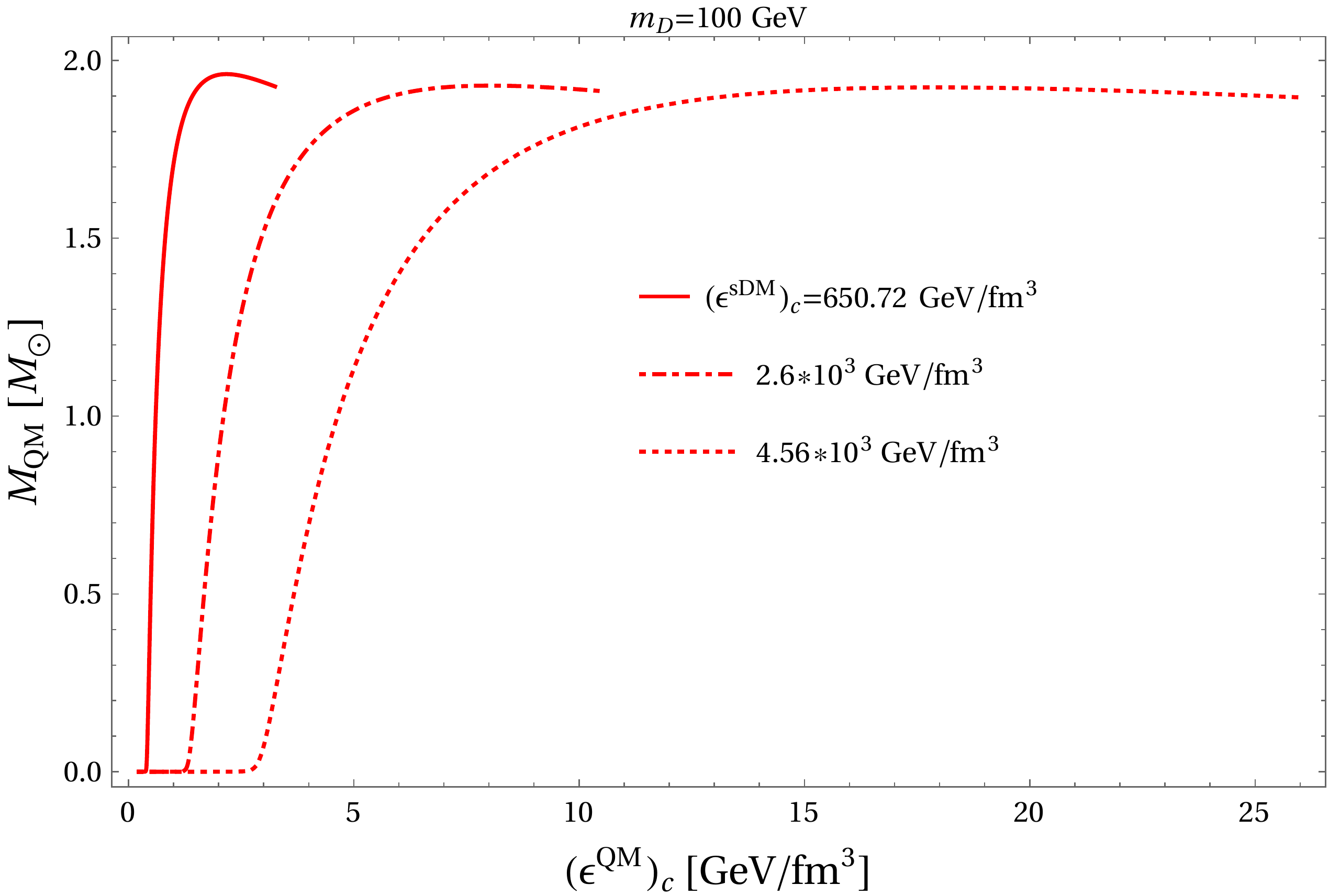}}\\\vspace{2pt}
{\includegraphics[scale=0.38]{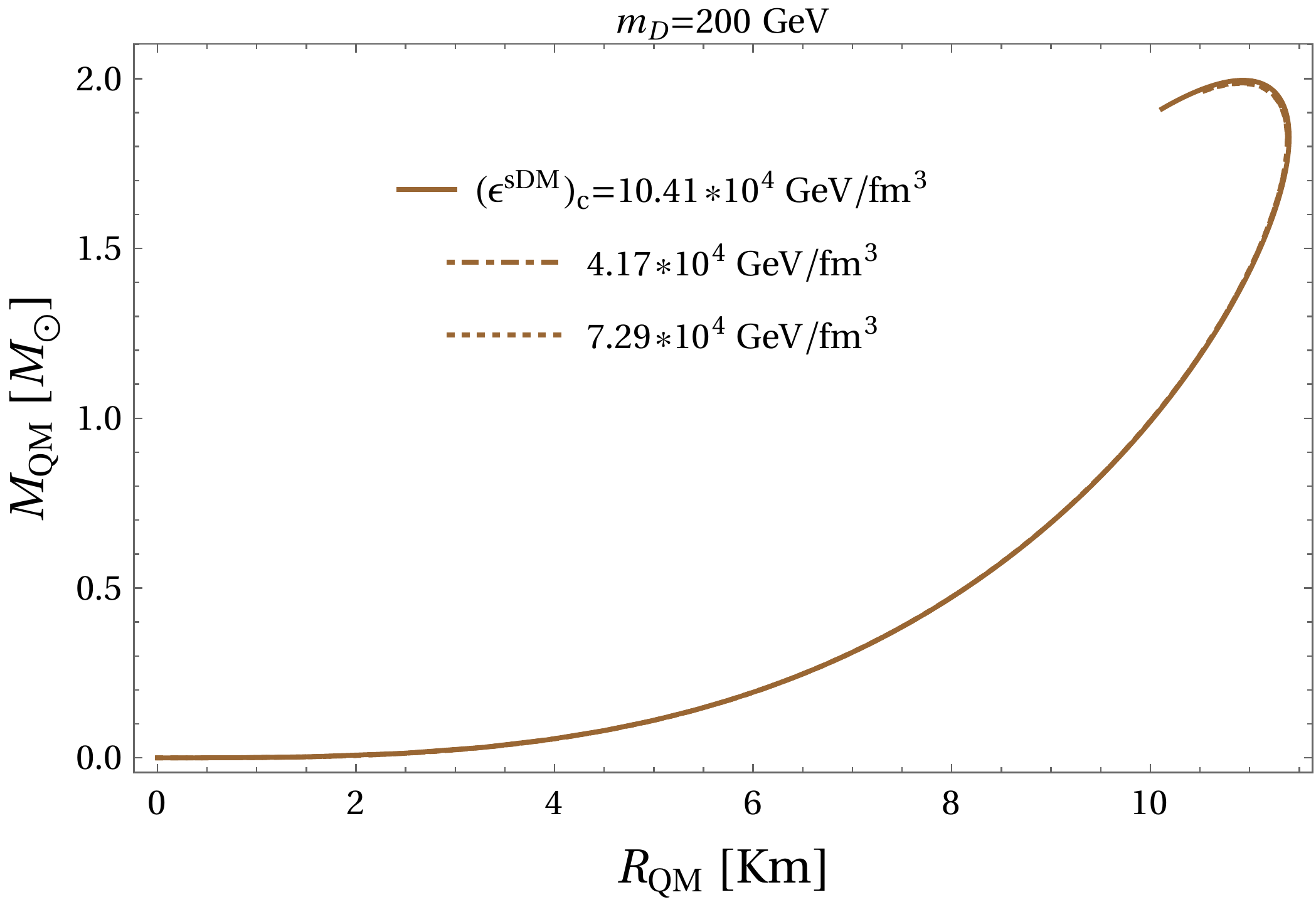}\vspace{3pt}
	  \includegraphics[scale=0.395]{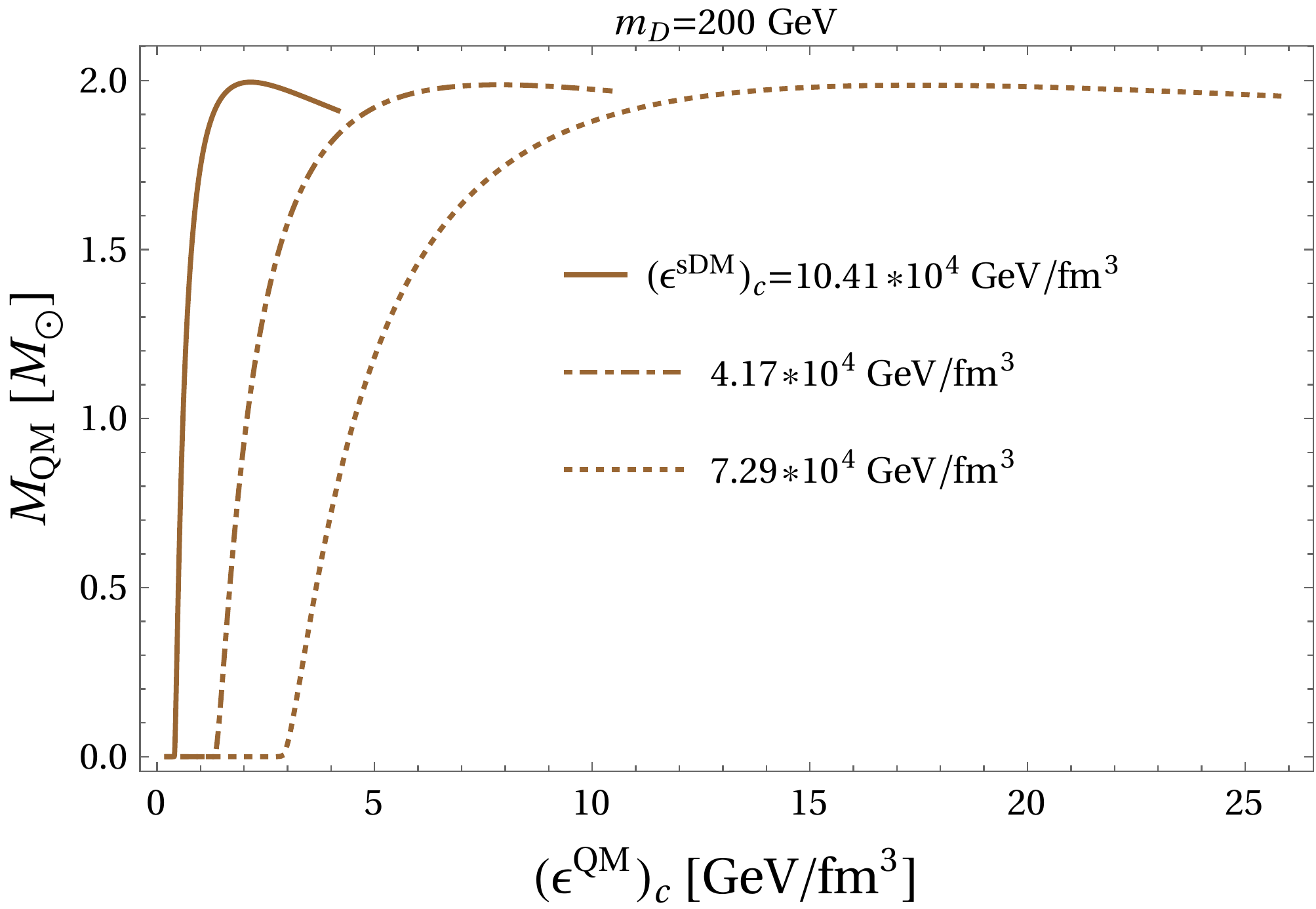}}\\
\end{adjustwidth}
\caption{\emph{Cont.}}
\label{fig:MRdens1000QM}

\end{figure}

\begin{figure}[H]\ContinuedFloat

\begin{adjustwidth}{-\extralength}{0cm}
\centering 

{\includegraphics[width=0.65\textwidth]{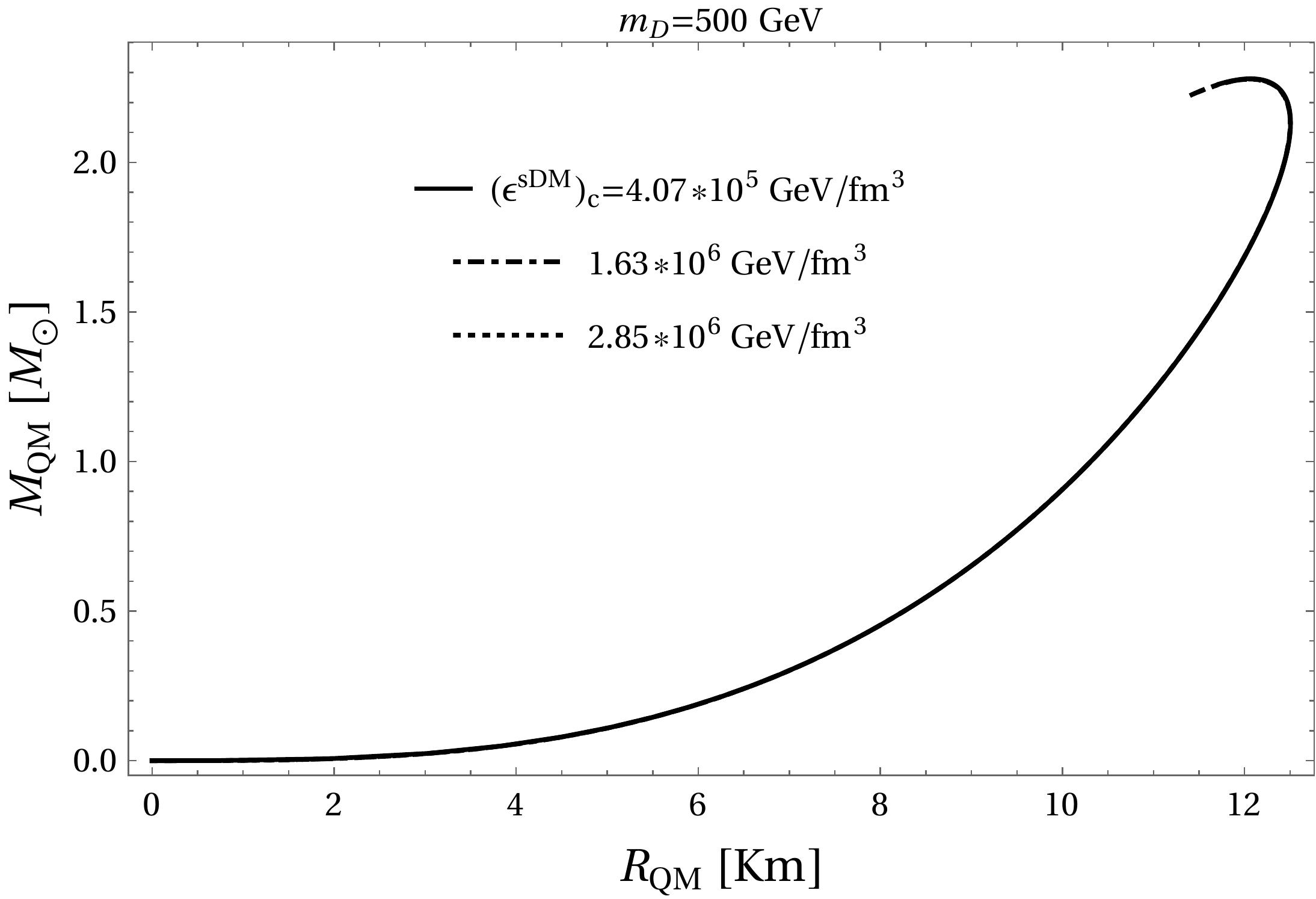}
	  \includegraphics[scale=0.4]{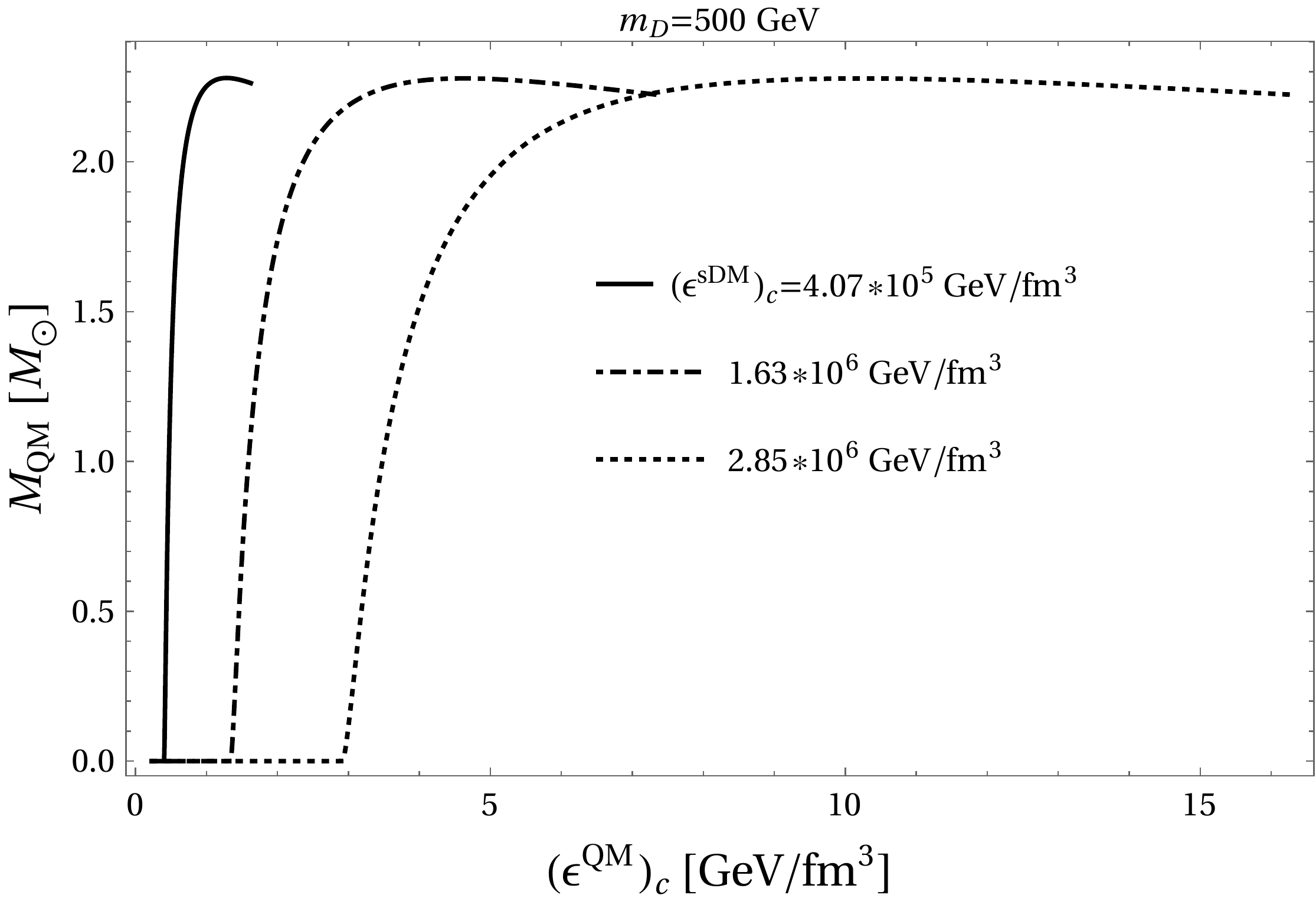}}
\end{adjustwidth}
\caption{Mass--radius and mass--energy density relationships for QM cores with different amounts of strongly ($y=10^{3}$) interacting dark matter ($s$DM) at the center of the admixed stars. Different values for the dark fermion masses $m_{D}$ are considered and results characterized by the same color. Notice that the effects of $s$DM are mainly for high-mass stars and especially marked for $m_{D}=10$ GeV.}
\label{fig:MRdens1000QM}

\end{figure}


In order to study the structure of the opposite case, we solved the two-fluid TOV Equations (\ref{eq:2TOV}) with the condition $p_{\rm DM}(R_{\rm DM})=0$ for different central energy densities of QM, as displayed in Figure \ref{fig:MRdens1000DM}. Pure $y=10^{3}$ DM stars display the same qualitative behavior as in the case of $m_{D}=500$ GeV, since, in this case, their masses and radii are almost unaffected by any amount of QM due to very large DM central energy densities. The same is true for $m_{D}=200$ GeV. See also Table \ref{tab:table1}.

Although not noticeable in Figure \ref{fig:MRdens1000DM}, when $m_{D}=1$ GeV, our calculations show that the QM in the DM core yields higher masses (not shown in the figure) that are in contradiction with the negative gradient of pressure required by the TOV equations. This happens, because we are considering unstable QM central energy densities for the DM star, as can be seen in Figure \ref{fig:MRdens1000QM}, which are manifested by producing increasing profiles of pressure then leading to mechanical instabilities associated with complex frequencies, thus destabilizing the whole admixed star. When $m_{D}=10$ GeV, we find a self-bound-like behavior for the DM star. This occurs since the DM and QM central energy densities are almost equal in the admixed star, and, in this case, the QM component dominates, modifying the behavior in the mass--radius diagram. As before, the central DM densities are increased by a factor of $10$. The cases with $m_{D}=50, 100$ GeV display a behavior that is a mixture of dark and quark matter, where QM mainly affects the sector of lower DM stellar masses, and the structure remains almost the same near the maximum mass . There, the central DM energy densities are almost the same since the DM energy densities are enormous compared the QM ones.

\begin{figure}[H]

\begin{adjustwidth}{-\extralength}{0cm}
\centering 

{\includegraphics[scale=0.39]{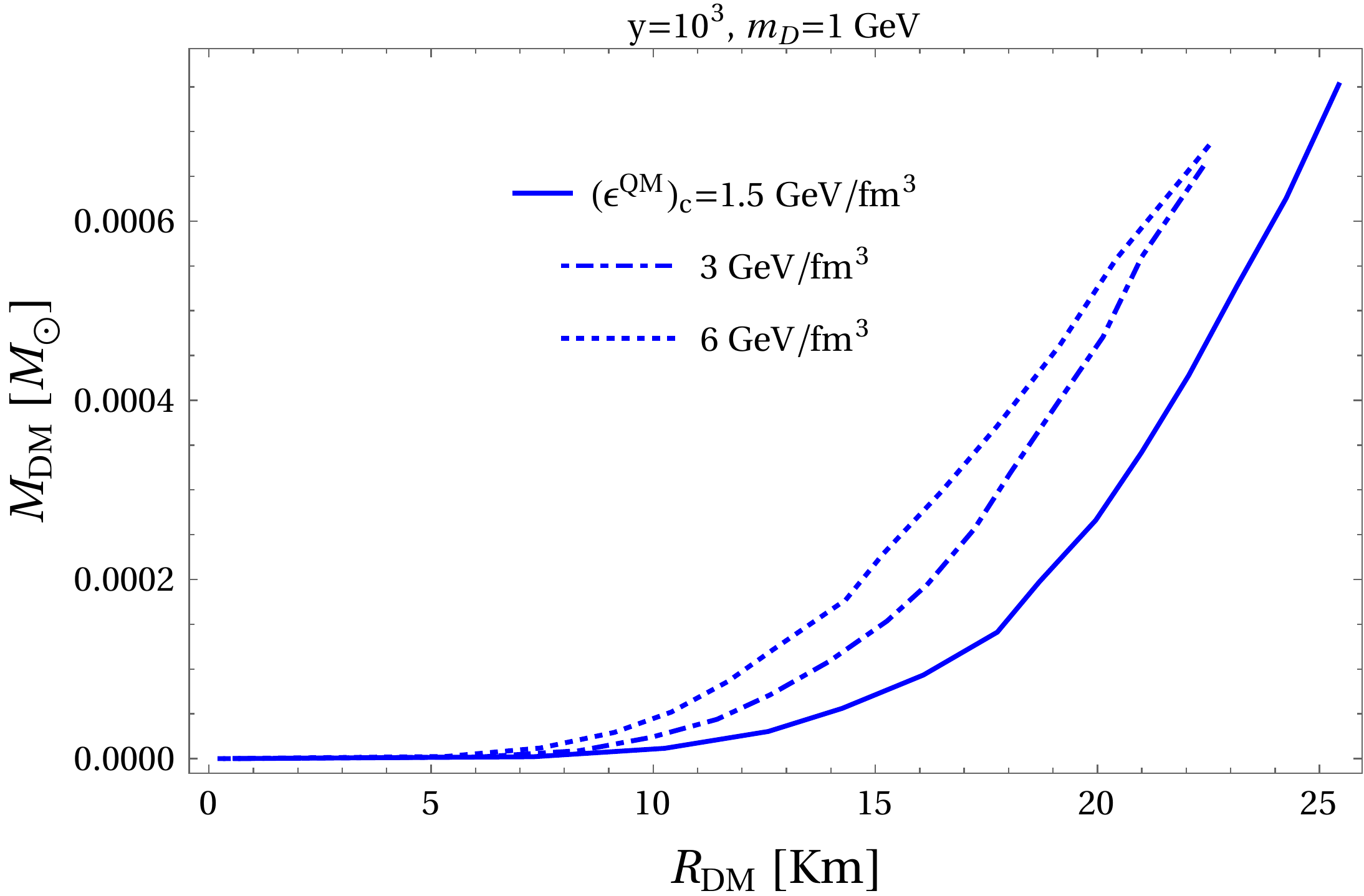}\vspace{3pt}
	  \includegraphics[scale=0.39]{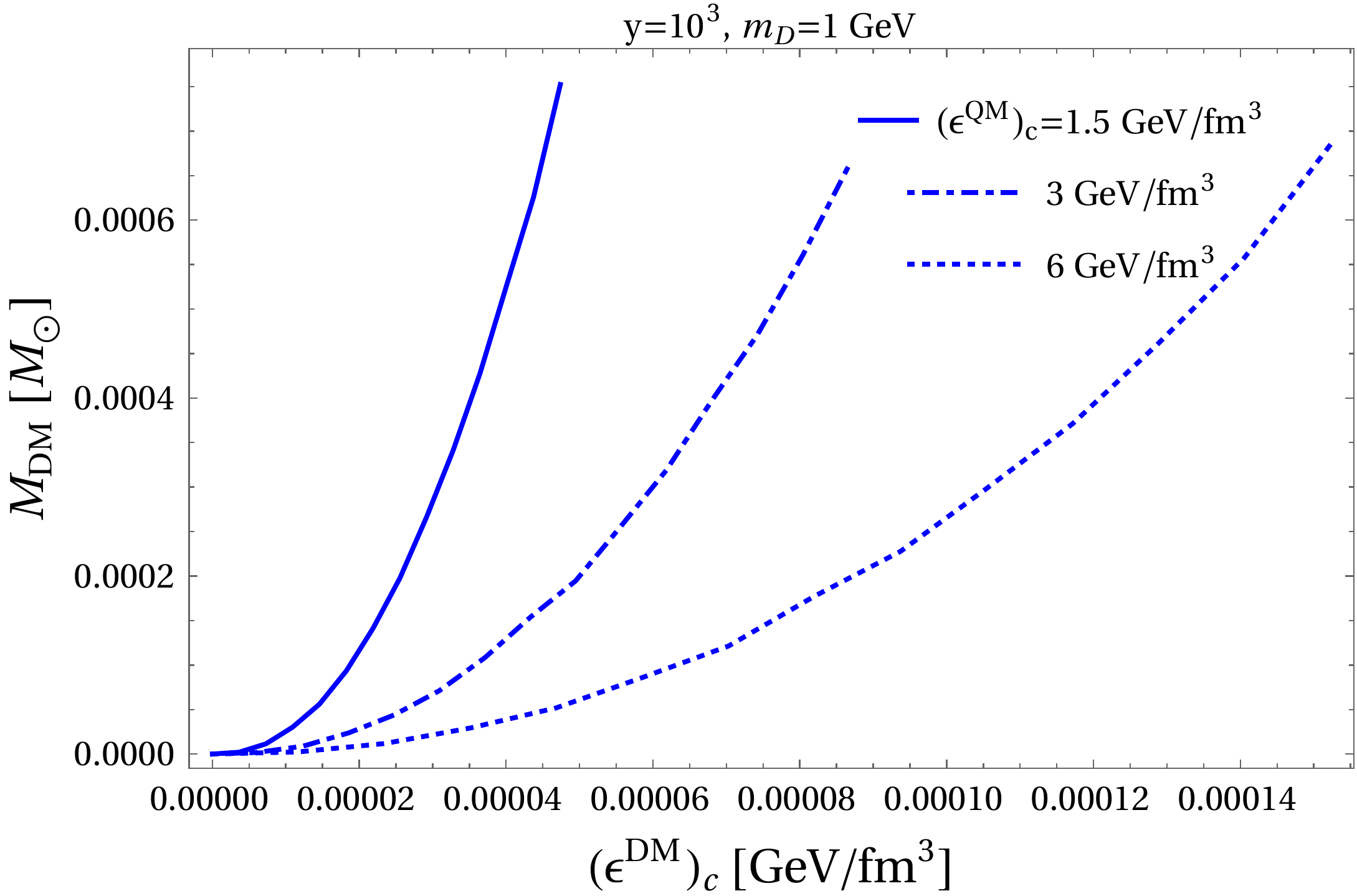}}\\
{\includegraphics[scale=0.408]{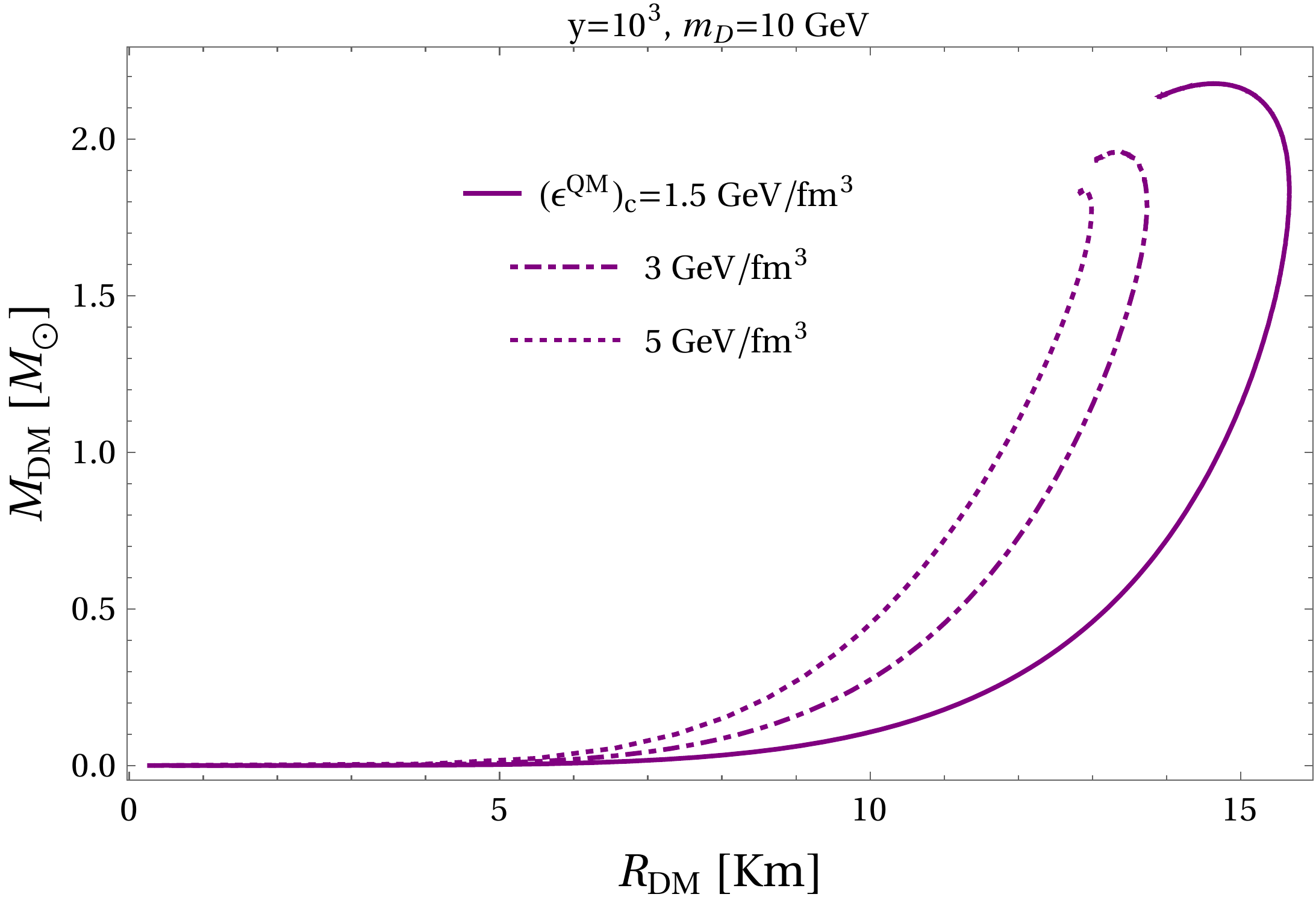}\vspace{3pt}
	  \includegraphics[scale=0.423]{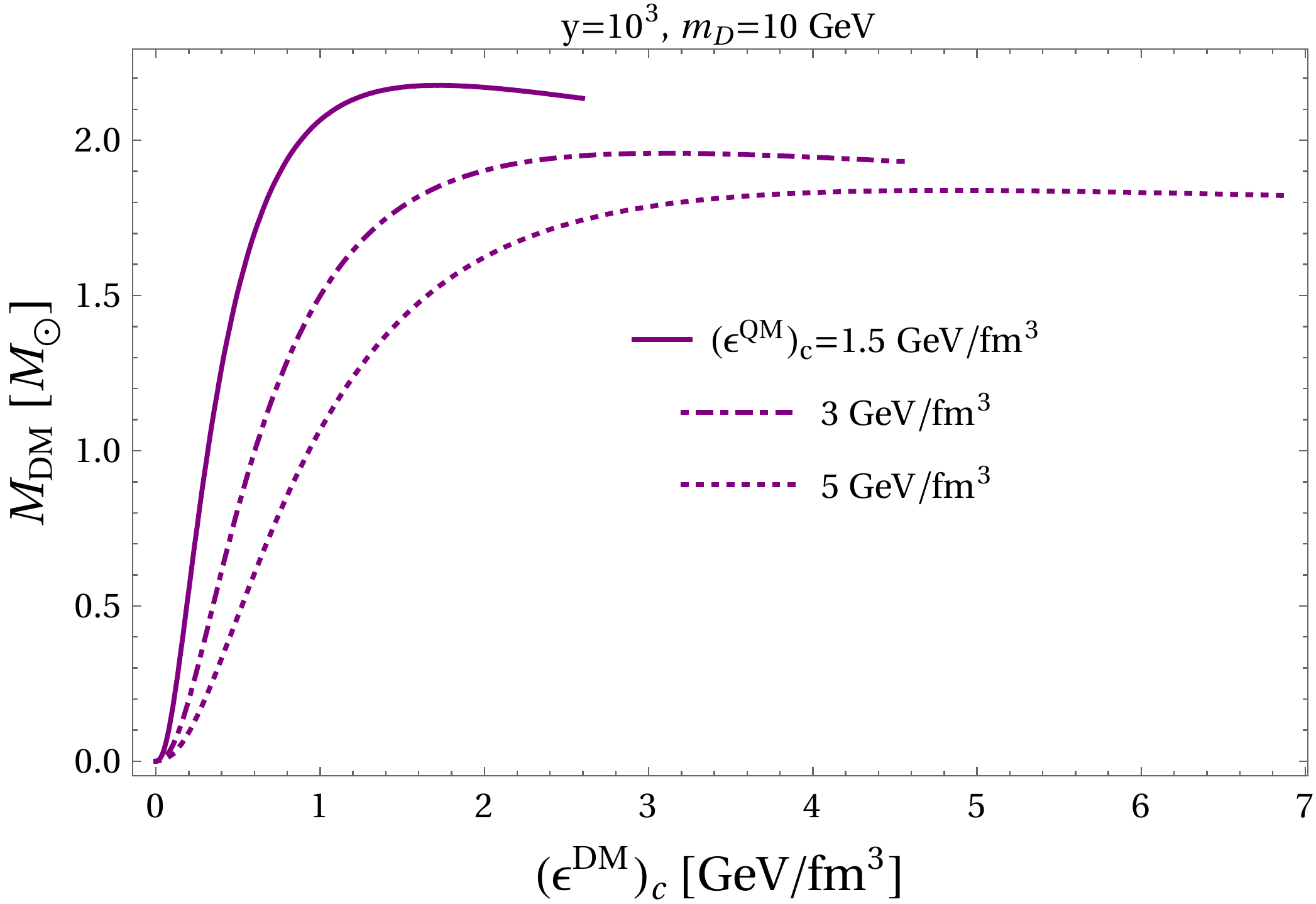}}\\
\end{adjustwidth}
\caption{\emph{Cont.}}
\label{fig:MRdens1000DM}

\end{figure}

\begin{figure}[H]\ContinuedFloat

\begin{adjustwidth}{-\extralength}{0cm}
\centering 

{\includegraphics[scale=0.387]{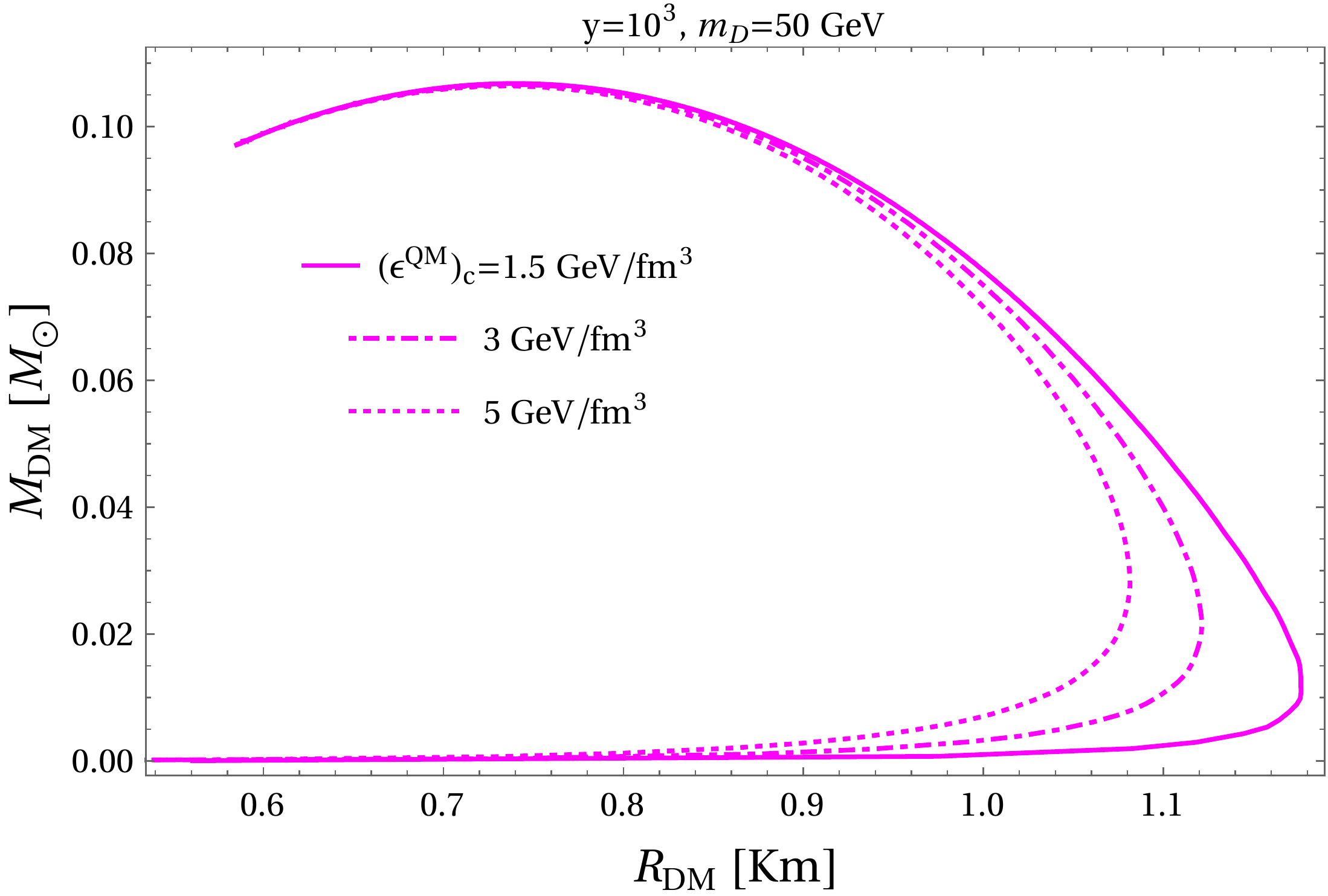}\vspace{3pt}
	  \includegraphics[scale=0.40]{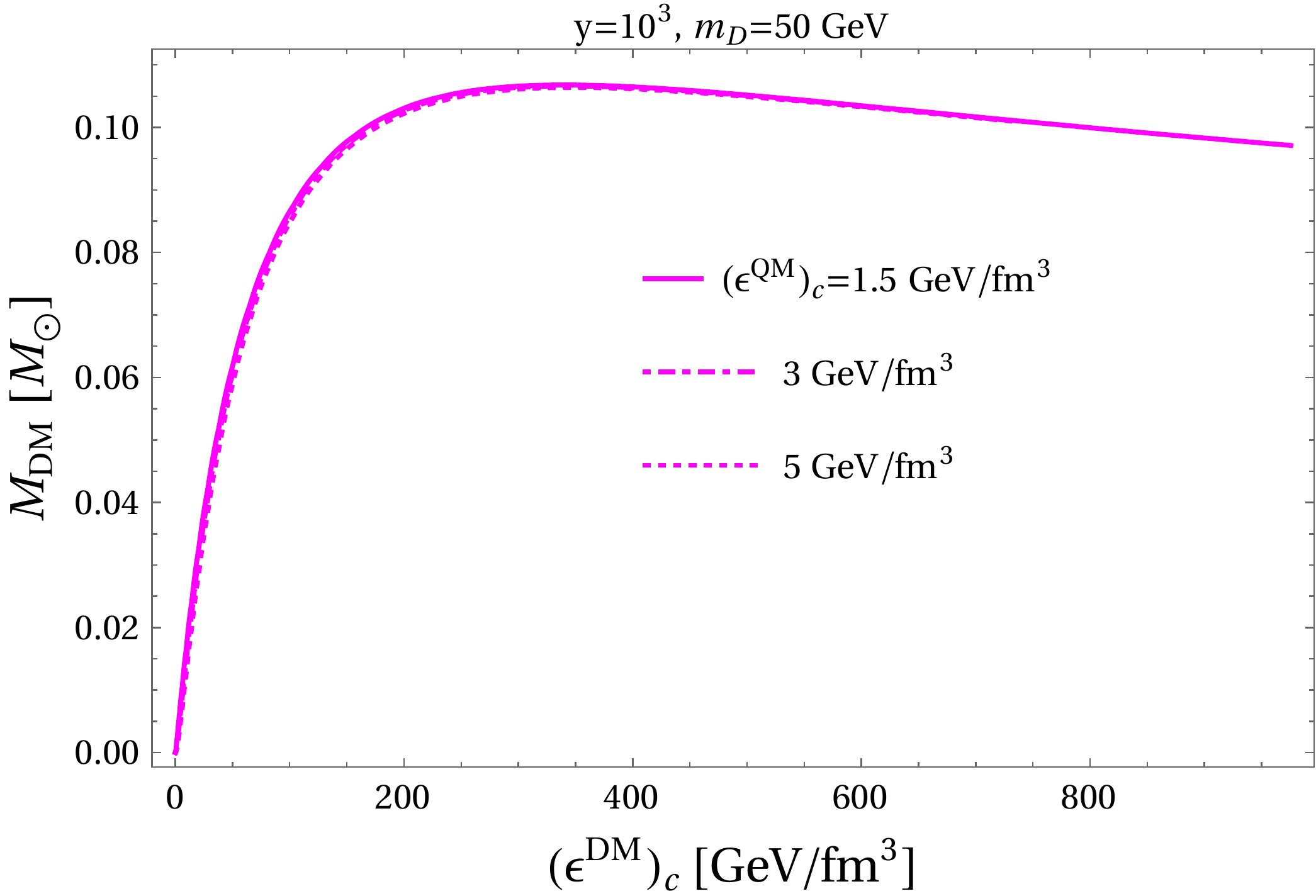}}\\\vspace{5pt}
{\includegraphics[scale=0.37]{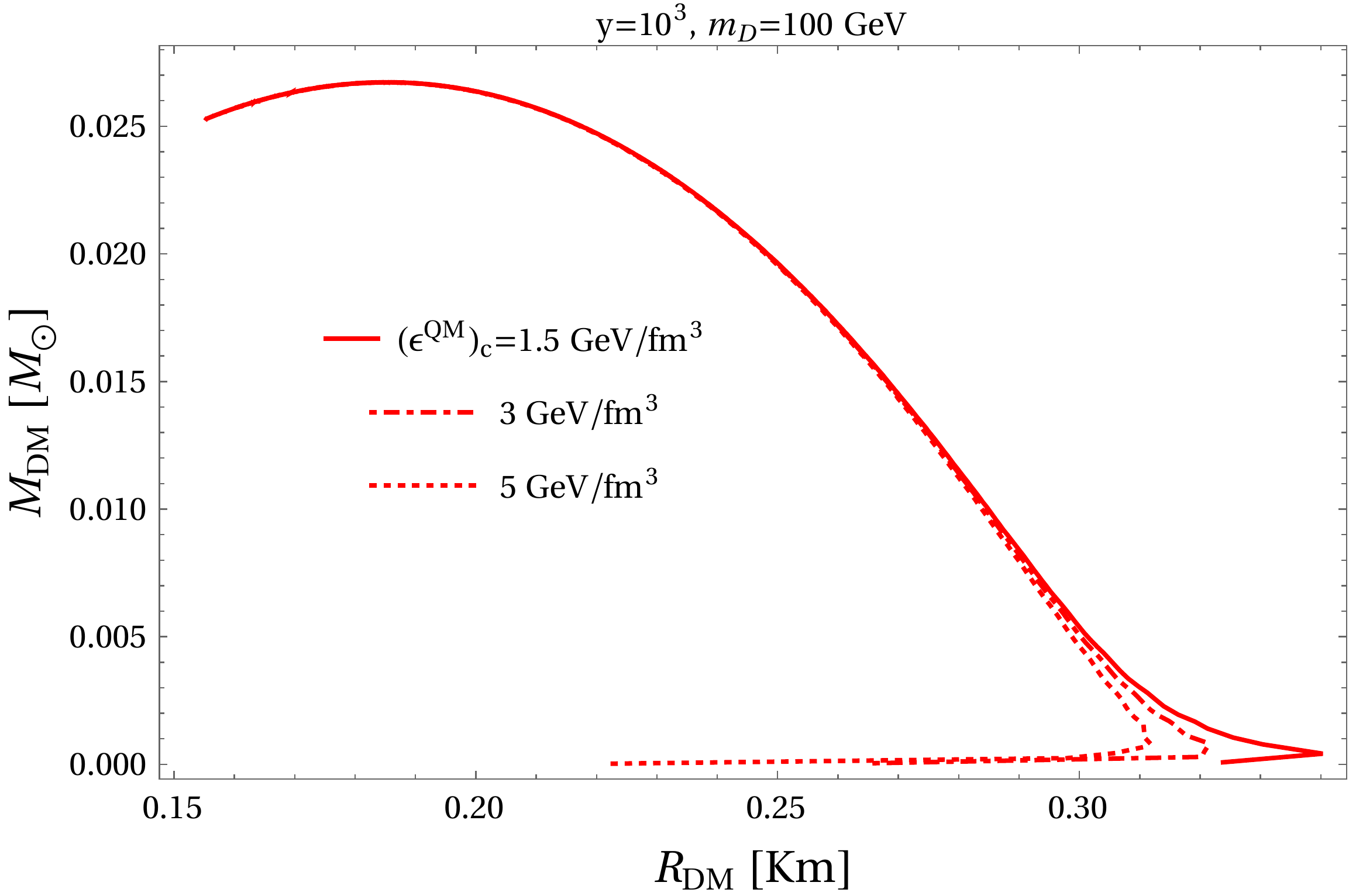}\vspace{3pt}
	  \includegraphics[scale=0.335]{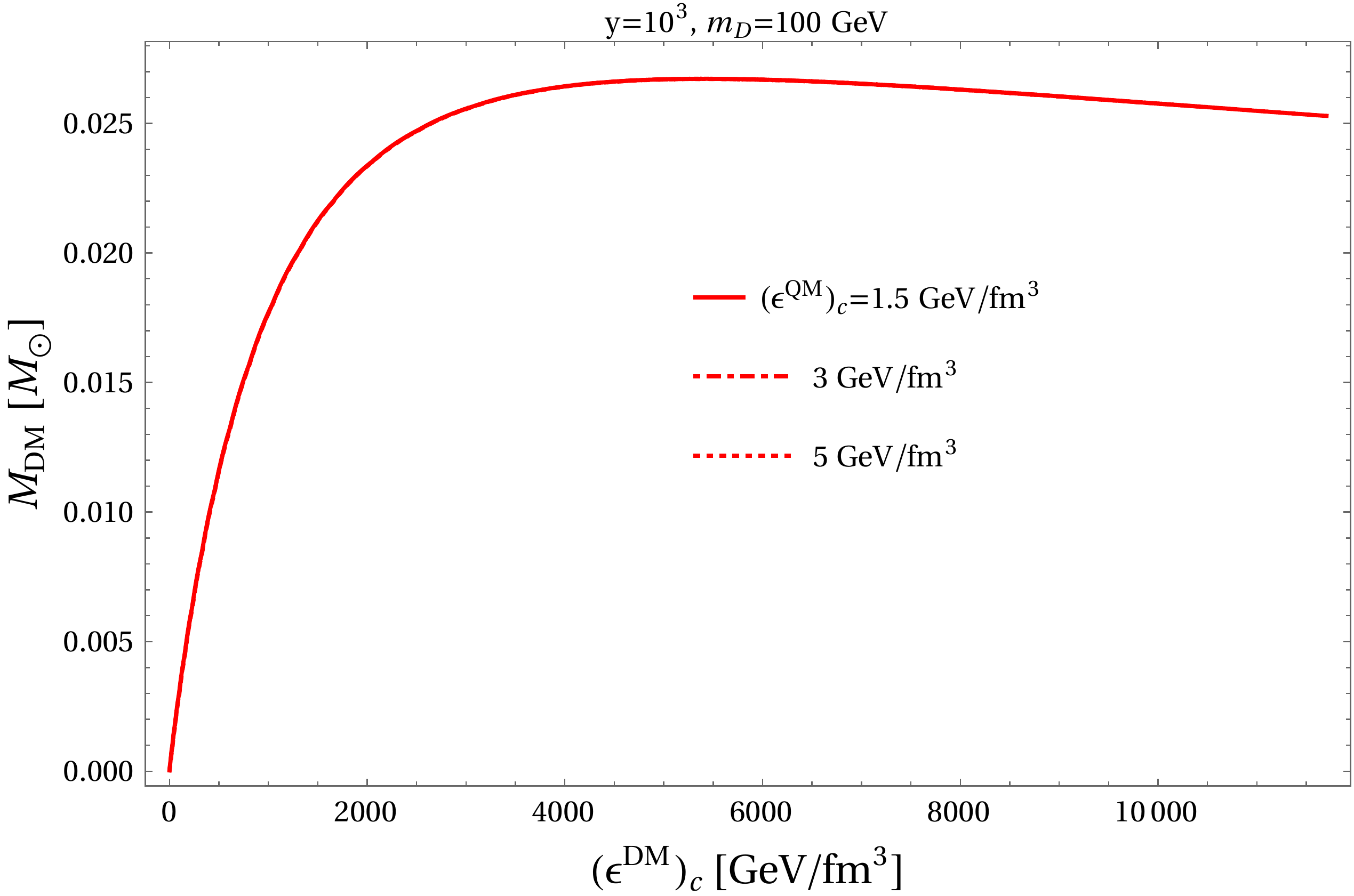}}\\\vspace{5pt}
{\includegraphics[scale=0.405]{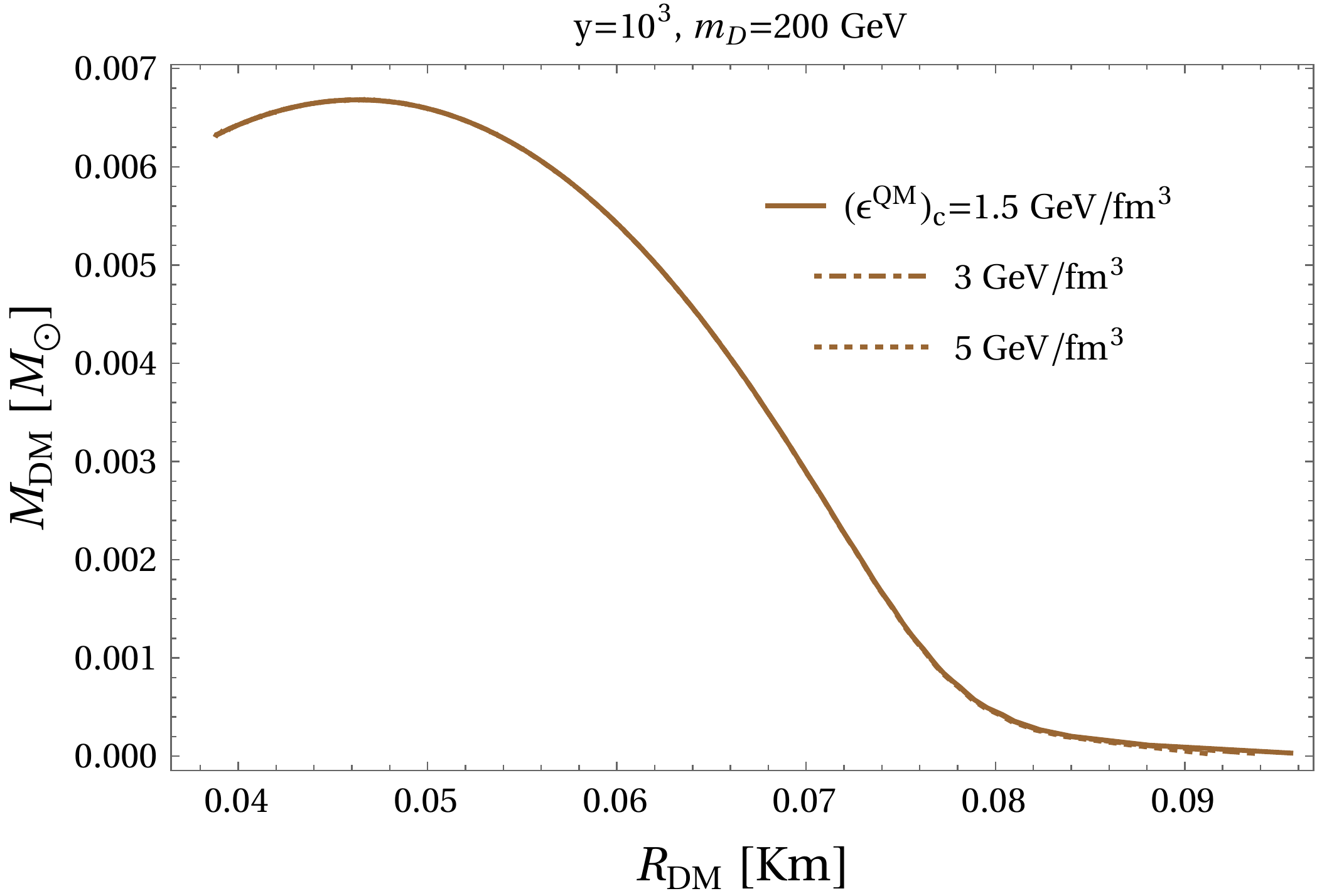}\vspace{3pt}
	  \includegraphics[scale=0.42]{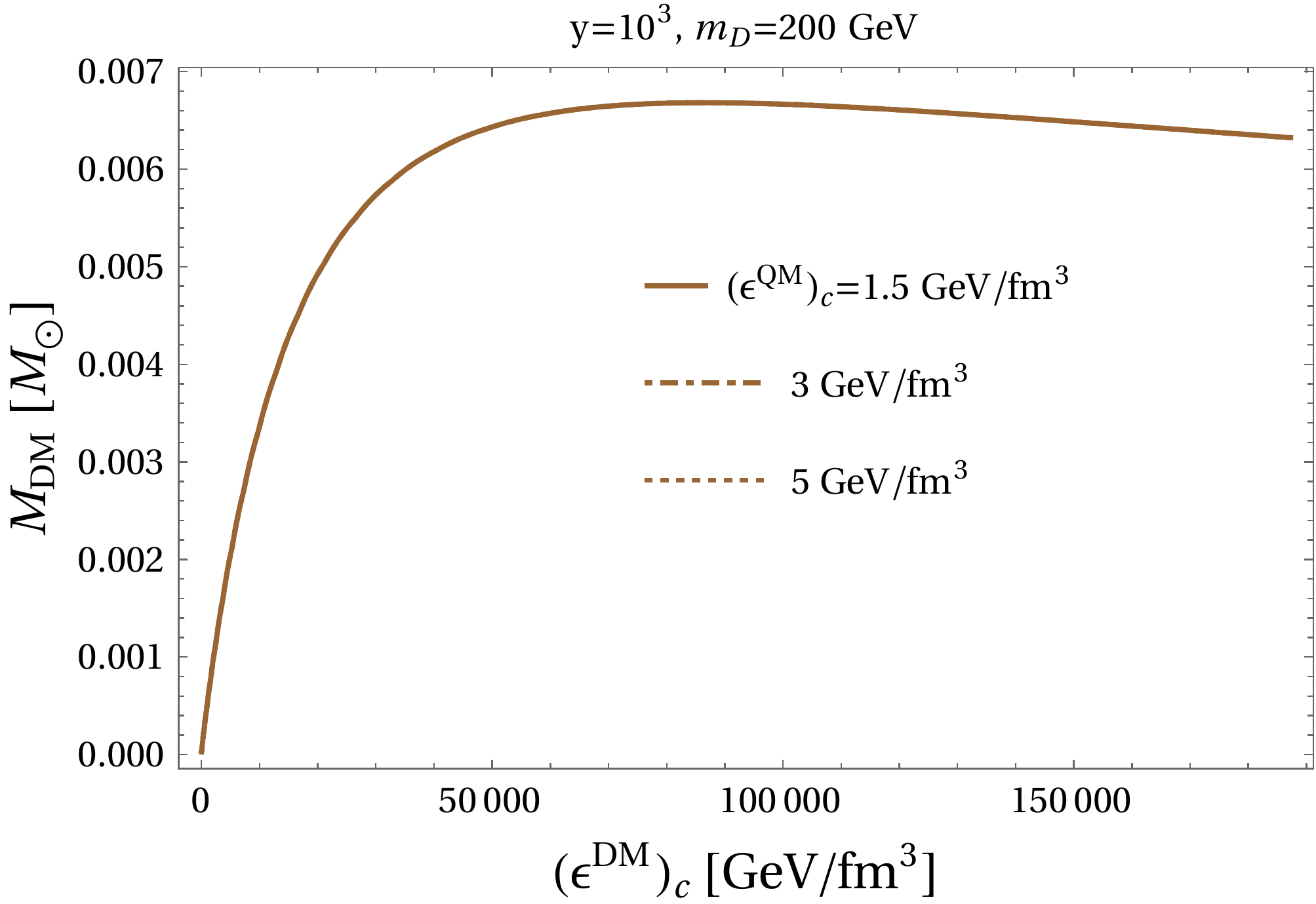}}\\
\end{adjustwidth}
\caption{\emph{Cont.}}
\label{fig:MRdens1000DM}

\end{figure}

\begin{figure}[H]\ContinuedFloat

\begin{adjustwidth}{-\extralength}{0cm}
\centering 

{\includegraphics[scale=0.41]{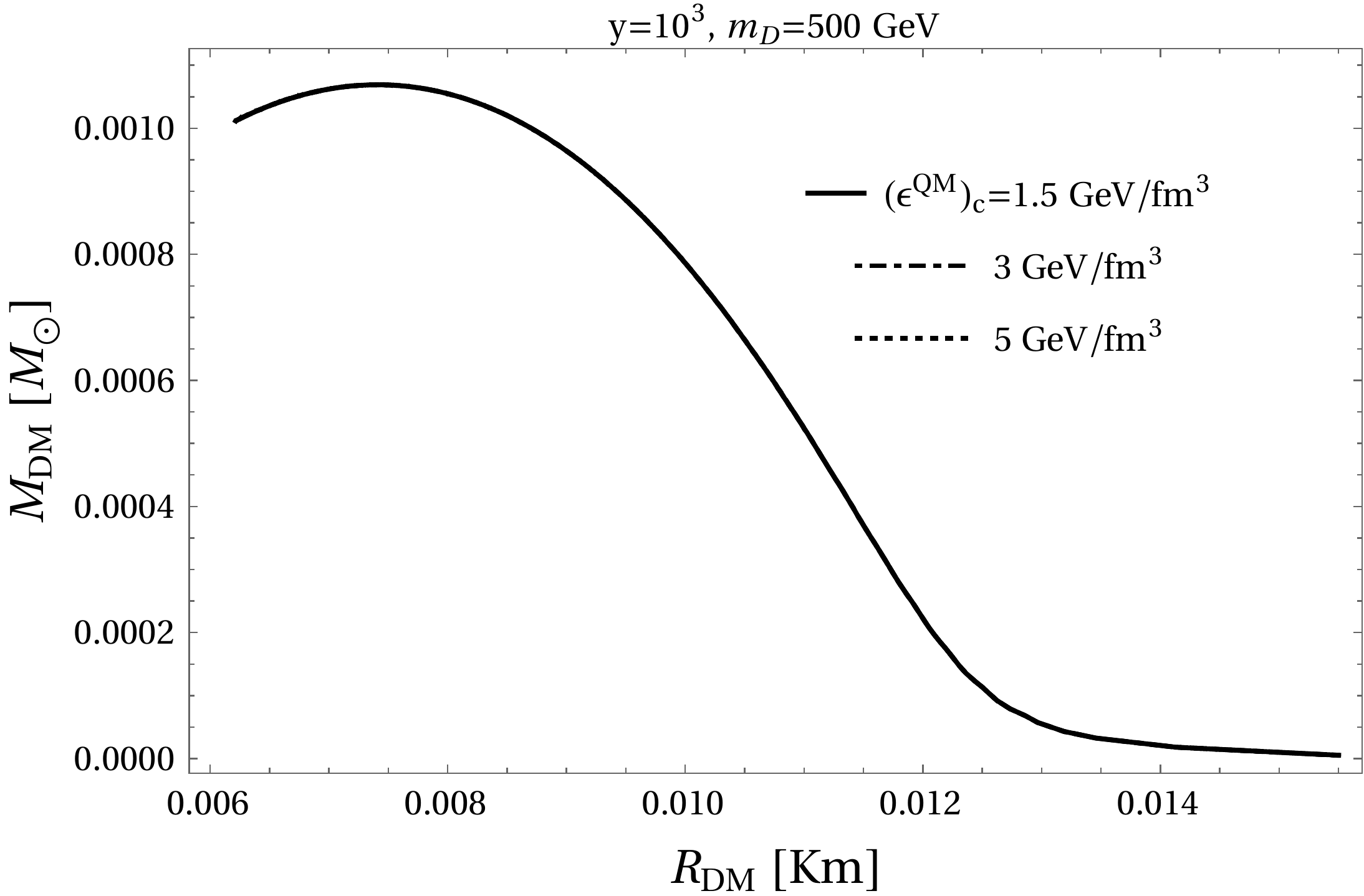}
	  \includegraphics[scale=0.385]{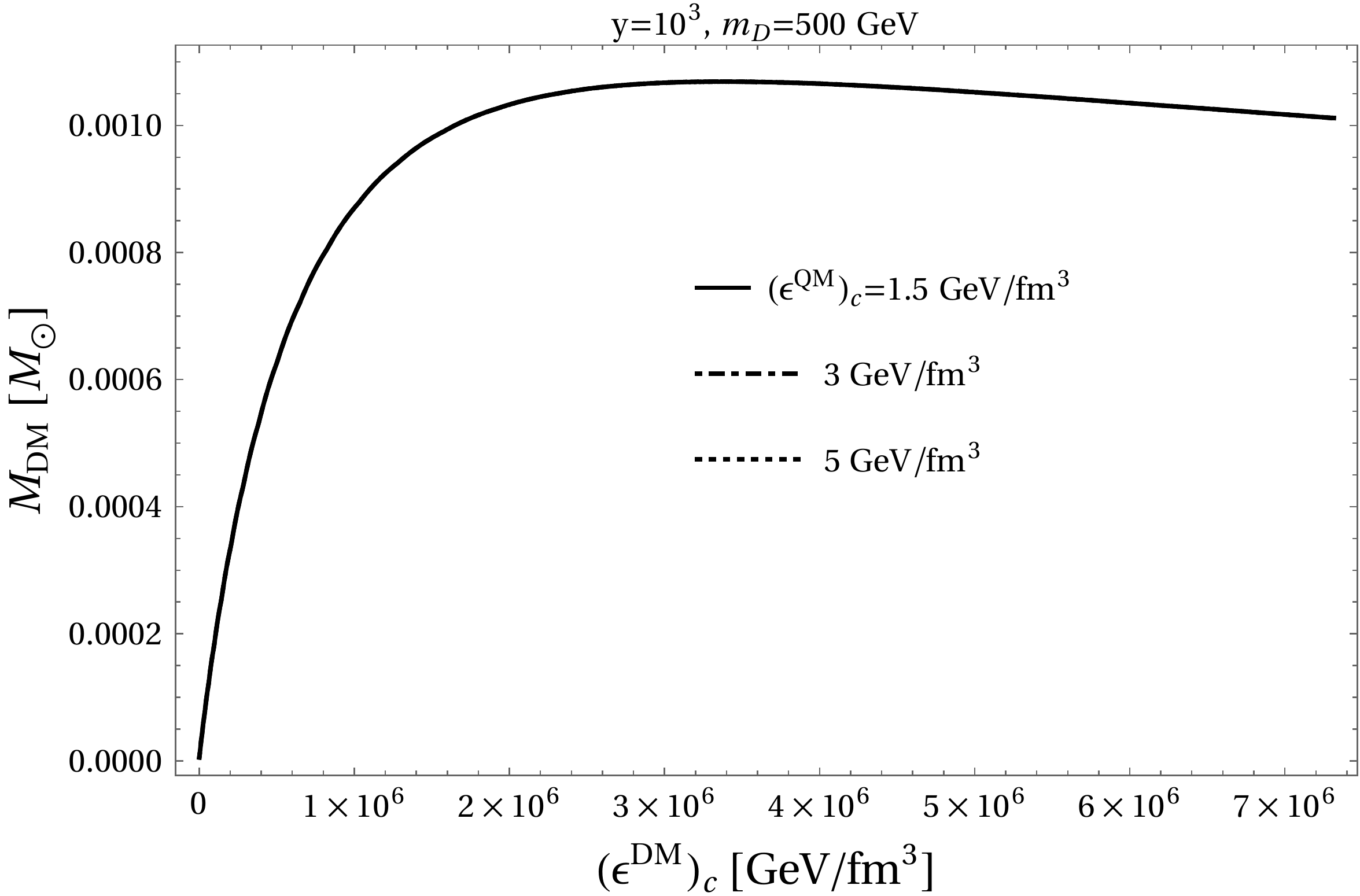}}
\end{adjustwidth}
\caption{Mass--radius and mass--energy density relations obtained for $s$DM cores for different fermion masses $m_{D}$ (indicated by different colors) with increasing amounts of QM at the centers of the admixed stars. Notice that the competition between $s$DM and QM densities in some cases allows for the presence of very small stars, which are not present in the one-fluid case, producing qualitatively different behavior in the mass--radius relations of $s$DM stars, especially for low $m_{D}$.}
\label{fig:MRdens1000DM}

\end{figure}


\subsubsection{Solving the Coupled Radial Oscillation Equations}

Finally, we solved the coupled radial pulsation Equations (\ref{Rad1}) and (\ref{Rad2}) assuming an oscillating QM core in the admixed star with the boundary conditions (\ref{BC1}) and (\ref{BC2}) for different fixed central energy densities of strongly interacting DM. The results are displayed in Figure \ref{fig:f0Mdens1000QM}. We found that only the case of $m_{D}=1$ GeV was unaffected by strongly self-interacting DM. As we increased $m_{D}$, the fundamental frequency was strongly affected. In fact, as occurred in the QM cases with weakly self-interacting DM, only low-mass QM stars survived radial oscillations and behaved as strange quark planets and strangelets. The oscillation frequencies of these objects can reach $\sim$${10^{5}}$ kHz for $m_{D}=500$ GeV.

\begin{figure}[H]

\begin{adjustwidth}{-\extralength}{0cm}
\centering 

{\includegraphics[scale=0.41]{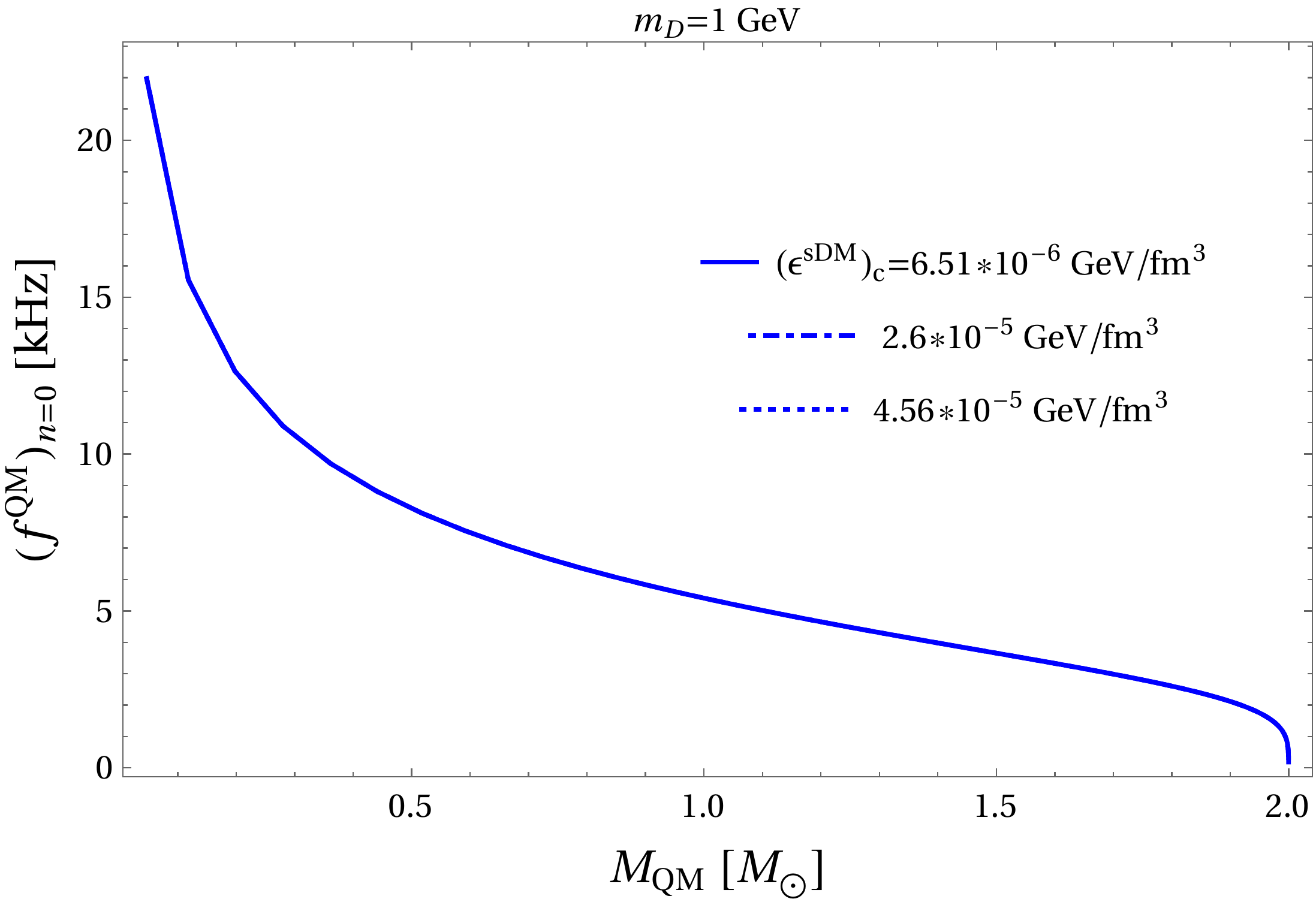}\vspace{3pt}
	  \includegraphics[scale=0.39]{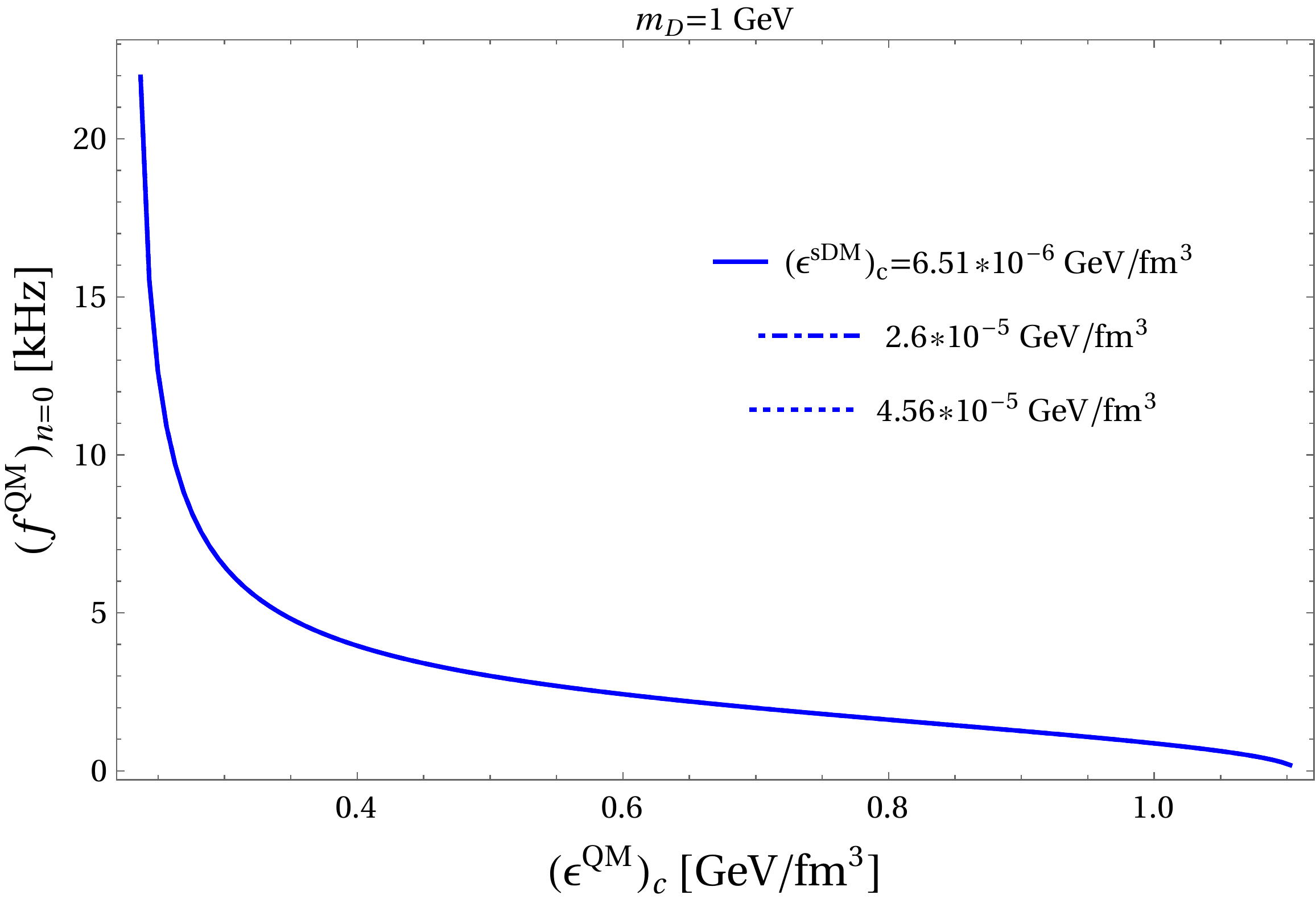}}\\
\end{adjustwidth}
\caption{\emph{Cont. }}
\label{fig:f0Mdens1000QM}

\end{figure}

\begin{figure}[H]\ContinuedFloat

\begin{adjustwidth}{-\extralength}{0cm}
\centering 

{\includegraphics[scale=0.365]{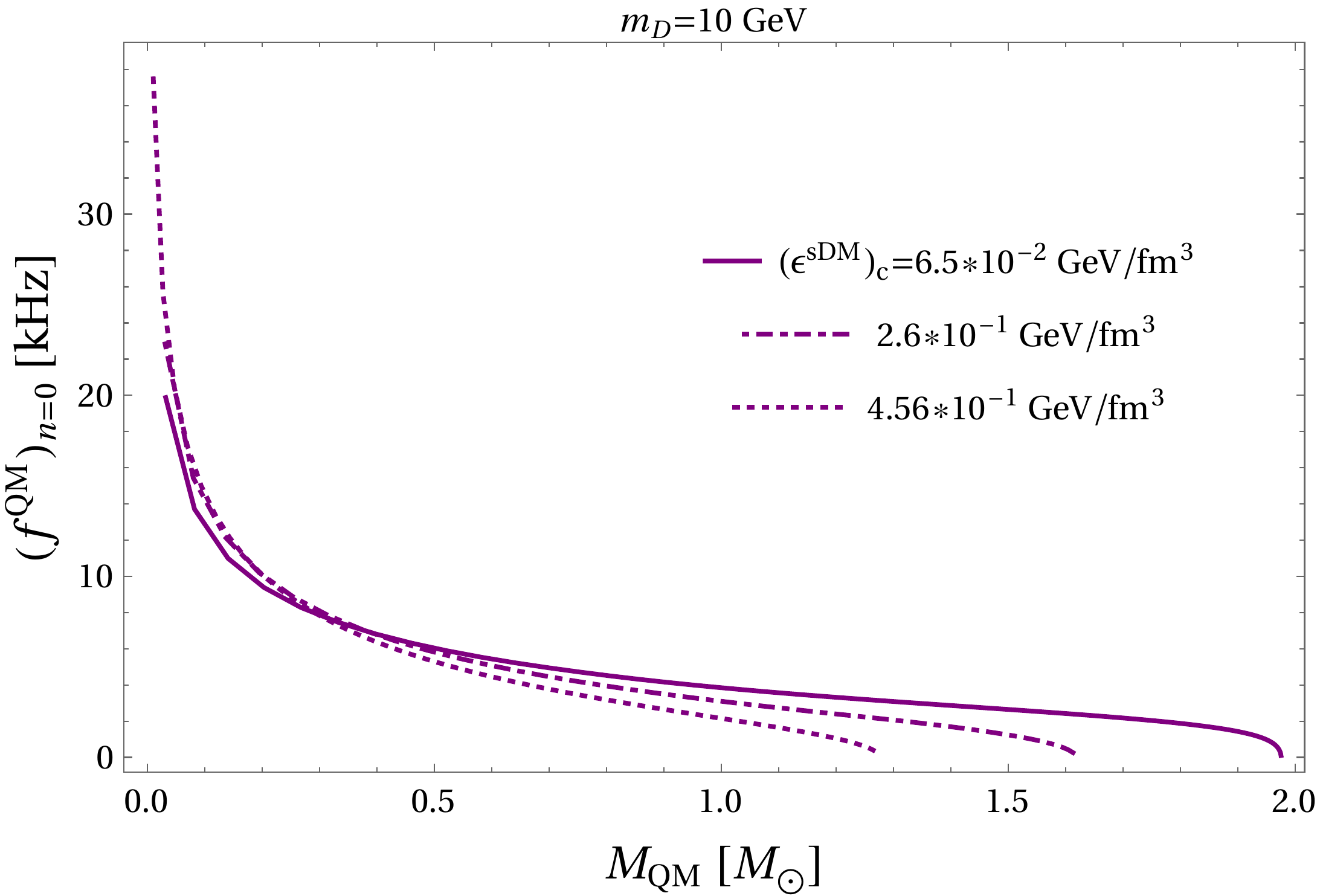}\vspace{3pt}
	  \includegraphics[scale=0.375]{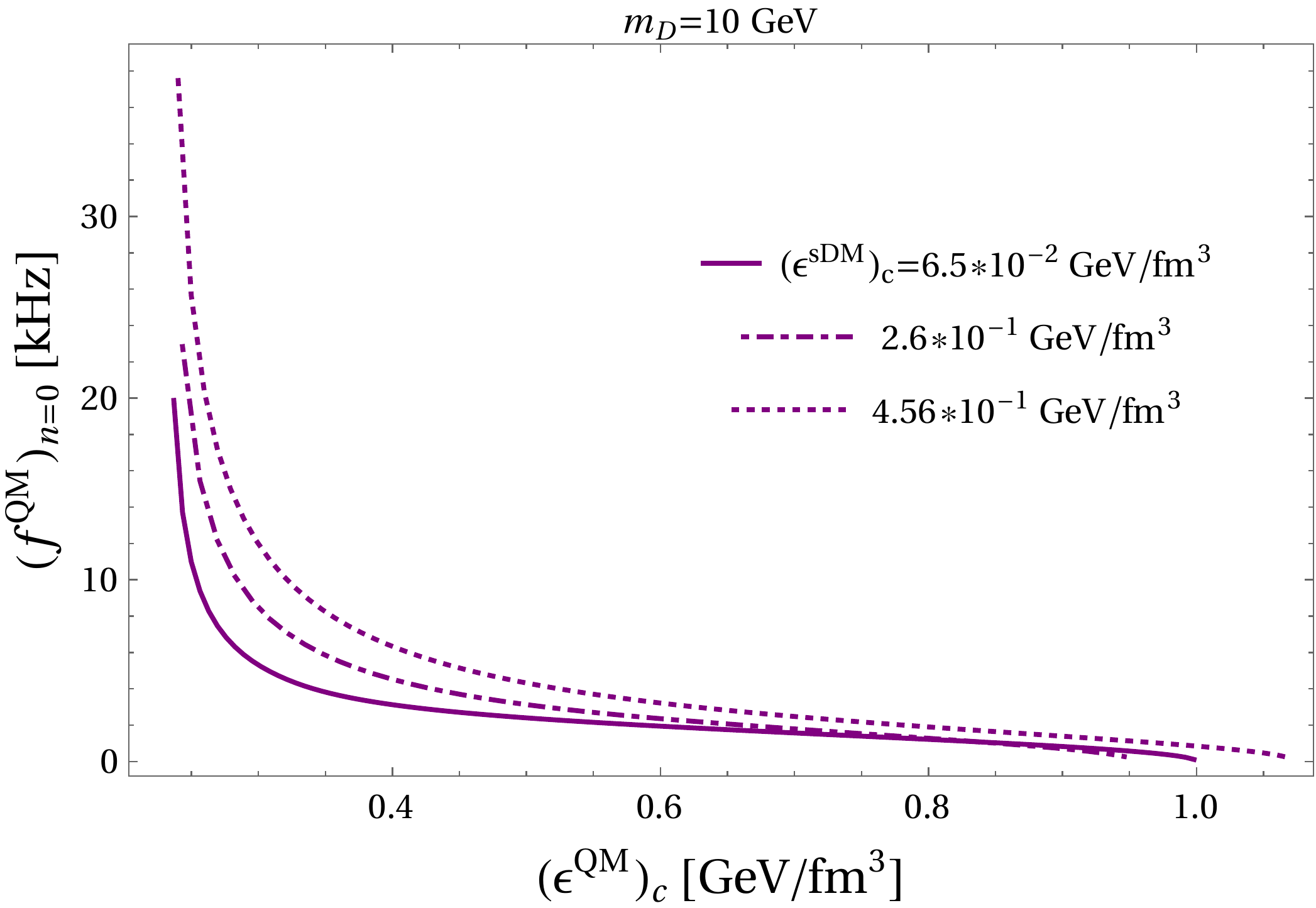}}\\\vspace{5pt}
{\includegraphics[scale=0.36]{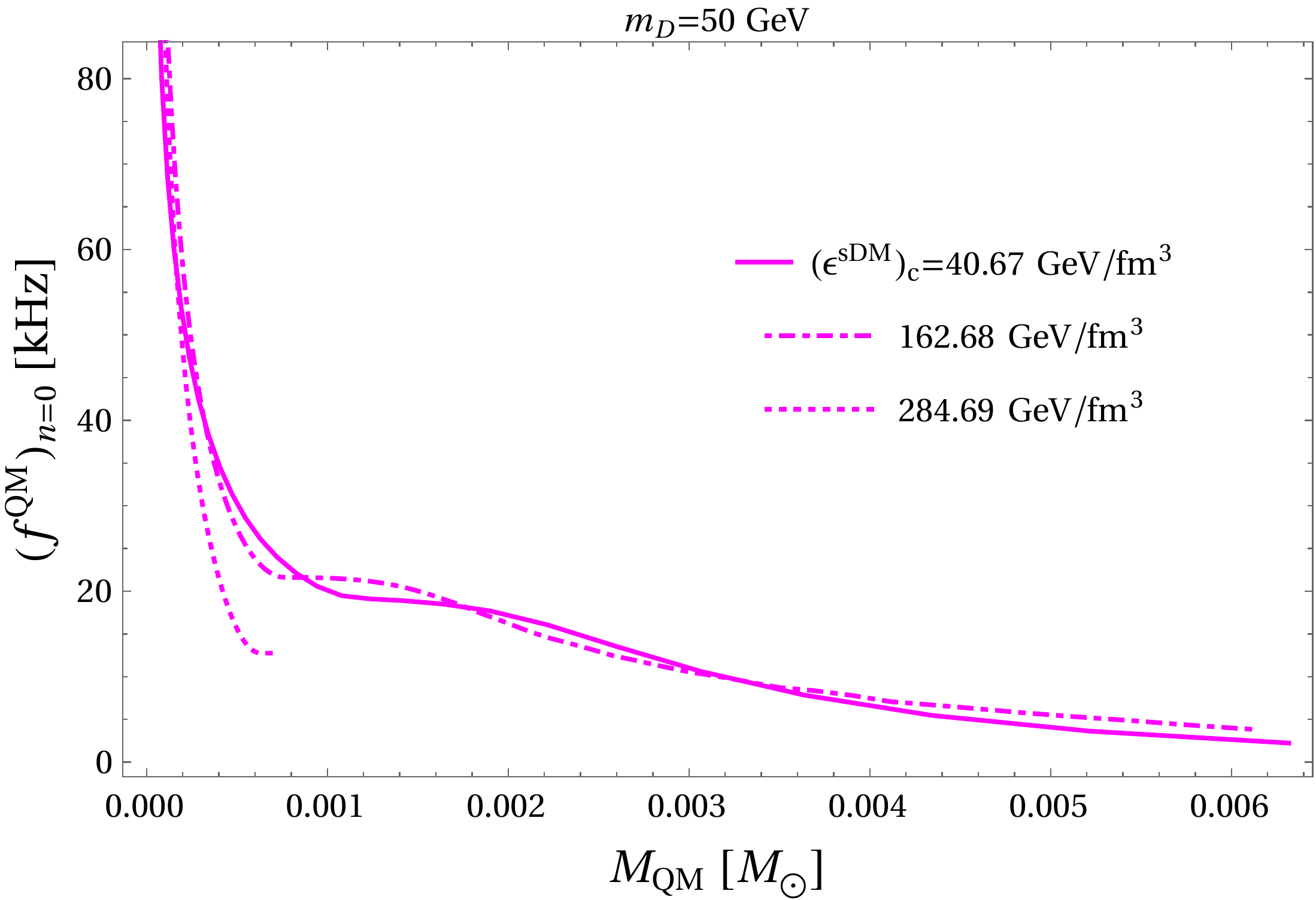}\vspace{3pt}
	  \includegraphics[scale=0.388]{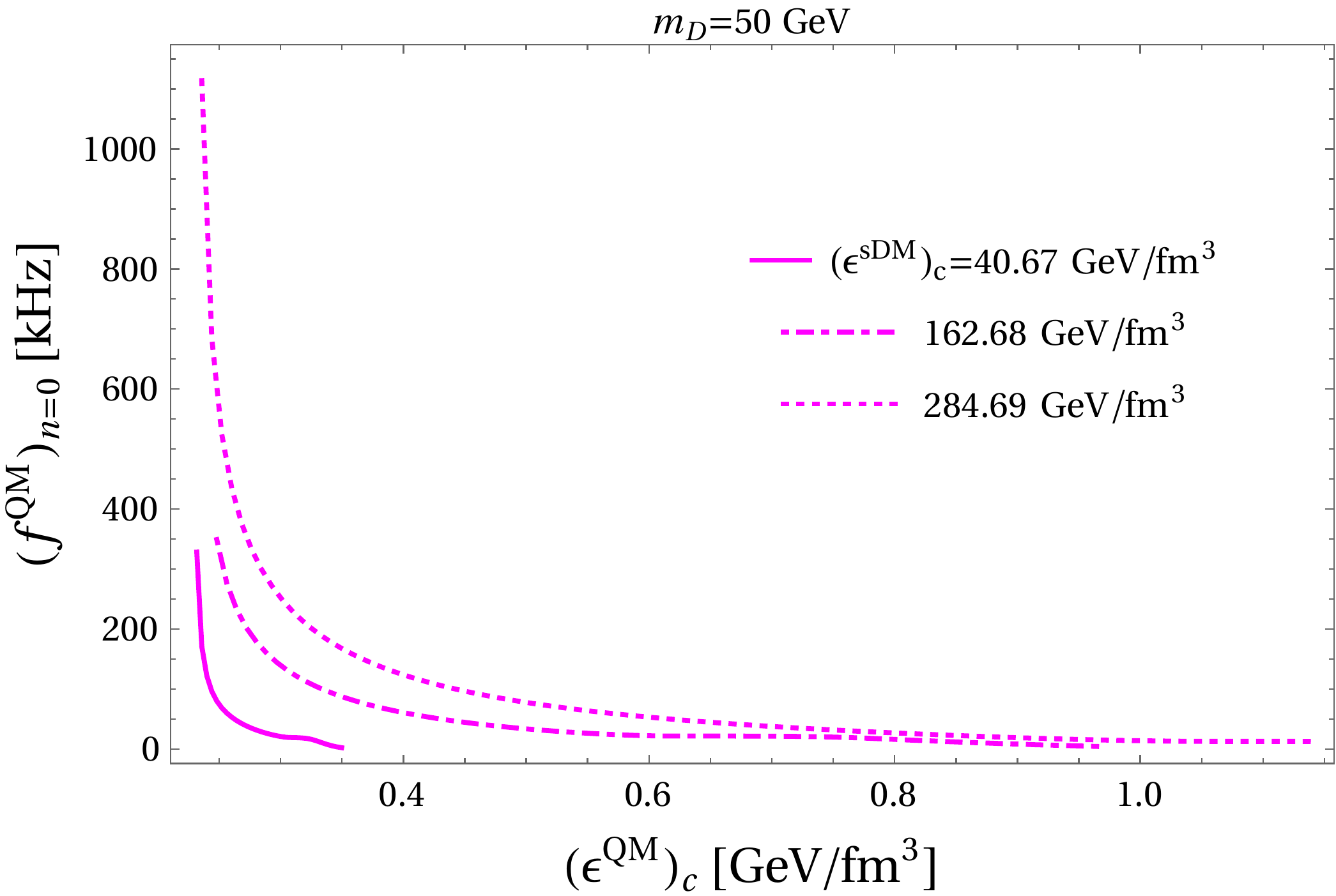}}\\\vspace{5pt}
{\includegraphics[scale=0.343]{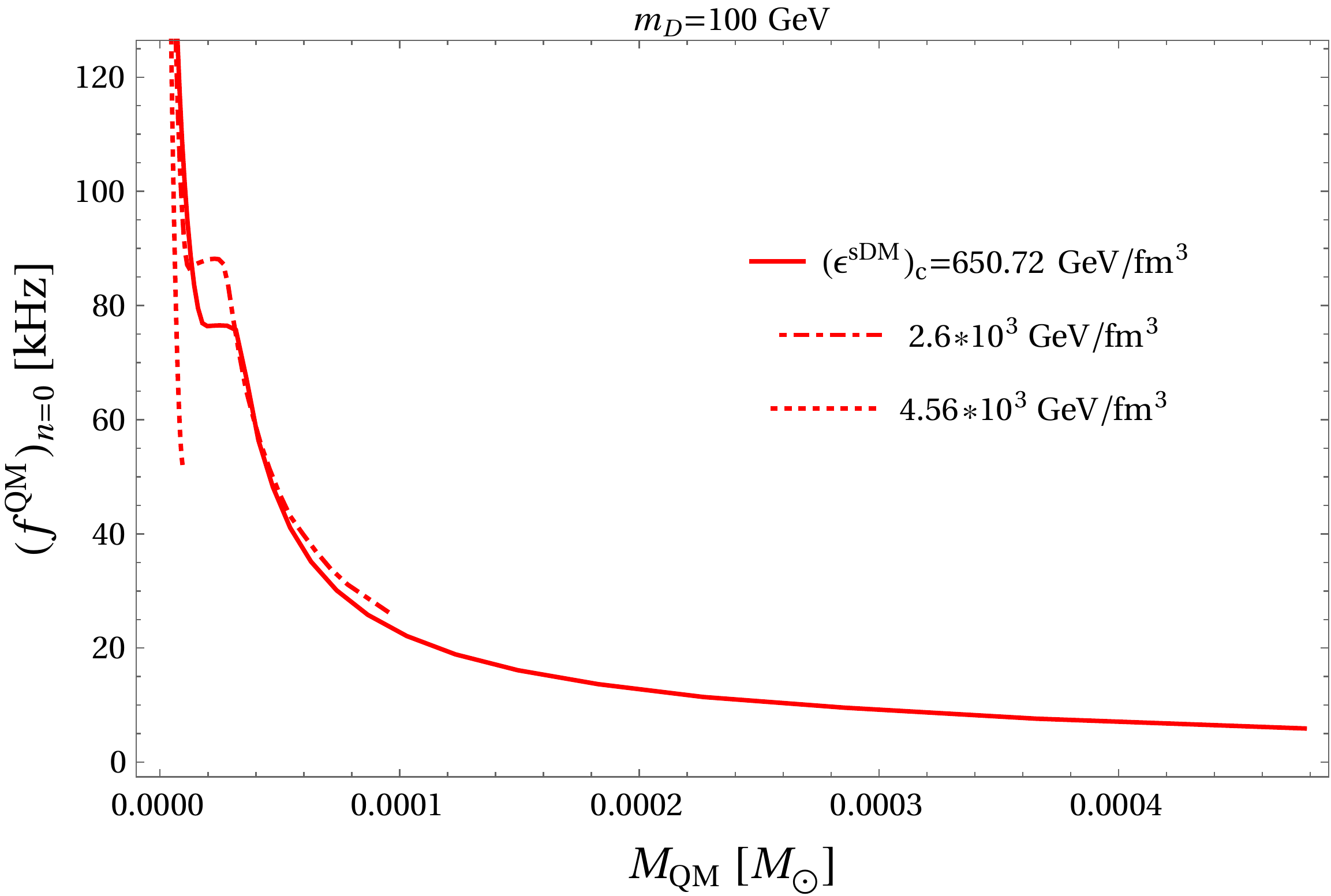}\vspace{3pt}
	  \includegraphics[width=0.588\textwidth]{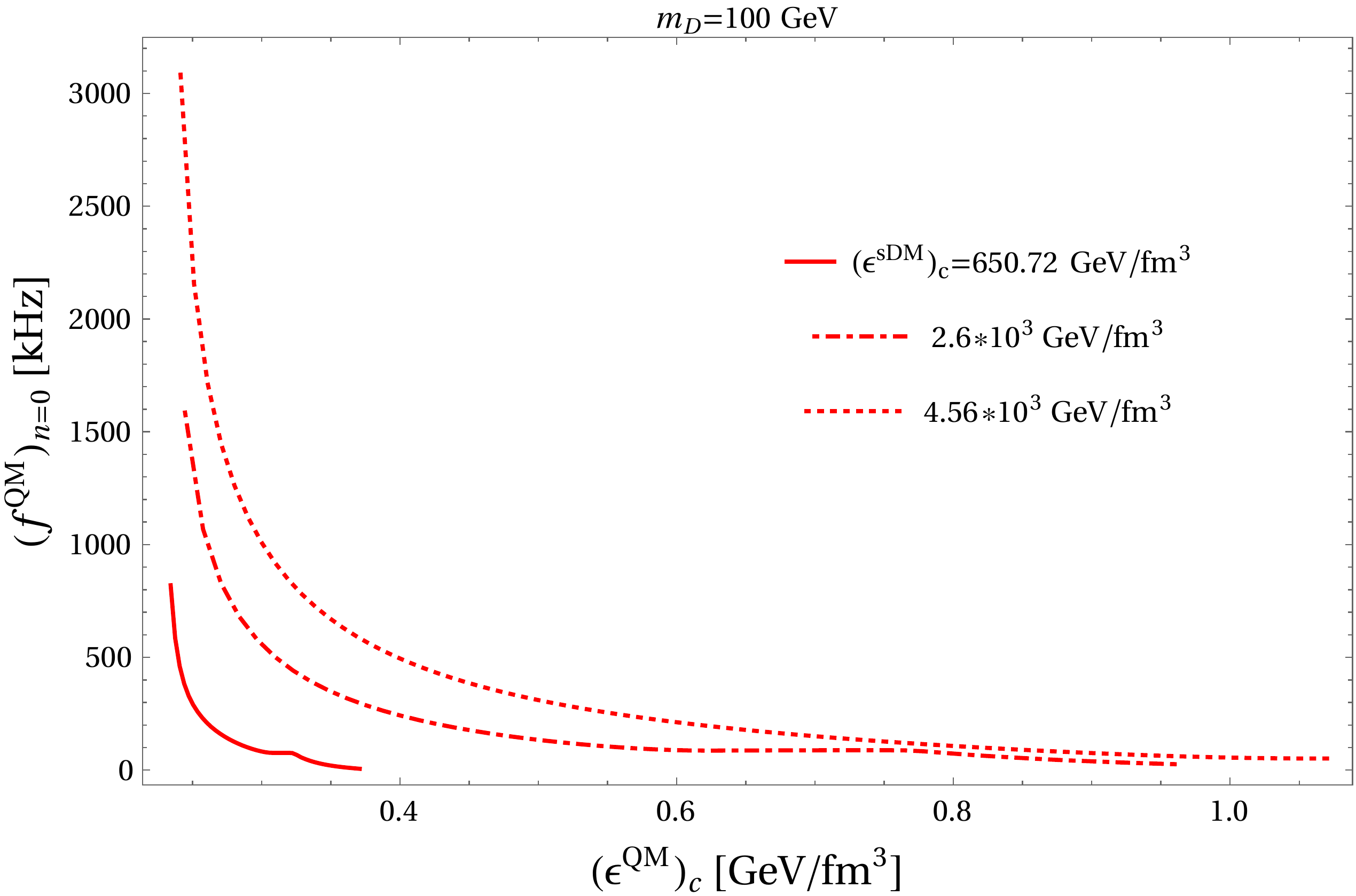}}\\\vspace{5pt}
{\includegraphics[scale=0.345]{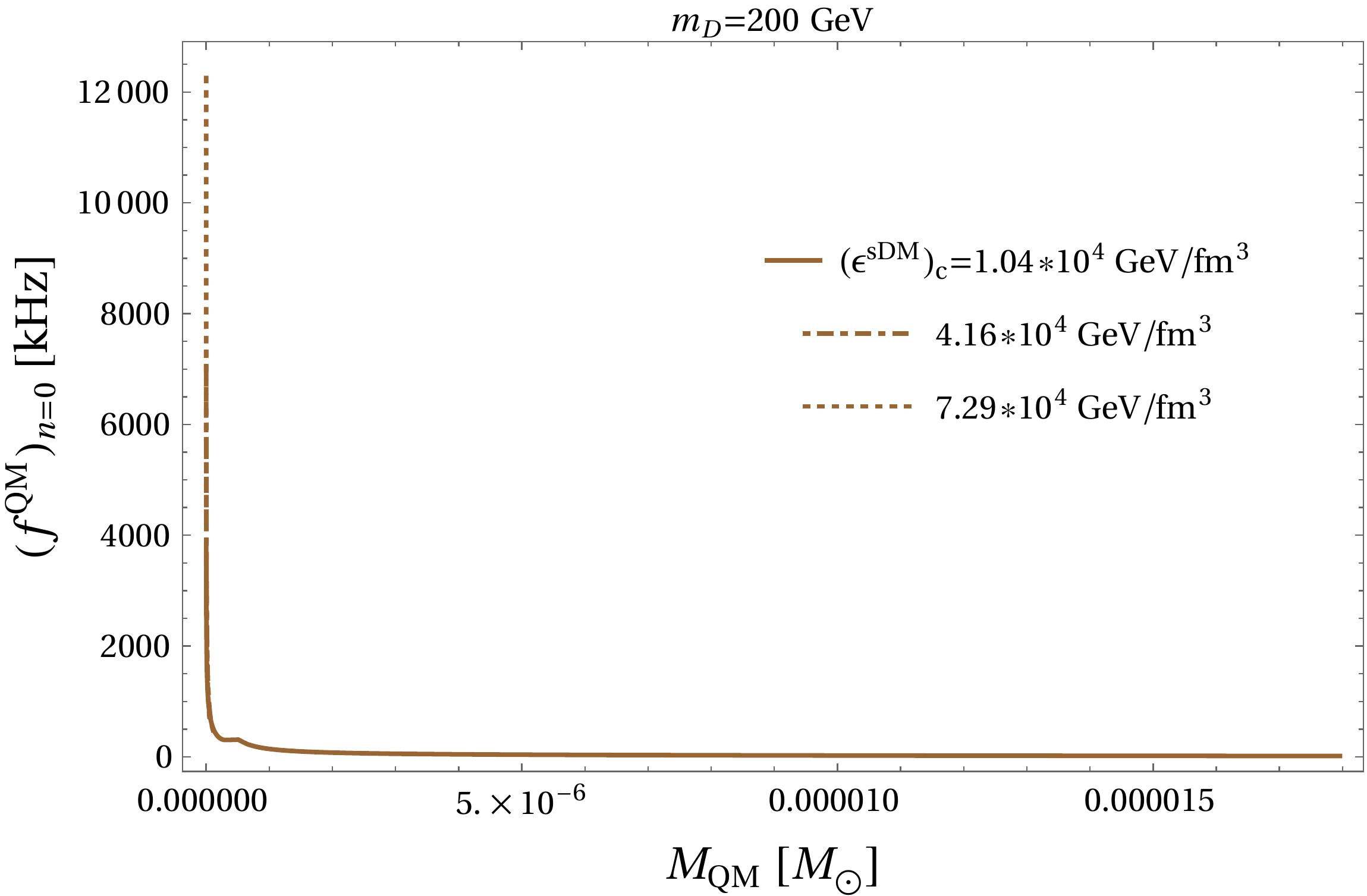}\vspace{3pt}
	  \includegraphics[scale=0.345]{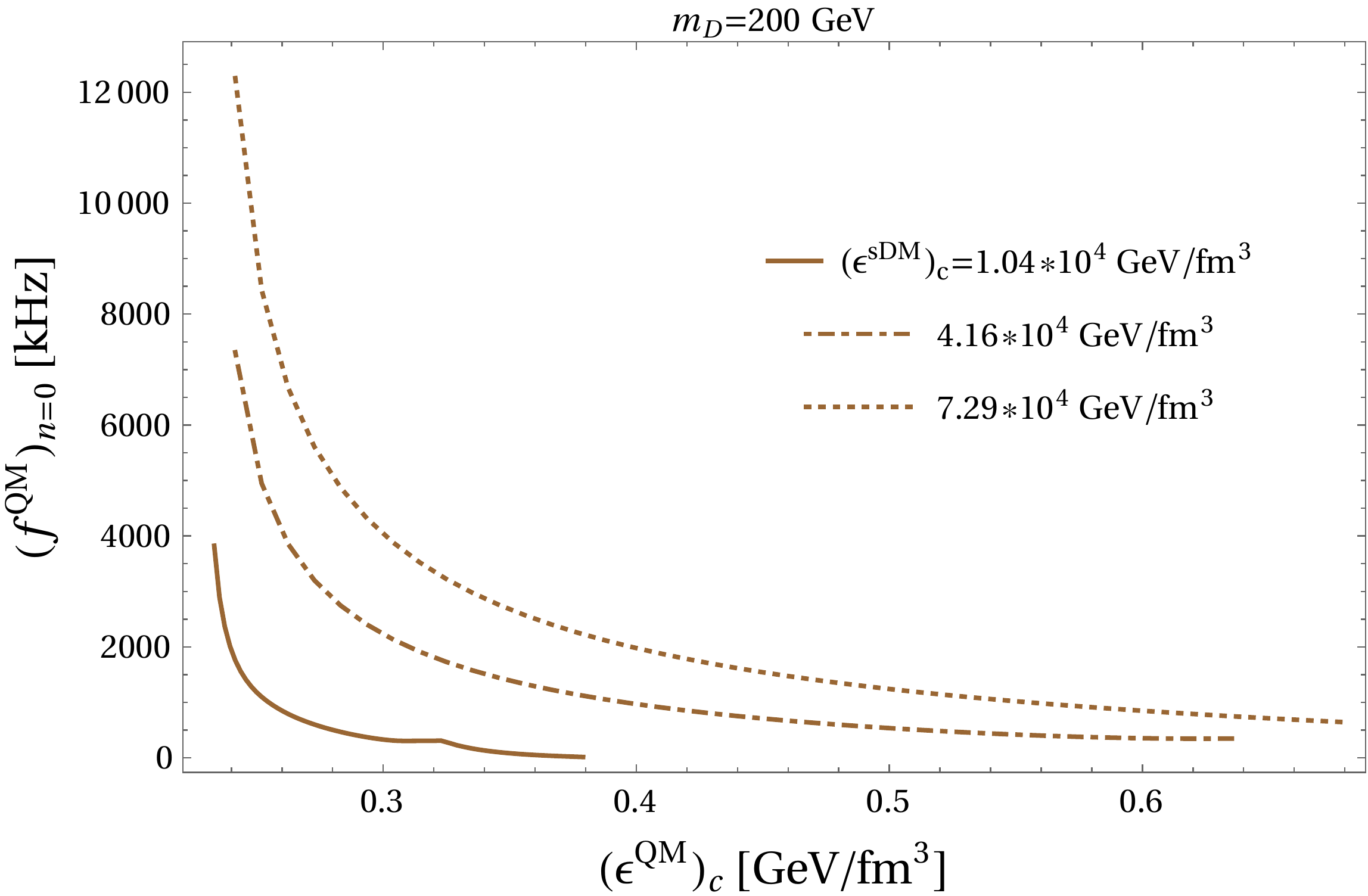}}\\
\end{adjustwidth}
\caption{\emph{Cont.} }
\label{fig:f0Mdens1000QM}

\end{figure}

\begin{figure}[H]\ContinuedFloat

\begin{adjustwidth}{-\extralength}{0cm}
\centering 

{\includegraphics[scale=0.413]{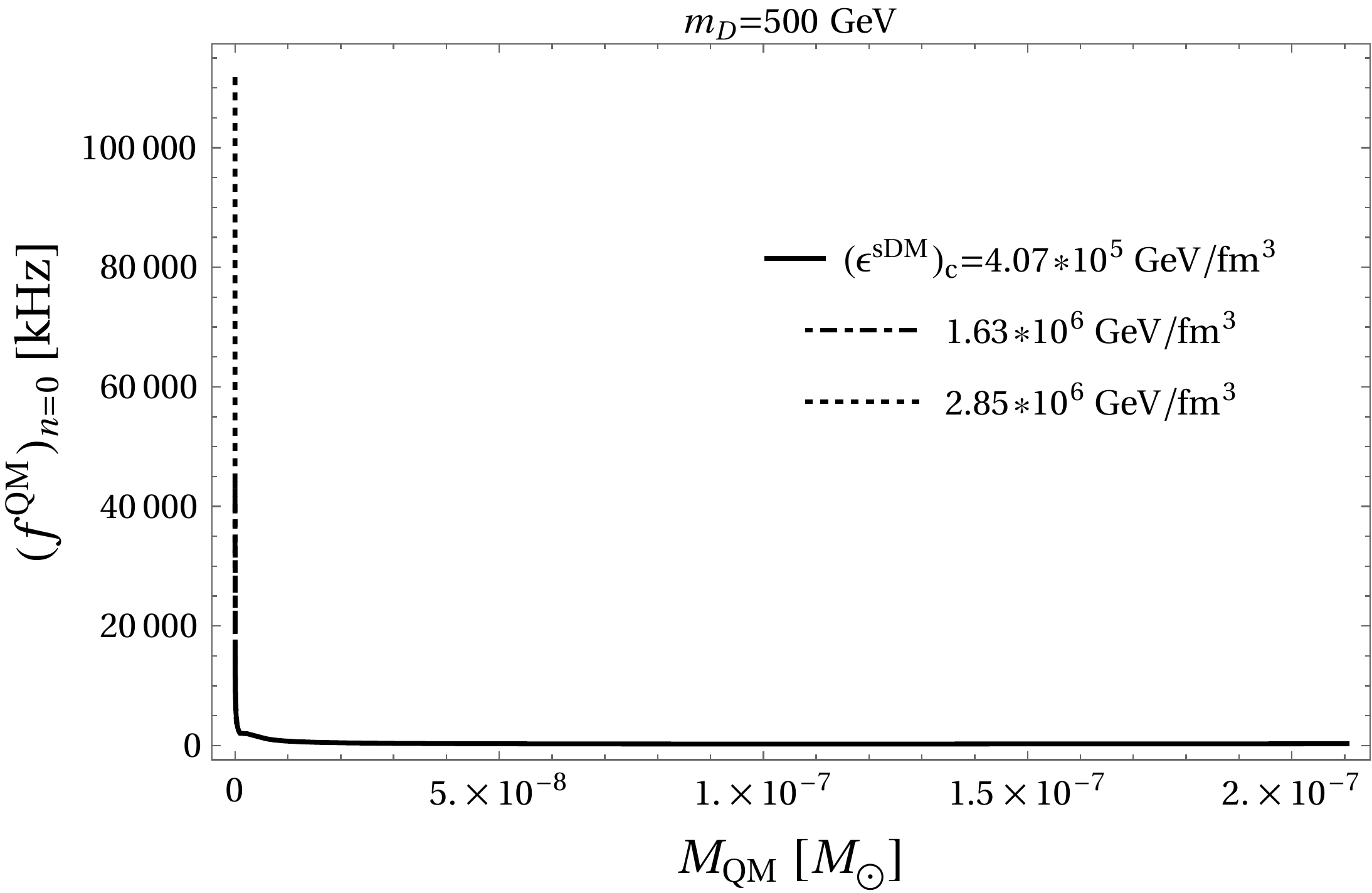}\vspace{3pt}
	  \includegraphics[scale=0.353]{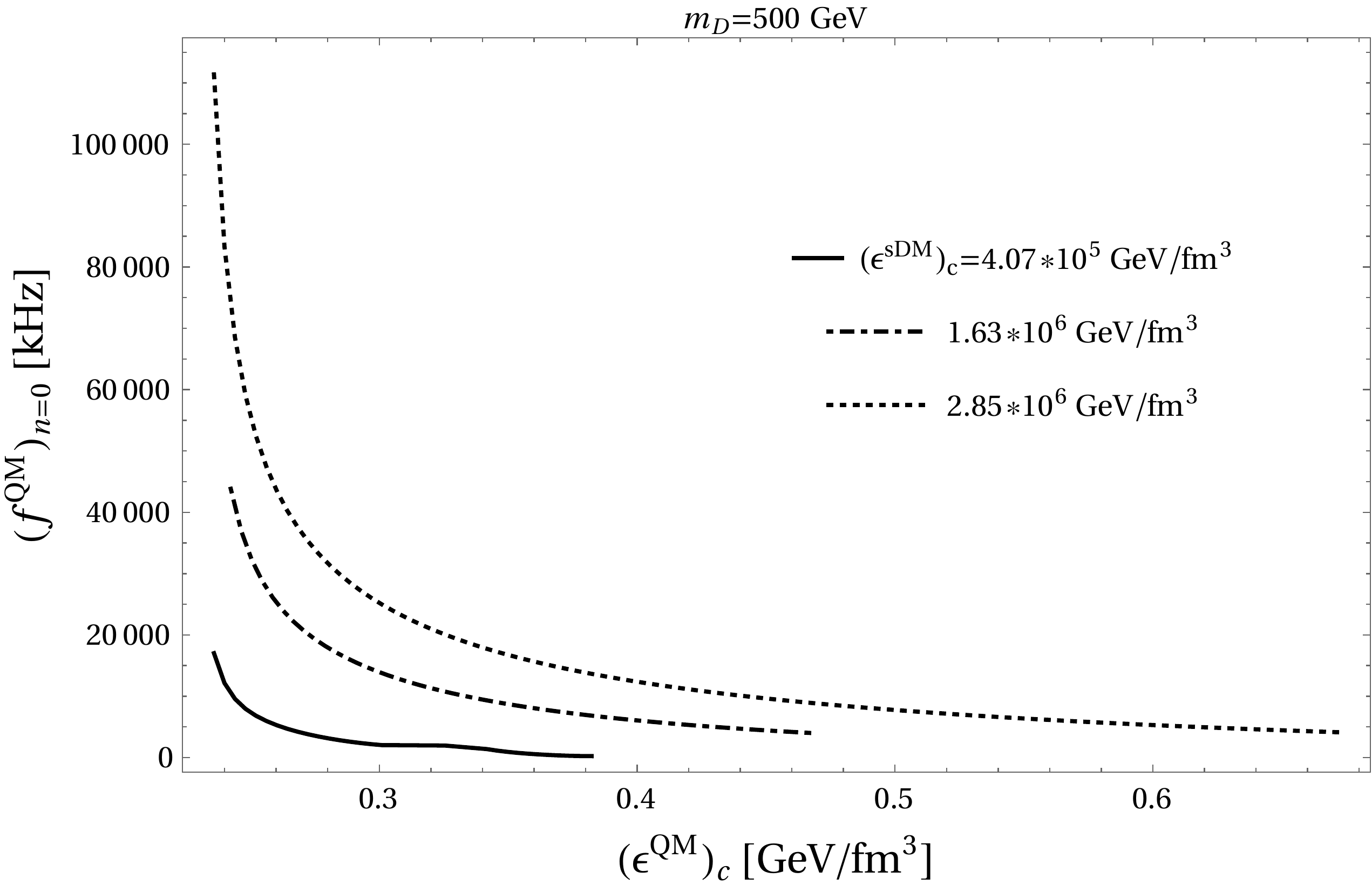}}
\end{adjustwidth}
\caption{Fundamental-mode frequency, $f_{n=0}$, versus QM core masses and central energy densities with different amounts of $s$DM for increasing values of the dark fermion masses, $m_{D}$, denoted by different colors. It can be seen that the densities of $s$DM for$m_{D}=1$ GeV have almost no effect on the stability of the corresponding QM cores. Nevertheless, as one increases $m_{D}$, the stable QM core masses are reduced to lower and lower values and require higher QM central densities. }
\label{fig:f0Mdens1000QM}

\end{figure}


In Figure \ref{fig:f0Mdens1000DM}, we show our results\endnote{Some results in this figure display a non-smooth behavior associated only with numerical limitations when using standard root-finding routines to obtain the frequencies for dimensionless central $s$DM ($\sim$$10^{-8}$) and QM (reaching $\sim$${10}^{-14}$ for $m_{D}=500$ GeV) values, i.e., being different by many orders of magnitude, very small, and sensitive to different numerical methods 
	. For these reasons, systematic differences are introduced and manifest as non-smooth curves due to variations in numerical precision when obtaining the dimensionless frequencies with values around $10^{-4}$ for large $m_{D}$, which otherwise would require very time-consuming computations. It should be noted that the same root-finding routines work very well when obtaining the other results shown in this work. In future studies, we propose 
	 to use improved theoretical and numerical approaches.} after solving the coupled radial pulsation Equations (\ref{Rad1}) and (\ref{Rad2}), assuming an oscillating DM core with boundary conditions (\ref{BC1}) and (\ref{BC2}) for different central energy densities of QM. In correspondence with the results of Figure \ref{fig:MRdens1000DM} for the cases $m_{D}=1,10$ GeV, only a small family of DM stars survived the radial oscillation analysis for low mass stars. These DM stars increase their stability as long as one increases the QM component. The qualitative behavior resembles that of a strange star. This occurs due to the high QM central energy densities compared to the DM ones, with the QM component dominating the stability of the admixed star. On the other hand, the cases with $m_{D}=50, 100, 200, 500$ GeV display the standard behavior of pure $y=10^{3}$ DM stars due to the very high DM central energy densities. Interestingly, the same phenomenon of increasing stability for higher central QM energy densities occurs in all these cases. The physical picture indicates that low QM central energy densities support a small subset of DM stars against gravitational collapse. As we increase the central energy densities, the admixed star supports higher and higher central DM energy densities. 

\begin{figure}[H]

\begin{adjustwidth}{-\extralength}{0cm}
\centering 

{\includegraphics[scale=0.423]{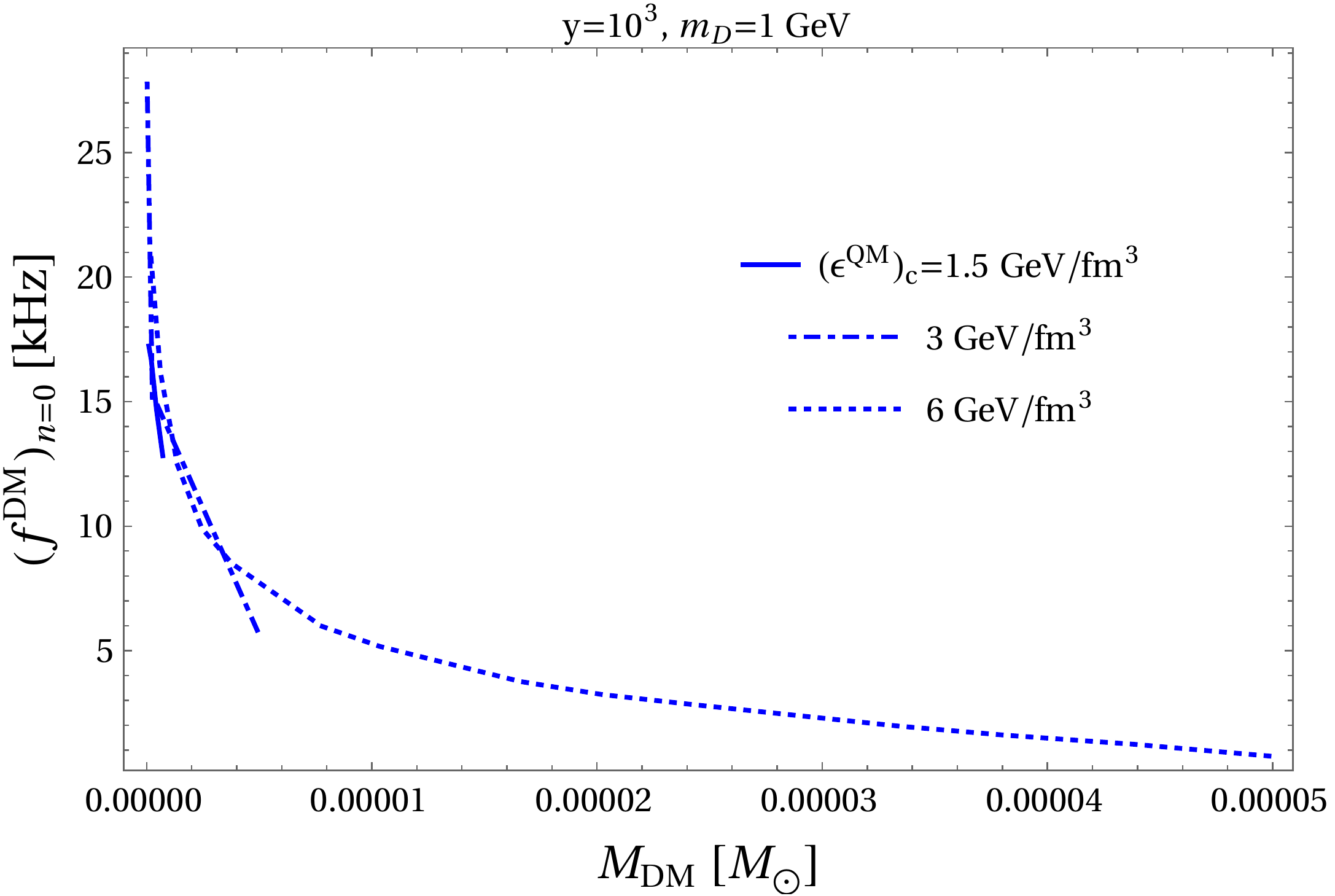}\vspace{3pt}
	  \includegraphics[scale=0.433]{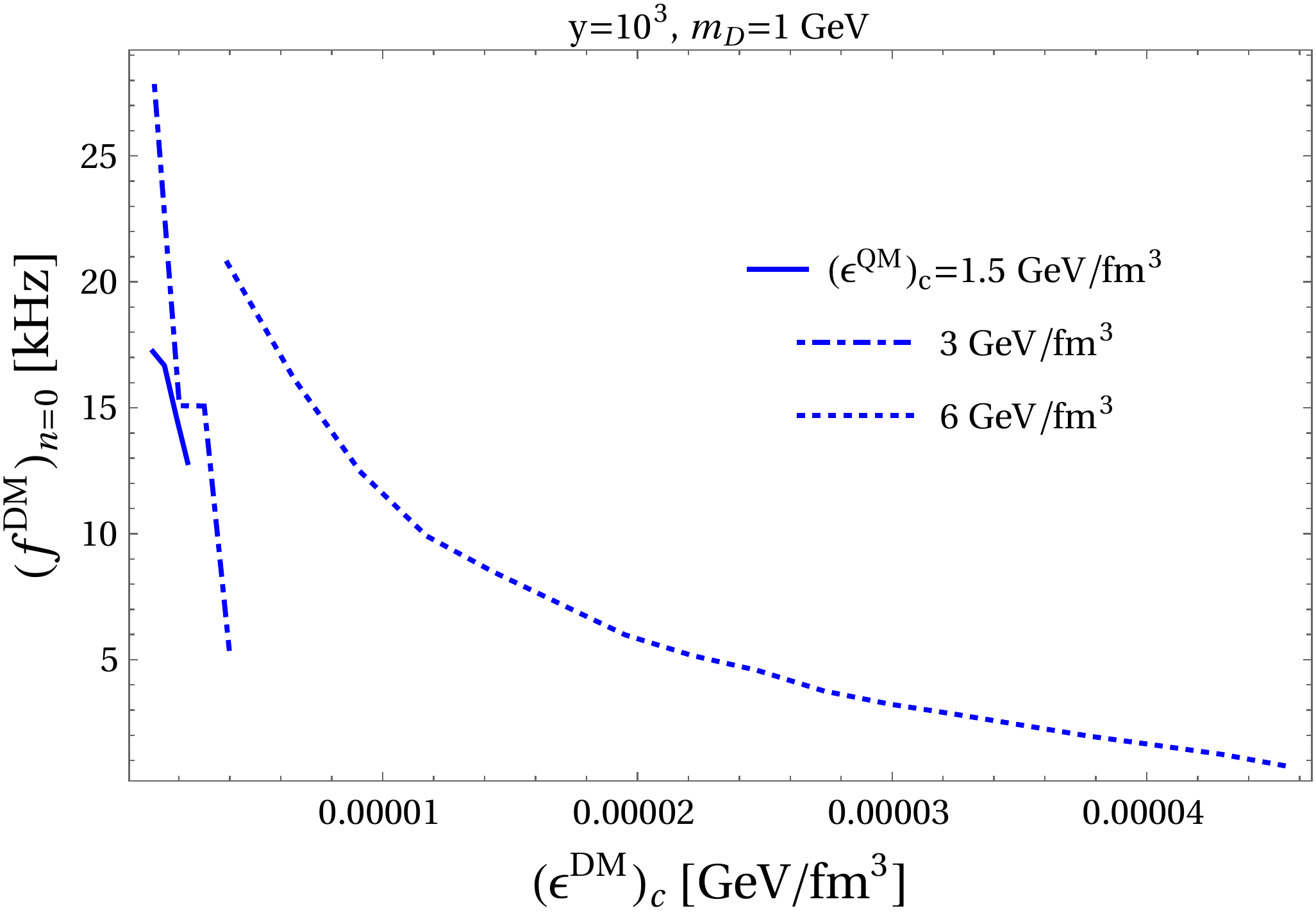}}\\
\end{adjustwidth}
\caption{\emph{Cont.}}
\label{fig:f0Mdens1000DM}

\end{figure}

\begin{figure}[H]\ContinuedFloat

\begin{adjustwidth}{-\extralength}{0cm}
\centering 

{\includegraphics[scale=0.37]{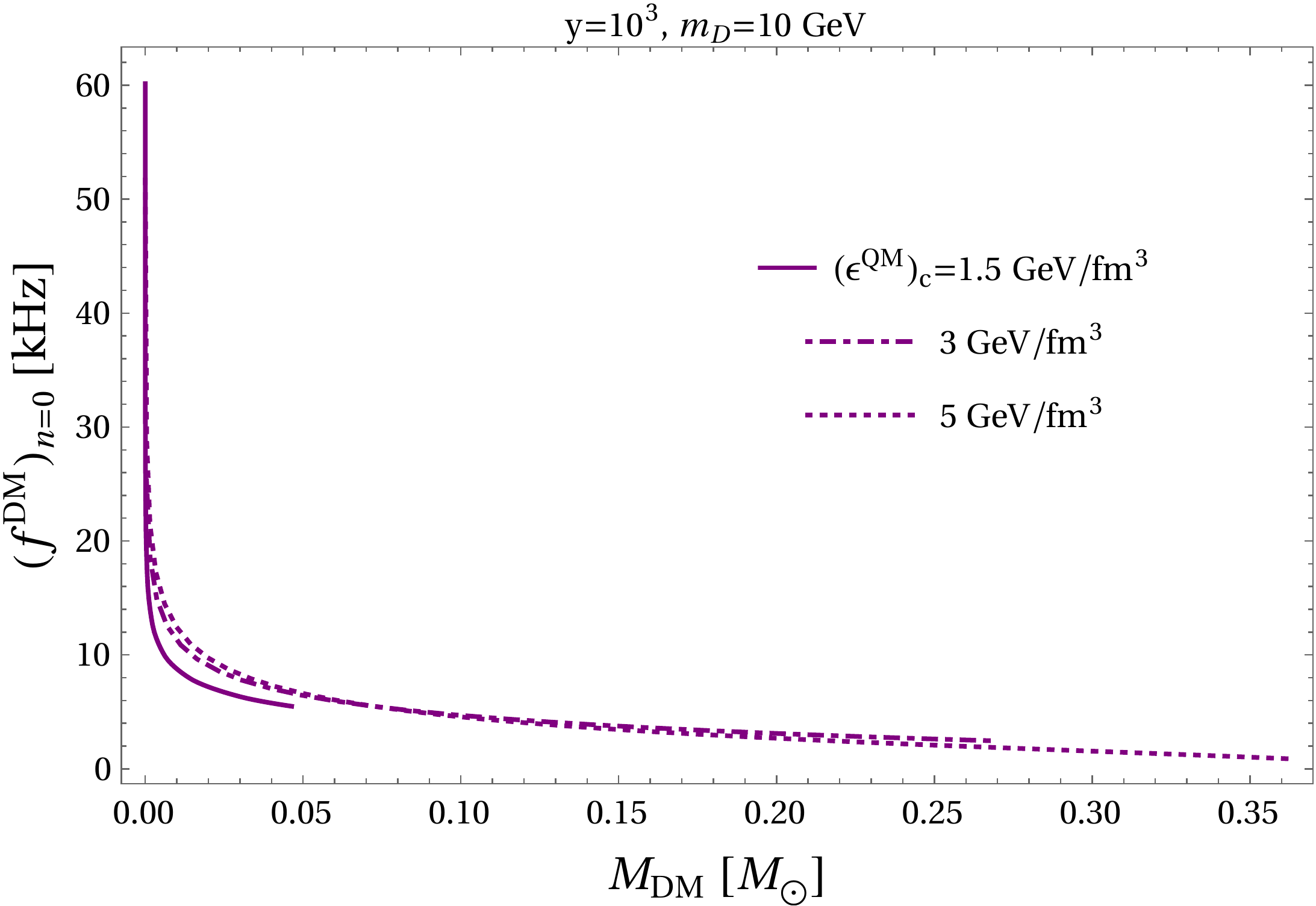}\vspace{3pt}
	  \includegraphics[scale=0.38]{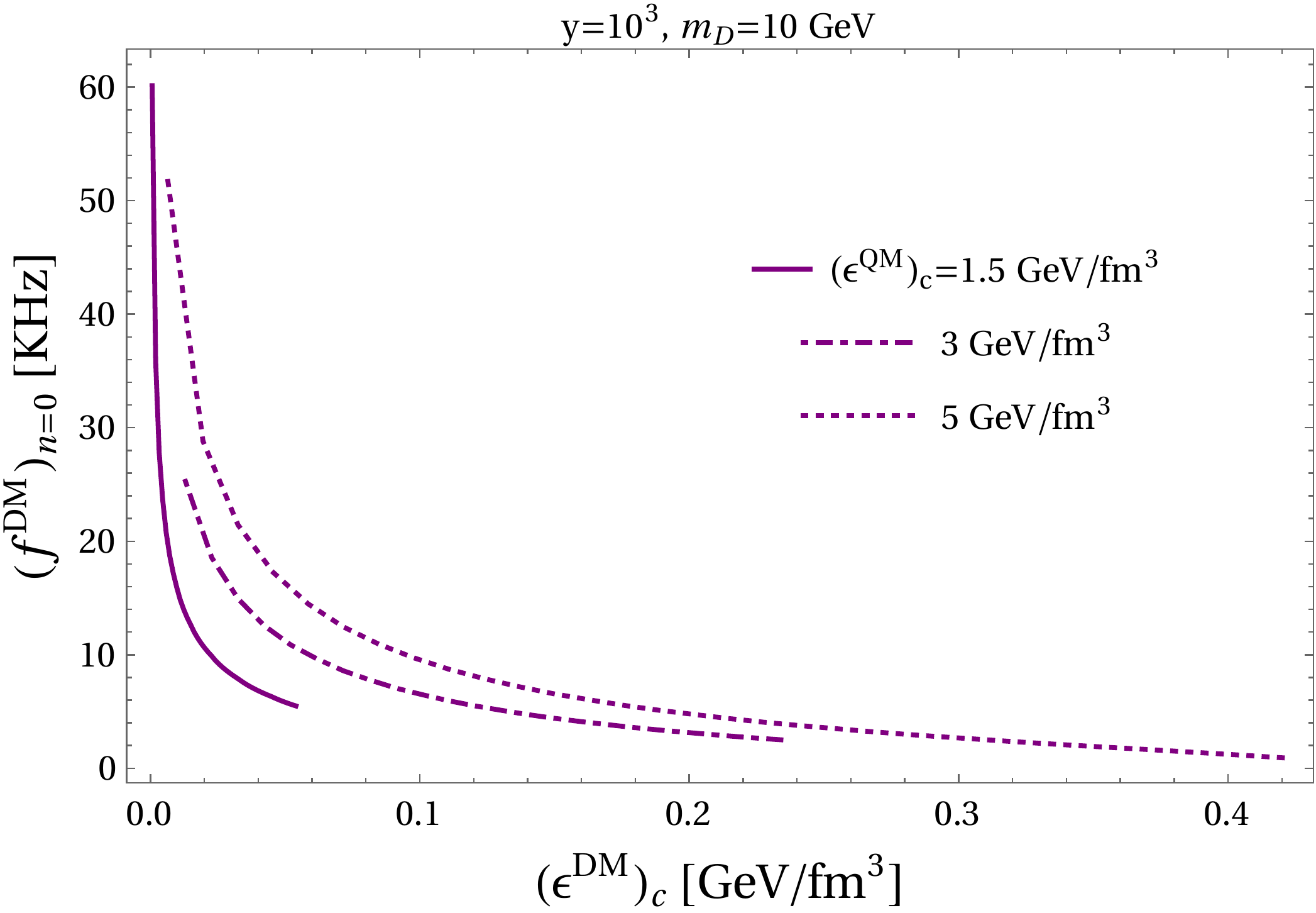}}\\\vspace{5pt}
{\includegraphics[scale=0.364]{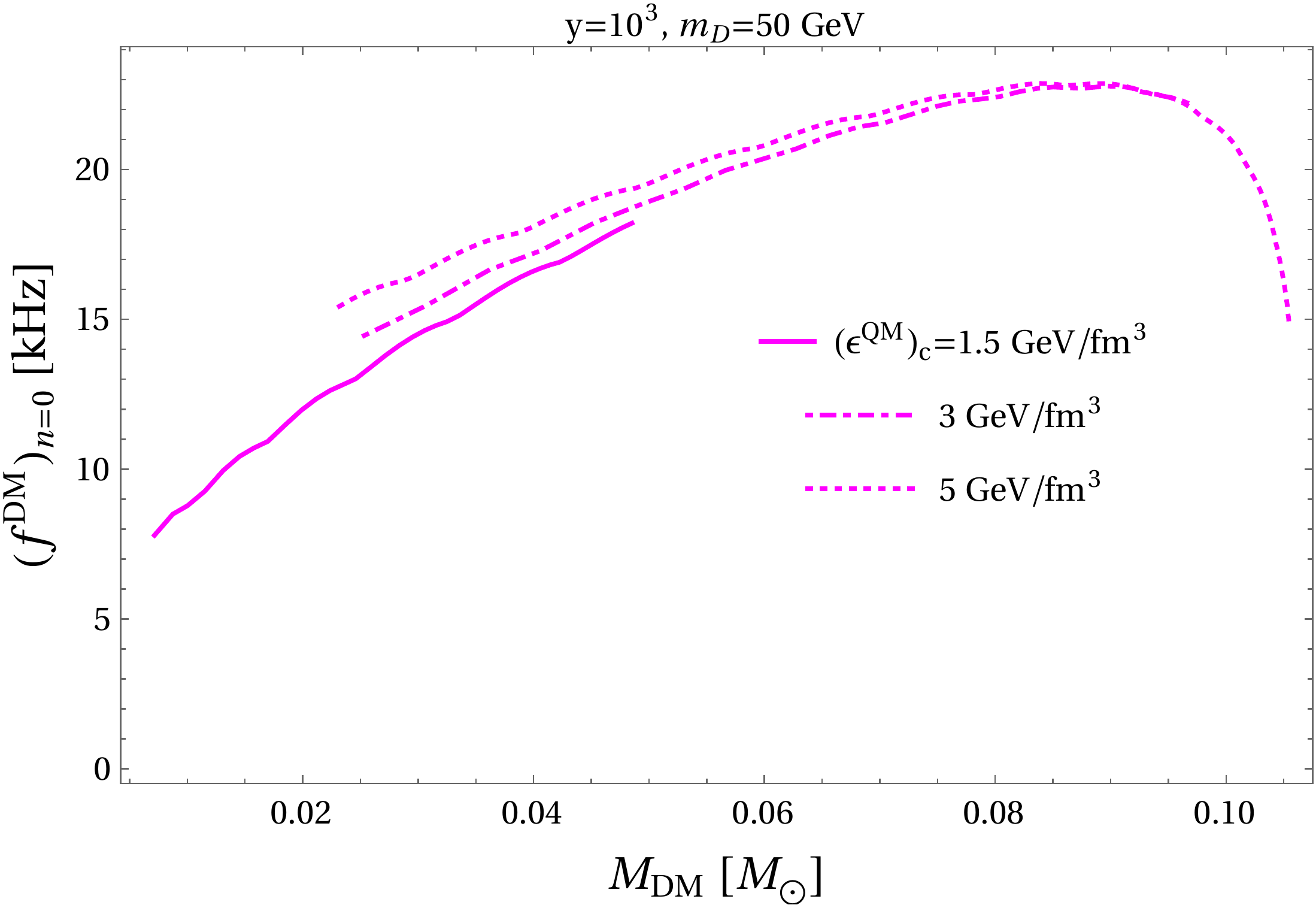}\vspace{3pt}
	  \includegraphics[scale=0.38]{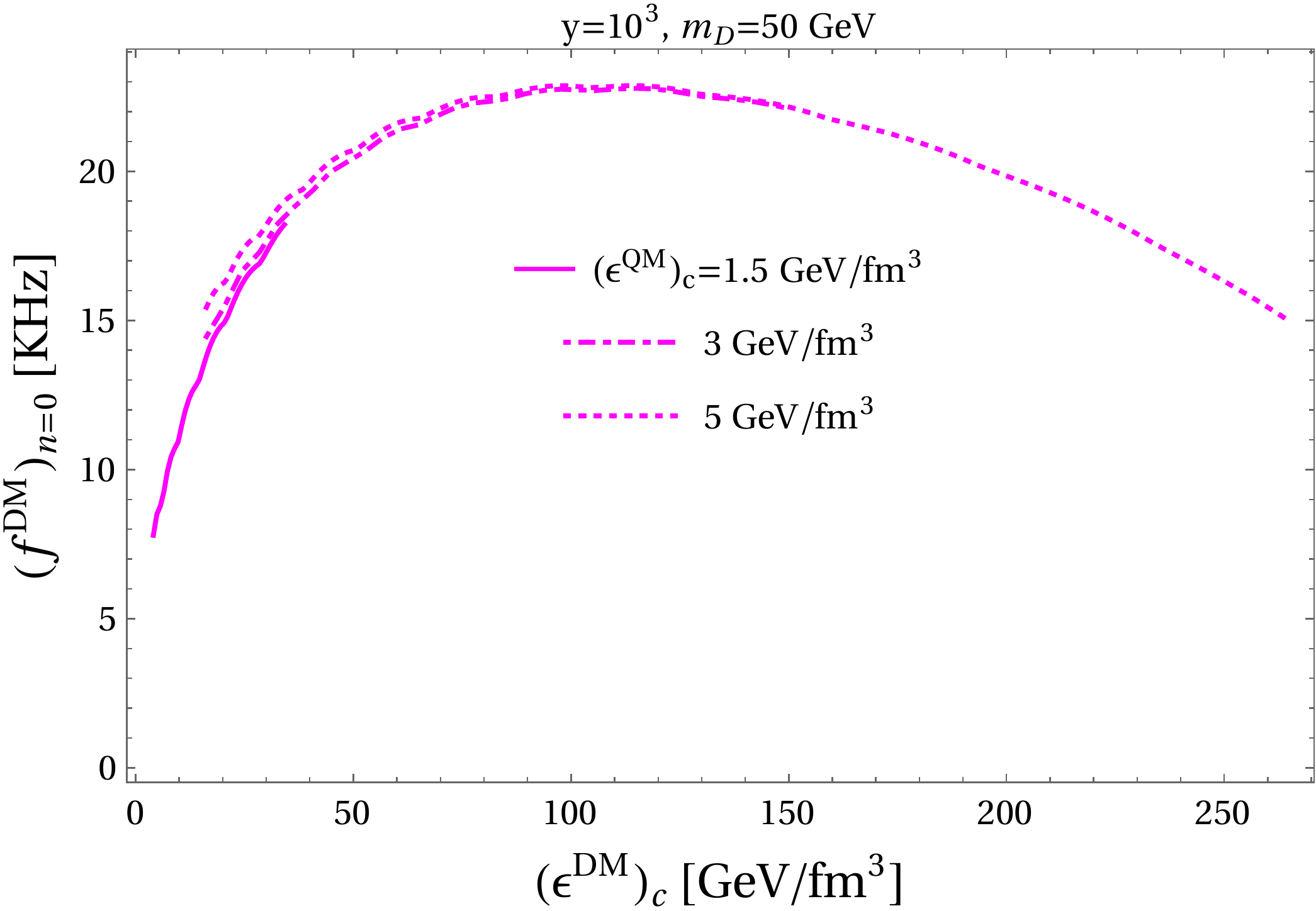}}\\\vspace{5pt}
{\includegraphics[scale=0.345]{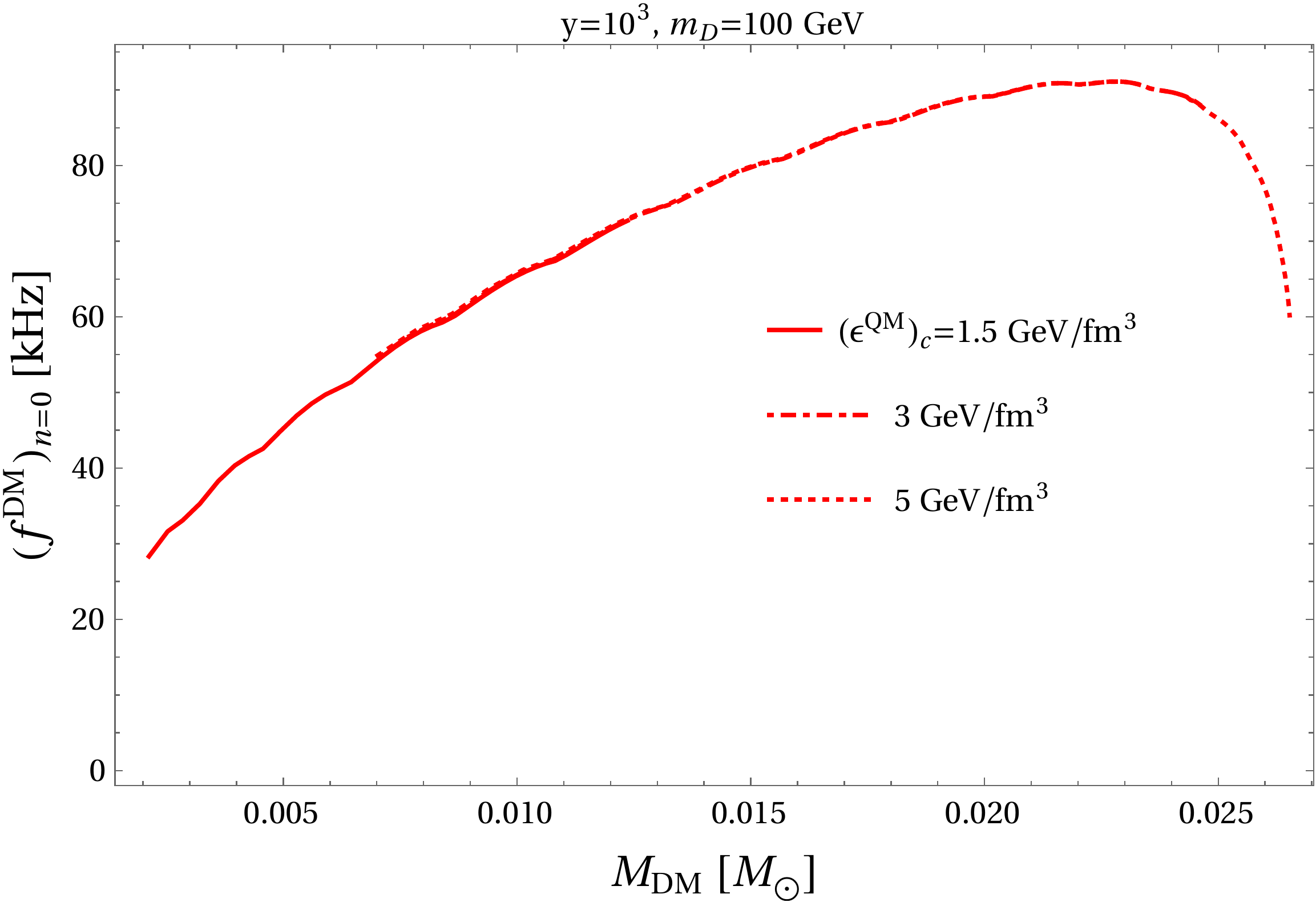}\vspace{3pt}
	  \includegraphics[scale=0.315]{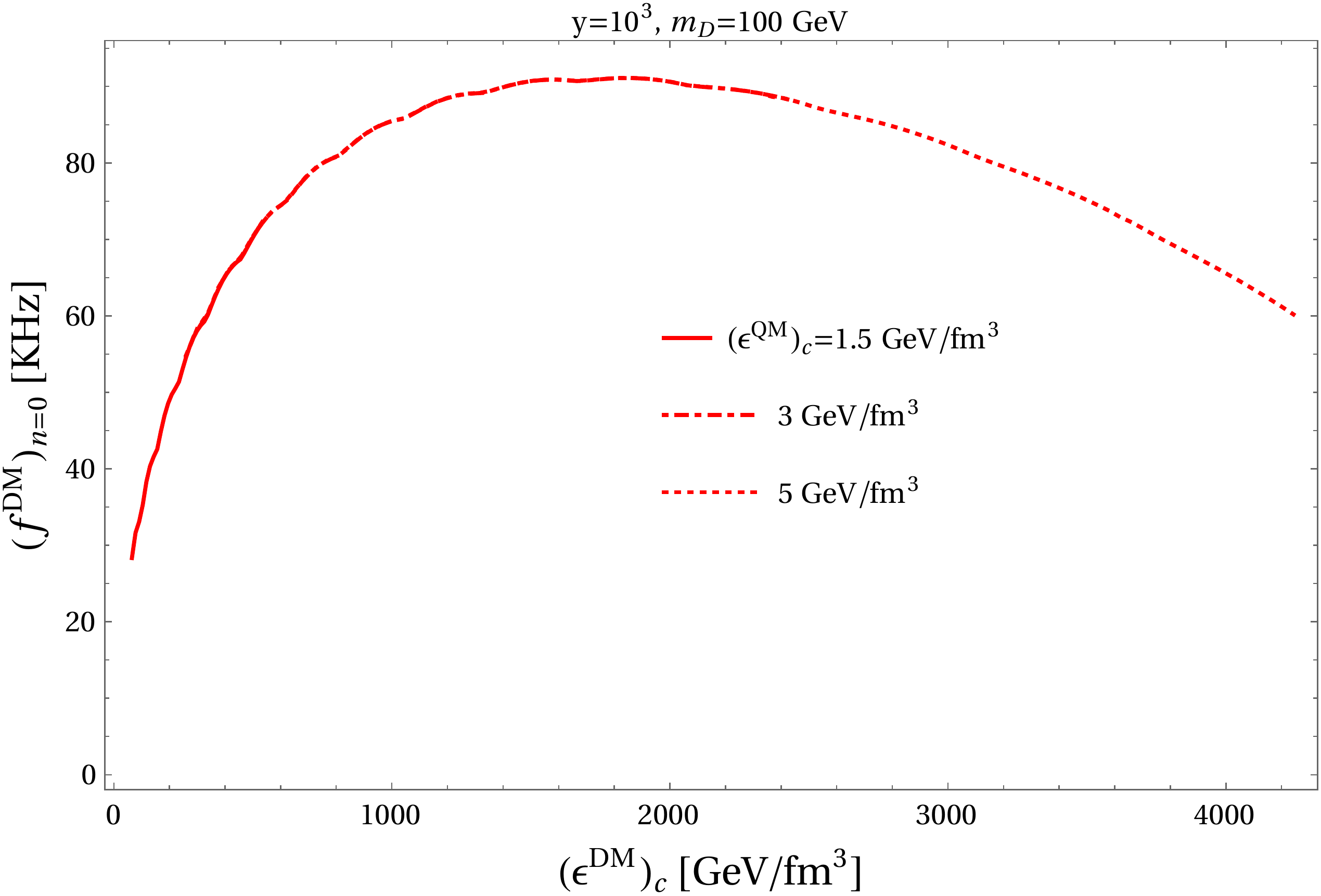}}\\\vspace{5pt}
{\includegraphics[scale=0.38]{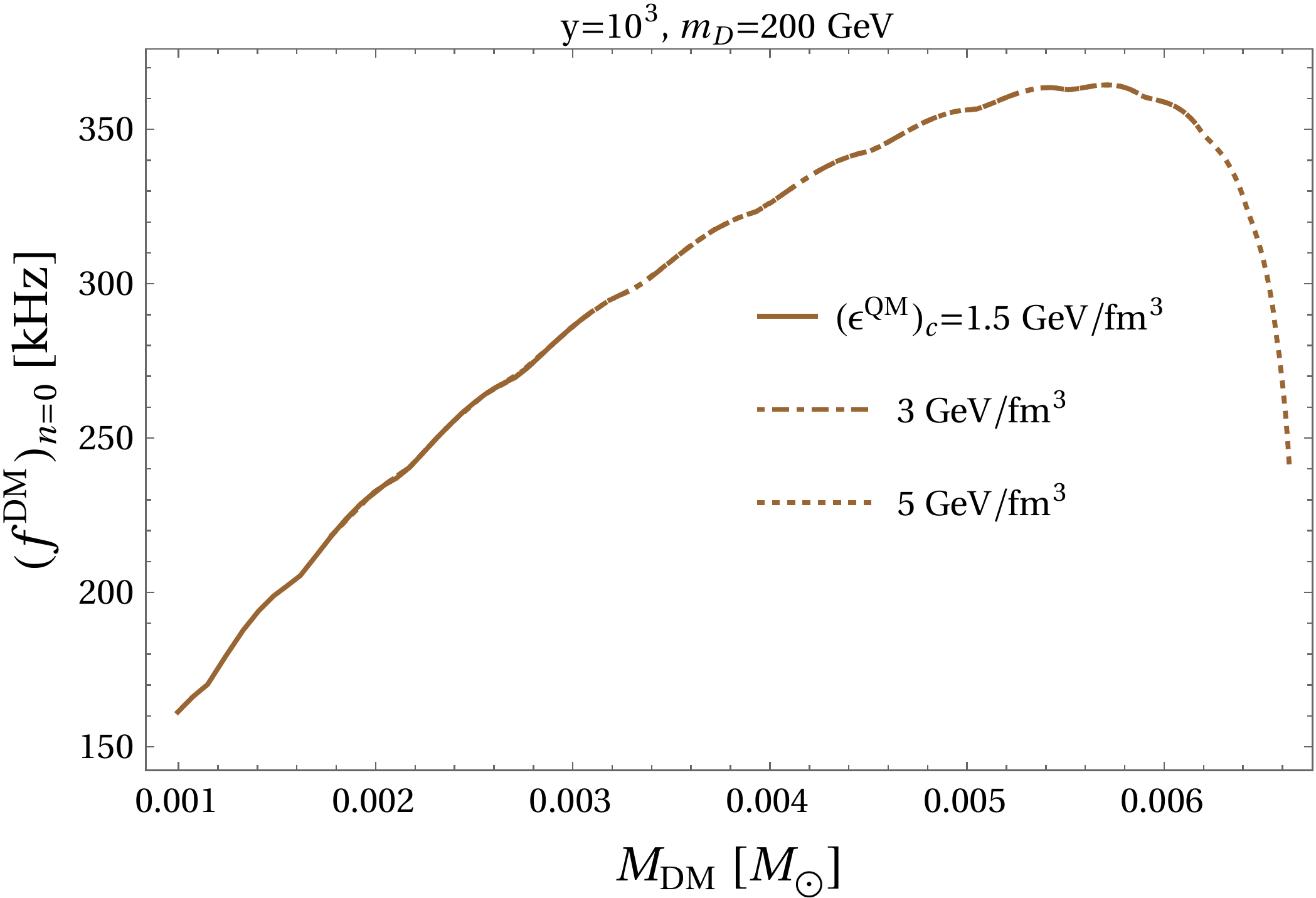}\vspace{3pt}
	  \includegraphics[scale=0.398]{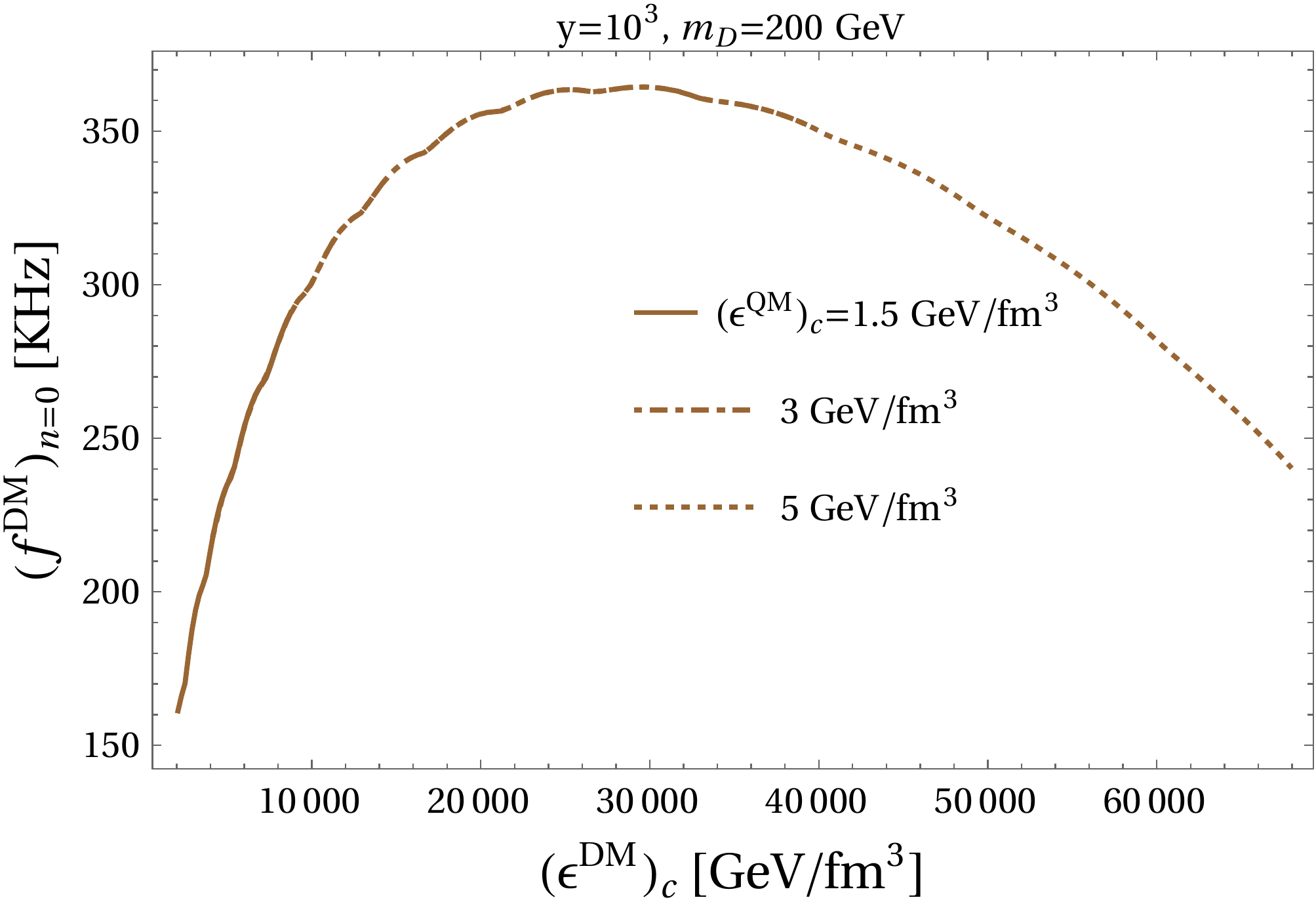}}\\
\end{adjustwidth}
\caption{\emph{Cont.}}
\label{fig:f0Mdens1000DM}

\end{figure}

\begin{figure}[H]\ContinuedFloat

\begin{adjustwidth}{-\extralength}{0cm}
\centering 

{\includegraphics[scale=0.41]{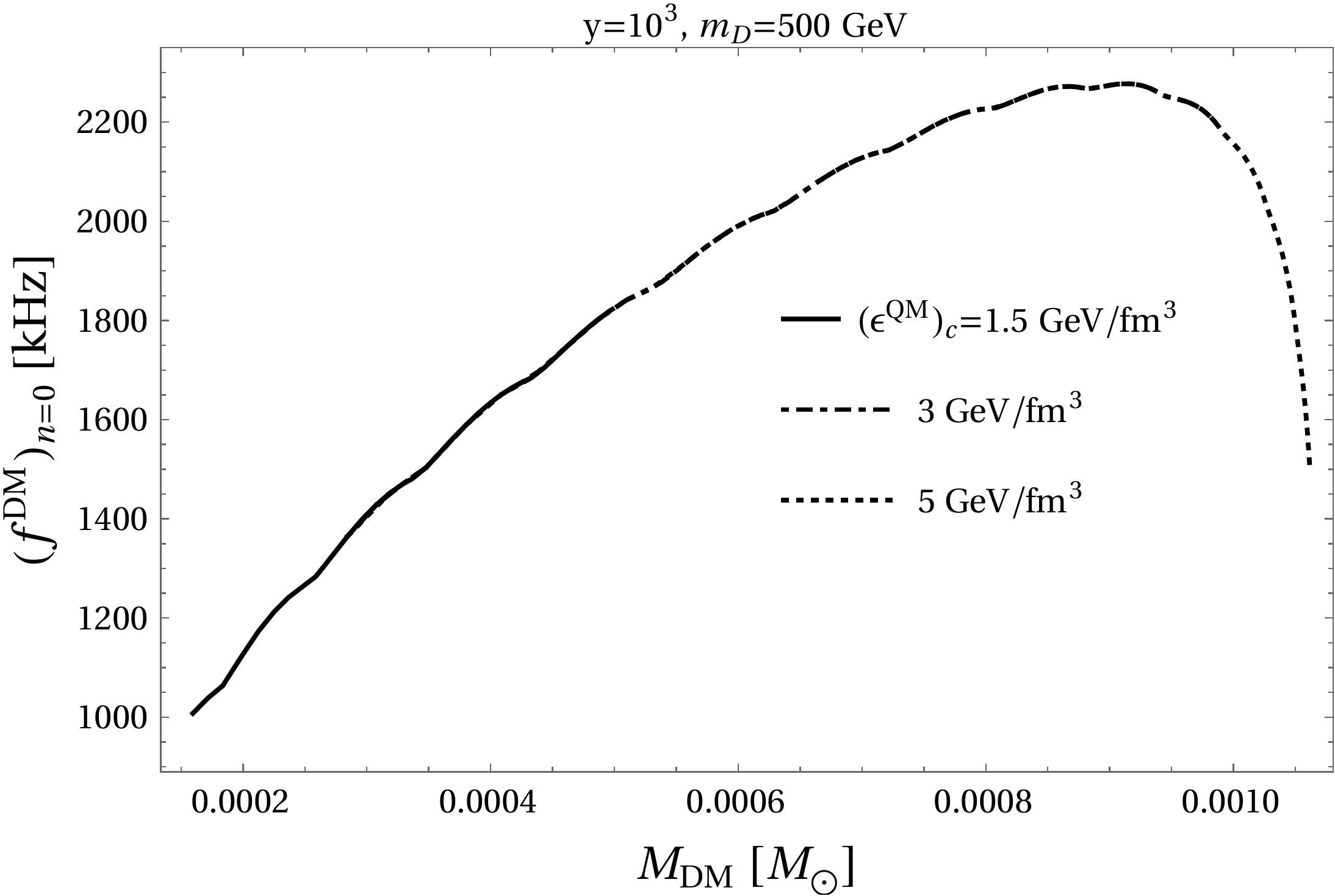}\vspace{3pt}
	  \includegraphics[scale=0.40]{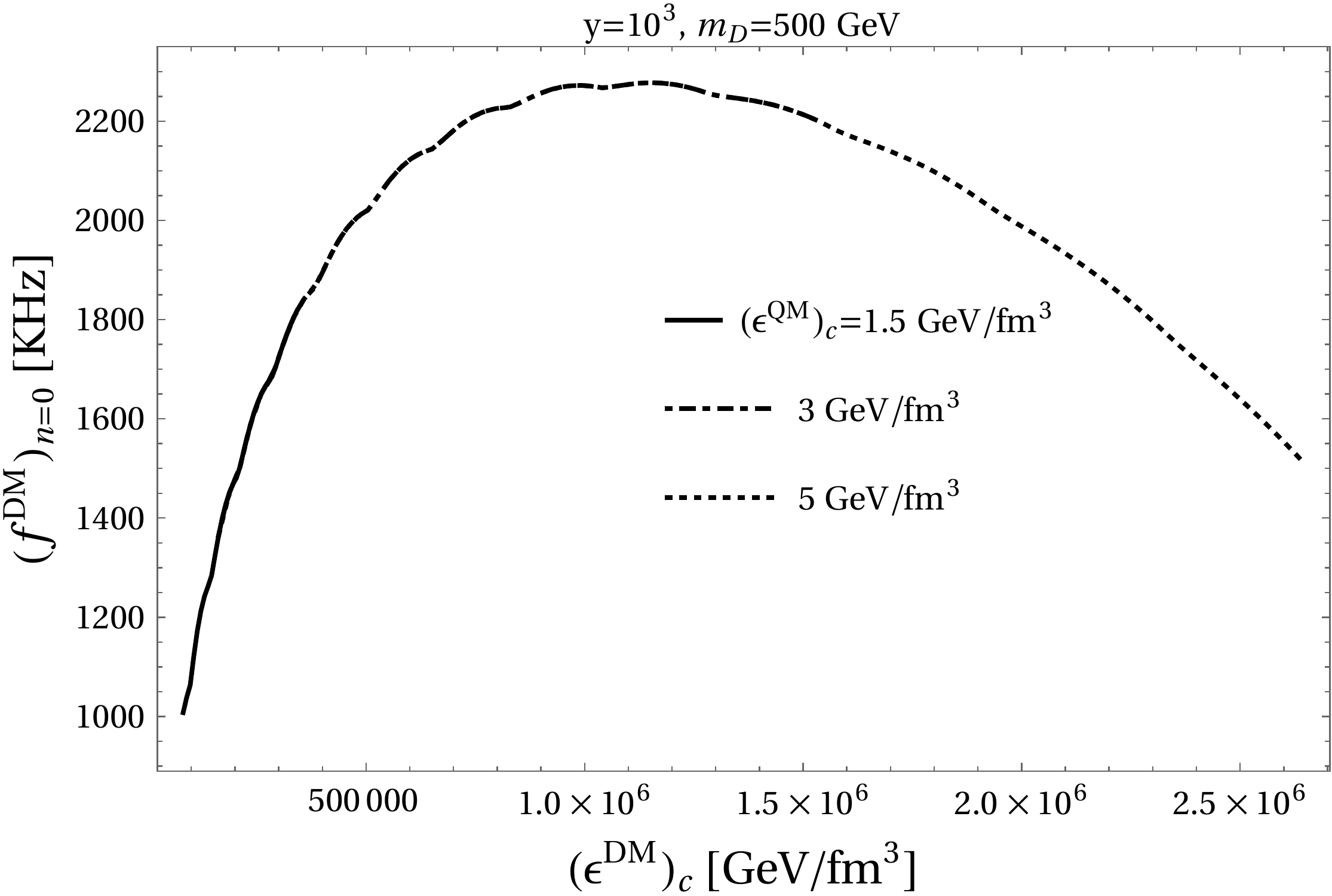}}
\end{adjustwidth}
\caption{Same notation as in Figure \ref{fig:f0Mdens1000QM}, but now for the oscillating core composed of $s$DM corresponding to increasing $m_{D}$ (with different colors) and increasing values of central QM densities. Notice the changing qualitative behavior when QM densities dominate over $s$DM for low $m_{D}=1,10$ GeV but the opposite happening for larger $m_{D}$, where QM only allows for more stable $s$DM cores.}
\label{fig:f0Mdens1000DM}

\end{figure}

\section{Summary and Outlook}
  \label{sec:conclusion}

We have studied the stability and main global features of strange quark stars admixed with fermionic dark matter in a large parameter space, allowing dark fermion masses from $1$ to $500$ GeV and considering weakly and strongly self-interacting dark matter. For simplicity, cold quark matter was described within the MIT bag model with $B^{1/4}=145$ MeV which, in the one-fluid case, produces strange quark stars.

After solving the two-fluid TOV equations, we computed the associated stellar structure. We found that, depending on $m_{D}$ and the interaction parameter $y$, some of the obtained QM and DM stars display significant modifications of their stellar masses and radii, whereas others show no change at all. Furthermore, some of the DM stars display a self-bound-like behavior in the mass--radius diagram. In most situations, the central QM and DM densities are increased by the presence of the other component in the admixed~star.
  
For the radial pulsation analysis, a full general-relativistic two-fluid Sturm--Liouville-like problem should, in principle, be solved. Instead, inspired by the way one usually solves the two-fluid TOV equations by separating the total pressure and energy density into QM and DM components, we developed a framework where we separate the total Lagrangian variables entering the oscillation equations into QM and DM contributions. This method allowed us to solve the problem, assuming that we disturb only one component and the other is affected only indirectly.

Our calculations indicate that the static stability criterion $\partial{M}_{\rm QM/DM}/\partial{\epsilon^{\rm QM/DM}_{c}}\geq0$ alone might produce misleading and incomplete results when applied to two-fluid stars. We found that, in the case of QM stars admixed with DM, predominantly in the case of $y=0.1$, only very small QM stellar masses are dynamically stable leading to {\it dark strange planets} and {\it dark strangelets}. On the other hand, DM stars are mainly affected when small values of $m_{D}$ are considered, since larger dark fermion masses induce ultra-dense cores for which the QM contribution is almost negligible. 

Although our results are still very sensitive, both to the dark fermion mass $m_D$ and the kind of self-interaction involved in the dark sector, whose scale is encoded in the dimensionless variable $y$, there is hope that the parameter space can be dramatically constrained by gravitational wave events, as discussed recently in Ref. \cite{Wystub:2021qrn}. 

In a future publications, we plan to refine our description using an equation of state obtained from perturbative QCD \cite{Fraga:2013qra}. It would also be interesting to explore the effects of adding a nuclear mantle to our dark strange planets.
  
%
%



\authorcontributions{J.C.J. and E.S.F. contributed significantly to this work. All authors have read and agreed to the published version of the manuscript.}

\funding{This work is a part of the project INCT-FNA Proc. No. 464898/2014-5. 
J.C.J. acknowledges the support of Fundação de Amparo à Pesquisa do Estado de São Paulo (FAPESP), Grants No. 2020/07791-7 and 2018/24720-6. E.S.F. is partially supported by Coordenação de Aperfeicoamento de Pessoal de Nível Superior (CAPES), Finance Code 001; Conselho Nacional de Desenvolvimento Científico e Tecnológico (CNPq); Fundação de Amparo à Pesquisa do Estado do Rio de Janeiro~(FAPERJ).}

\conflictsofinterest{The authors declare no conflict of interest.} 

\begin{adjustwidth}{-\extralength}{0cm}

\printendnotes[custom] 

\reftitle{References}

\end{adjustwidth}

%


\end{document}